\documentclass[twocolumn,english,prl,aps,noshowpacs]{revtex4-1}

\usepackage[LGR,T1]{fontenc}
\usepackage[utf8]{inputenc}
\usepackage{standalone}
\usepackage{babel}
\usepackage{setspace}

\usepackage{amsmath}
\usepackage{amssymb}
\usepackage{mathtools}
\usepackage{units}

\usepackage{graphicx}
\usepackage{array}
\usepackage{textcomp}
\usepackage{multirow}

\usepackage[unicode=true,pdfusetitle,
bookmarks=true,bookmarksnumbered=false,bookmarksopen=false,
breaklinks=false,pdfborder={0 0 0},pdfborderstyle={},backref=false,colorlinks=false]
{hyperref}

\usepackage[frozencache,cachedir=.]{minted}
\usepackage{lmodern}

\DeclareRobustCommand{\greektext}{%
	\fontencoding{LGR}\selectfont\def\encodingdefault{LGR}}
\DeclareRobustCommand{\textgreek}[1]{\leavevmode{\greektext #1}}
\newcommand{\lyxdot}{.} 


\begin{document}

\title{\textit{Sparse chaos} in cortical circuits}
\author{Rainer Engelken$\textsuperscript{1,2,3,4}$}
\email{re2365@columbia.edu}

\author{Michael Monteforte$\textsuperscript{4}$}
\author{Fred Wolf$\textsuperscript{3,4,5,6,7}$}
\affiliation{Department of Neuroscience, Zuckerman Institute, Columbia University,
	New York, NY, United States of America$\textsuperscript{1}$,\\
	Kavli Institute of Brain Science, Columbia University, New York, NY,
	United States of America$\textsuperscript{2}$,\\
	Göttingen Campus Institute for Dynamics of Biological Networks, Göttingen,
	Germany$\textsuperscript{3}$,\\
	Max Planck Institute for Dynamics and Self-Organization, Göttingen,
	Germany$\textsuperscript{4}$,\\
	Faculty of Physics, Georg-August-Universität Göttingen, Göttingen,
	Germany$\textsuperscript{5}$,\\
	Cluster of Excellence "Multiscale Bioimaging: from Molecular Machines to Networks of Excitable Cells" (MBExC), University of Göttingen, Göttingen, Germany$\textsuperscript{6}$,\\
	Bernstein Center for Computational Neuroscience, Göttingen, Germany$\textsuperscript{7}$}
\begin{abstract}
Nerve impulses, the currency of information flow in the brain, are
generated by an instability of the neuronal membrane potential dynamics.
Neuronal circuits exhibit collective chaos that appears essential
for learning, memory, sensory processing, and motor control. However,
the factors controlling the nature and intensity of collective chaos
in neuronal circuits are not well understood. Here we use computational
ergodic theory to demonstrate that basic features of nerve impulse
generation profoundly affect collective chaos in neuronal circuits.
Numerically exact calculations of Lyapunov spectra, Kolmogorov-Sinai-entropy,
and upper and lower bounds on attractor dimension show that changes
in nerve impulse generation in individual neurons moderately impact
information encoding rates but qualitatively transform phase space
structure. Specifically, we find a drastic reduction in the number
of unstable manifolds, Kolmogorov-Sinai entropy, and attractor dimension.
Beyond a critical point, marked by the simultaneous breakdown of the
diffusion approximation, a peak in the largest Lyapunov exponent,
and a localization transition of the leading covariant Lyapunov vector,
networks exhibit \textit{sparse chaos}: prolonged periods of near stable dynamics
interrupted by short bursts of intense chaos. Analysis of large, more
realistically structured networks supports the generality of these
findings. In cortical circuits, biophysical properties appear tuned
to this regime of \textit{sparse chaos}. Our results reveal a close link between
fundamental aspects of single-neuron biophysics and the collective
dynamics of cortical circuits, suggesting that nerve impulse generation
mechanisms are adapted to enhance circuit controllability and information
flow.
\end{abstract}
\maketitle

Information is processed in the brain by the spatiotemporal activity
of large spiking neural circuits. Only the information that is encoded
in spike trains can be utilized by local networks, subsequent processing
stages, and ultimately to guide behavior. The spiking output of a
cortical neuron contains twenty- to hundred-fold less information
about synaptic input than its membrane potential does \citep{key-Polavieja},
making spike initiation a critical bottleneck for neural information
transmission. Experiments have revealed that neocortical neurons possess
a surprisingly broad encoding bandwidth, reliably encoding high-frequency
stimulus components in outgoing spike trains \citep{key-K=0000F6ndgen,key-Higgs,key-Tchumatchenko,key-Ilin}.
As predicted theoretically and observed experimentally, small changes
in spike onset rapidness can have a great impact on information encoding
bandwidth in feedforward architectures \citep{key-Fourcaud-TrocmeA,key-Fourcaud-TrocmeC,key-Naundorf2005,key-Naundorf2006,key-Naundorf2007,key-Wei,key-Ilin}. Here, spike onset rapidness refers to the steepness of the membrane potential change at the initiation of an action potential \cite{key-Naundorf2006}.
However, the influence of spike onset rapidness on recurrent network
dynamics has not been systematically studied.

One might expect that collective dynamics are insensitive to cellular
details, as the effect of single-cell properties can often become
negligible at the macroscopic circuit level. For example, asynchronous
irregular activity in idealized cortex models emerges robustly in
inhibition-dominated circuits and can be described by mean-field theory,
which is largely independent of the neuron model, such as the specific spike initiation mechanism \citep{key-VreeswijkSompolinsky,key-Renart}.
Instead, collective dynamics are expected to be primarily shaped by
the wiring diagram, or connectome, with most biological and artificial
learning algorithms operating at this level \citep{key-BCM,key-SussilloAbbott}.
Rate networks provide an example in which single-element input-output
functions determine critical properties of collective dynamics \citep{key-CSS}.
These analytical approaches have recently been extended to heterogeneous
networks, networks with bistable units, and spiking networks with
slow synaptic dynamics \citep{key-Aljadeff,key-Stern}.

\begin{figure*}
	\includegraphics[clip,width=2\columnwidth]{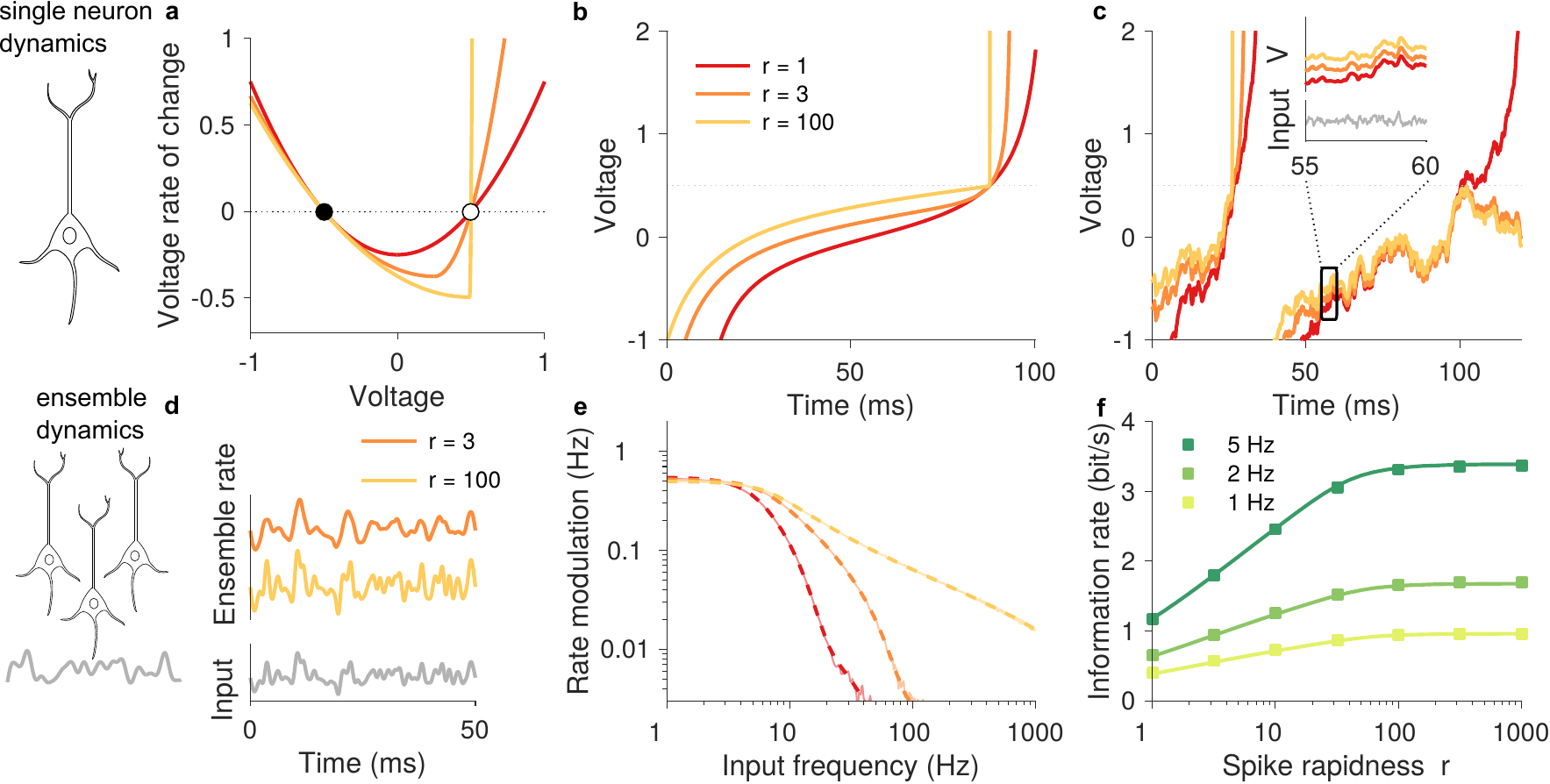} 
	\vspace{-0.5em}\caption{\label{fig:fig1}\textbf{High spike onset rapidness $r$ increases
			population encoding bandwidth.} \textbf{a} Single neuron dynamics
		have two fixed points: a stable fixed point (filled circle, resting
		potential) and an unstable fixed point (white circle, spike threshold).
		The slope at resting potential is $-1/\tau_{m}$, and the slope at
		the spike threshold $r/\tau_{m}$.\textbf{ b} Voltage traces of the
		neuron model with constant input currents for varying rapidness \textbf{$r$}.
		\textbf{c} Same as in panel b, but with fluctuating input currents.
		Note that the spike waveform and initiation depend strongly on \textbf{$r$},
		while the subthreshold dynamics are insensitive to \textbf{$r$}.
		The inset shows a magnified window of the voltage traces and the corresponding
		fluctuating input (gray) \textbf{d} Firing rates of an ensemble of
		rapid theta neurons with low and high rapidness for fluctuating input
		currents. Note that high rapidness enables the ensemble to accurately
		track the high-frequency components of the input. \textbf{e} Linear
		firing rate response for different values of rapidness, with direct
		numerical simulations (shaded lines) and Fokker Planck solutions (dashed
		lines) superimposed. ($\nu_{0}=1\textrm{\,\ Hz}$). \textbf{f} Mutual information rate in the Gaussian channel approximation based on spectral
		coherence, comparing Fokker Planck solutions (solid lines) and direct
		numerical simulations (squares) for different mean ensemble firing
		rates ($\nu_{0}=1$, $2$, $5$~Hz) (parameters: $\tau_{m}=10\textrm{ ms}$).}\vspace{-1em}
\end{figure*}
Here, we show that, surprisingly, spike onset rapidness fundamentally reshapes the
nature of chaos in large-scale spiking networks, such as the canonical balanced network model where excitatory currents are dynamically balanced by fast and strong recurrent inhibition \citep{key-VreeswijkSompolinsky,key-Renart}. Leveraging concepts from ergodic theory, we demonstrate that increasing the onset rapidness drives a transition from a \textit{dense} chaotic state—where chaotic fluctuations
are continuously fueled by many neurons—to a \textit{sparse} chaos regime characterized by intermittent, highly localized instability events. At the heart of this transition is the breakdown of the diffusion approximation, a standard assumption that treats synaptic input as Gaussian-like noise  (see Supplementary Information for details). 
In the high-rapidness regime, shot noise effects dominate, invalidating diffusion-based descriptions and revealing
a new dynamical regime. 
This transition, which occurs at a critical rapidness $r_{breakdown}$ that we first analyze in a feedforward setting, is driven by two opposing effects of spike onset rapidness. While faster onset increases single-neuron instability, it also reduces the time window during which a neuron is highly sensitive to input. Like a camera shutter that's open for a fraction of a second, the neuron only captures a few 'photons' of synaptic input during this window of instability. In this regime, the discrete nature of these inputs (shot noise) becomes prominent, and the central limit theorem underlying the diffusion approximation no longer applies, invalidating the diffusion approximation. As rapidness increases, the latter effect surprisingly dominates, shifting the network from \textit{dense} to \textit{sparse chaos}. These findings establish a direct link between
single-neuron spike initiation properties and the global chaotic dynamics of recurrent circuits. Beyond clarifying the fundamental biophysical origins of cortical chaos, our results suggest that neurons’ spike onset rapidness may be tuned to operate in this regime. Such tuning could enhance the controllability of network states and optimize information transmission across cortical layers, revealing a mechanism by which biophysical details shape large-scale brain function.

In the following, we will first introduce a novel, analytically solvable neuron model with an adjustable spike onset rapidness. We will then investigate how rapidness affects information transmission in a feedforward setting, identifying a critical rapidness value, $r_\mathrm{breakdown}$, where the diffusion approximation breaks down. Subsequently, we will analyze the full Lyapunov spectrum of recurrent networks of these neurons to characterize their dynamical stability and attractor properties. We will show that increasing rapidness leads to a transition from dense to sparse chaos and significantly reduces the attractor dimension. Finally, we will demonstrate that these findings are robust in more realistic cortical circuit models.

\begin{figure*}
	\includegraphics[clip,width=2\columnwidth]{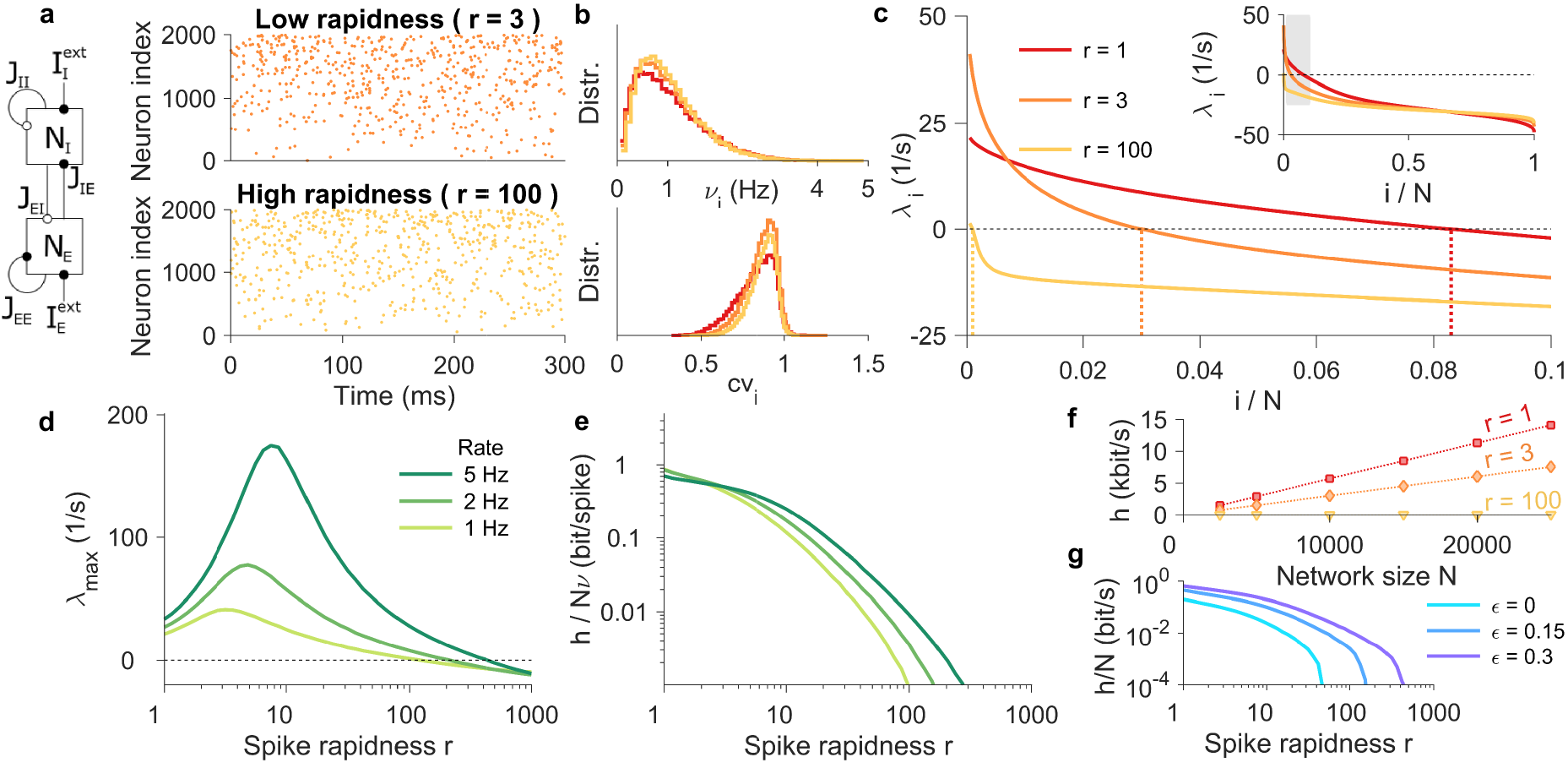} \hfill{} 
	\vspace{-0.5em}
	\caption{\label{fig:fig2}\textbf{High spike onset rapidness spike $r$ dramatically
			reduces chaos and dynamical entropy rate.} \textbf{a} Spike trains
		of 50 random neurons for low (upper panel) and high (lower panel)
		rapidness. \textbf{b} Distribution of firing rates (upper panel) and
		coefficients of variation (lower panel) for different values of rapidness
		(ordered by time-averaged single neuron firing rate) \textbf{c} Lyapunov
		spectra reorganize with increasing rapidness (inset: full Lyapunov
		spectra) \textbf{d} Largest Lyapunov exponent and \textbf{e} entropy
		rate $h$ as a function of rapidness for different mean firing rates
		($\bar{\nu}=1$, $2$, $5$~Hz), \textbf{f} Entropy rate $h$ for
		different network sizes (upper panel) and \textbf{g} different strengths
		of the scaling $\epsilon$ of the excitatory couplings (parameters:
		$N_{I}=2000$, $N_{E}=8000$, $K=100$, $\bar{\nu}=1\textrm{ Hz}$,
		$J_{0}=1$, $\tau_{m}=10\textrm{ ms}$).}\vspace{-1em}
\end{figure*}
\vspace{-2em}
\subsection*{Neuron model with tunable spike onset}
\vspace{-1em}To study the impact of spike onset rapidness on the collective network
dynamics, we constructed a novel, analytically solvable neuron model
with an adjustable spike onset rapidness $r$, as shown in Figure~\ref{fig:fig1}a
(see Methods for equations). The model, which we call the rapid theta model, is a modification of the theta neuron \citep{key-thetaneuron,key-thetanets,key-Monteforte2010}, a canonical model of neuronal excitability, and allows us to tune the rapidness of spike initiation while keeping other properties, such as the resting potential and subthreshold dynamics, largely unchanged. The model's membrane potential dynamics exhibit two fixed points: a stable fixed point representing the resting potential and an unstable fixed point corresponding to the spike threshold. The slope of the membrane potential dynamics, while at the spike threshold, it is $r/\tau_{m}$.
Increasing $r$ decreases the time constant
at the unstable fixed point $V_{U}$, resulting in greater instability
and sharper spike initiation (Fig.~\ref{fig:fig1}b,c). Thus, neurons with high $r$ spend less time in the upswing towards an action potential but are more susceptible to input. Unlike many neuron models, including the exponential integrate-and-fire model \citep{key-Fourcaud-TrocmeA}, our model can be solved exactly between spikes—a crucial prerequisite for precise and efficient calculation of the Lyapunov spectrum, which quantifies the sensitivity of the system to initial conditions.
\vspace{-2em}
\subsection*{Rapid Spikes Enhance Information Transmission}
\vspace{-1em}
Before investigating the effects of spike onset rapidness in recurrent networks, we first examined its impact on information transmission in a feedforward setting. This allows us to isolate the effects of rapidness on single-neuron encoding from the complexities of recurrent dynamics. As illustrated in Figures \ref{fig:fig1}d and~e, high rapidness enables neurons to transmit high-frequency components of a time-varying input current in their ensemble-averaged firing rate, whereas low rapidness limits transmission to low and mid-range frequencies. These results are consistent with previous studies \citep{key-Fourcaud-TrocmeA,key-Fourcaud-TrocmeC,key-Naundorf2005,key-Naundorf2006,key-Naundorf2007,key-Wei,key-Ilin}, which have established that rapidness is a determining factor in a neuron's ability to transmit information about presynaptic signals embedded in noise.
We quantified the information transmission using the Gaussian channel approximation of the mutual information based on the spectral coherence between the input and output signals (see Supplementary Information). We found that the mutual information rate grows approximately logarithmically with rapidness (Fig.~\ref{fig:fig1}f and Supplementary
Information for analytical results). This logarithmic relationship arises because rapidness determines the cutoff frequency beyond which spectral coherence scales as $f^{-1}.$ 
\begin{figure*}
	\includegraphics[clip,width=2\columnwidth]{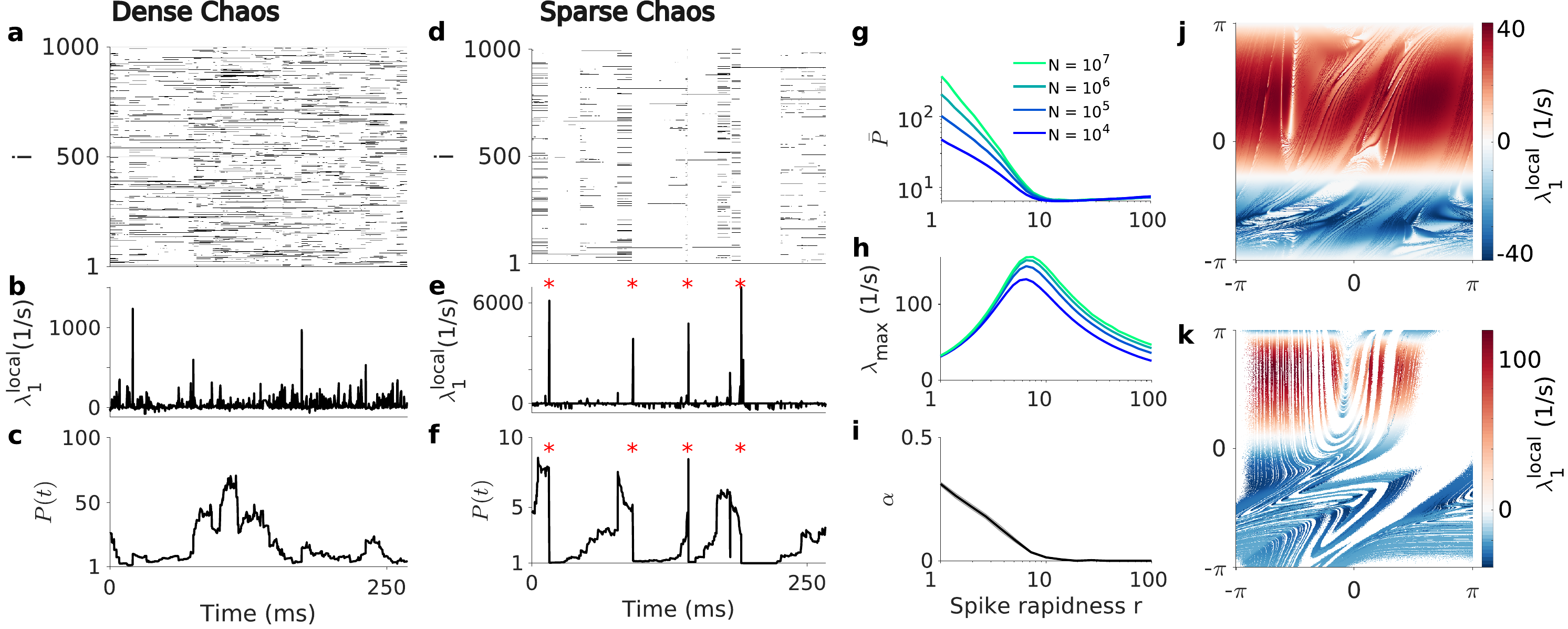}
	\vspace{-0.5em}
	\caption{\textbf{\label{fig:fig2-NEW}Localization of Lyapunov vectors for
			high spike onset rapidness $r$ reveals two types of network chaos.}
		\textbf{}\protect \\
		\textbf{a} First Lyapunov vector (marked black if $|\delta\phi_{i}(t)|>\nicefrac{1}{\sqrt{N}}$)
		\textbf{b }First local Lyapunov exponent $\lambda_{1}^{\textrm{local}}(t)$
		and \textbf{c} participation ratio $P(t)$ of the first Lyapunov vector
		as a function of time for $r\approx1.33$. \textbf{d}, \textbf{e},
		\textbf{f} Same as \textbf{a}, \textbf{b}, \textbf{c} for $r\approx31.6$.
		Note that large local Lyapunov exponents are followed by low participation
		ratio (red stars).\textbf{ g} Average participation ratio $\bar{P}$
		and \textbf{h} largest Lyapunov exponent vs. spike rapidness $r$
		for different network size $N$, \textbf{i} Power-law scaling exponent
		$\alpha$ from $\bar{P}\sim N^{\alpha}$ decreases approximately logarithmically
		as a function of $r$ and shows localization above peak rapidness
		$r_{\mathrm{peak}}$. \textbf{j} Poincaré section of the phases of
		neurons 2 and 3 whenever neuron 1 spikes for $r=1$. The first local
		Lyapunov exponent (LLE) at each point is color-coded, red colors indicate
		local instability, blue indicates local stability\textbf{ k} same
		as \textbf{j} for $r=4$ (parameters: $N_{I}=1000$, $K=100$, $\bar{\nu}=3\textrm{ Hz}$,
		$J_{0}=1$, $\tau_{m}=10\textrm{ ms}$).}\vspace{-1em}
\end{figure*}
\vspace{-2em}
\subsection*{ Diffusion Approximation fails for Rapid Spike Onset}
\vspace{-1em}
Importantly, at high rapidness, we observed qualitative deviations from the predictions of the commonly used diffusion approximation. The diffusion approximation assumes that a high rate of weak synaptic input can be treated as Gaussian white noise. However, we observe a qualitative change in the high-frequency response when rapidness exceeds $r_{\mathrm{breakdown}}\propto\sqrt{K\nu_{0}\tau_{m}/J_{0}}$,
which indicates a breakdown of the diffusion approximation, different from previously observed shot noise effects \cite{key-Richardson-2010,key-Richardson2018,key-Richardson2024} (see Supplementary Information for analytical derivation).
 This breakdown is associated
with a qualitatively different dynamical regime of the network dynamics,
as we will discuss in subsequent sections.

Unlike many neuron models, including the exponential integrate-and-fire
model \cite{key-Badel-2008,key-Fourcaud-Trocm=0000E92003}, our model can be solved exactly between spikes—a crucial prerequisite
for precise and efficient calculation of the Lyapunov spectrum. To
analyze the impact of spike initiation on dynamical stability, we
calculate the full Lyapunov spectrum of a spiking network of rapid
theta neurons. Lyapunov exponents measure the rate of exponential
divergence and convergence of nearby trajectories. The Lyapunov spectrum
provides a good estimate of the attractor dimension, according to
the Kaplan-Yorke conjecture, including exact upper and lower bounds
\citep{key-KY,key-Ledrappier,key-EckmannRuelle}. In an $N$-dimensional
dissipative chaotic system, trajectories do not cover the entire phase
space. Instead, after a transient period, they relax onto a strange
attractor, which has a dimensionality $D\leq N$. The Kaplan Yorke
attractor dimension is given by the number of Lyapunov exponents whose
cumulative sum interpolates to zero. One can think of it as the largest-dimensional
infinitesimal hypersphere, whose volume does not shrink by the dissipative
system dynamics. A lower bound on the attractor dimension is given
by the number of positive Lyapunov exponents. The dynamical entropy
rate, another canonical measure in dynamical systems, is bounded above
by the sum of positive Lyapunov exponents. This bound is exact for
smooth densities of the physical measure along unstable directions,
as given by Pesin’s identity~\citep{key-LedrappierYoung}. We analytically
calculate the Jacobian of the flow, which determines how infinitesimal
perturbations of the network state evolve from one spike time to just
after the next. We evaluate these Jacobians in numerically exact,
event-based simulations. The product of the Jacobians gives the long-term
Jacobian, which yields the spectrum of all Lyapunov exponents based
on Oseledets' multiplicative ergodic theorem \citep{key-Benettin1980}.
\vspace{-2em}
\subsection*{Impact on Dynamical Entropy Rate}
\vspace{-1em}
We find that rapidness strongly influences the dynamical stability
of recurrent networks in the balanced state (Fig.~\ref{fig:fig2}).
We first discuss results from inhibitory random (Erdős–Rényi) networks,
followed by mixed excitatory-inhibitory random networks and more structured
network topologies. One might expect that increasing rapidness $r$
would intensify the collective chaos since individual neurons become
more unstable. This is indeed the case for low rapidness: We find
that the largest Lyapunov exponent grows approximately linearly with
rapidness (Fig.~\ref{fig:fig2}d) until reaching a peak value. However,
for high rapidness, the largest Lyapunov exponent decreases proportionally
to $1/r$ (Fig.~\ref{fig:fig2}d). Numerical simulations reveal that
the peak rapidness scales as $r_{\mathrm{peak}}\propto\sqrt{K\nu_{0}\tau_{m}/J_{0}}$.
Above this peak, we observe a localization of the first covariant
Lyapunov vector (Fig.~\ref{fig:fig2-NEW}), which follows the same
scaling as the peak rapidness $r_{\mathrm{peak}}$ (Supplementary
Section XI). At lower values of rapidness $r$, the largest Lyapunov
exponent remains independent of the connectivity $K$. Furthermore,
an analytical calculation reveals that the breakdown of the diffusion
approximation follows the same scaling as the peak rapidness $r_{\mathrm{peak}}$.
This alignment suggests a deep connection between the breakdown of
the diffusion approximation, the peak in the largest Lyapunov exponent,
and the localization of the first covariant Lyapunov vector. At a critical rapidness that scales with $r_{\mathrm{crit}}\propto N^{0.5}K^{0.5}\bar{\nu}^{0.8}\tau_{m}^{0.8}J_{0}^{-0.7}$,
the largest Lyapunov exponent reaches zero, and the network activity
becomes dynamically stable. However, for any finite network size $N$,
the largest Lyapunov exponent can be reduced arbitrarily by choosing
a sufficiently large $r$ (Supplementary Section X). The dynamical
entropy rate decreases monotonically with rapidness, reaching zero
at $r_{\mathrm{crit}}$ (Fig.~\ref{fig:fig2}e). The full spectrum
reveals that the monotonic reduction is due to a drastic reduction
in the number of positive Lyapunov exponents, which overcompensates
the increase of the first few Lyapunov exponents (Fig.~\ref{fig:fig2}c).
Despite such a drastic change in the collective dynamics, the statistics
of the spike trains remain essentially unaffected (Fig.~\ref{fig:fig2}a,
b). This is surprising since, in many other systems, a transition
from chaos to stability is strongly reflected in the autocorrelations
and pairwise cross-correlations of the activity \citep{key-CSS}.

The scaling of the entropy rate with network size $N$ reveals that
the network chaos is extensive: for sufficiently large $N$, the entropy
rate grows linearly with $N$ (Fig.~\ref{fig:fig2}f). Convergence
of the Lyapunov spectra is demonstrated in Supplementary Section VIII.
Using random matrix theory, we calculate the mean Lyapunov exponent
analytically (see Supplementary Section IX).

The transition from chaos to stability with increasing rapidness also
occurs for random networks with both excitatory and inhibitory coupling.
To isolate the effect of excitation, we parameterized the coupling
matrix such that the input variance to each population remains unchanged
for different scaling values $\epsilon$ of the excitatory couplings
(see Supplementary Information for the definition of $\epsilon$.)
As the scaling $\epsilon$ of the excitatory couplings increases,
the dynamical entropy rate also increases. If the excitation is strong
enough, $r_{\mathrm{crit}}$ diverges, and these networks remain chaotic
for any $r$.

\begin{figure*}[htbp!]
	\includegraphics[clip,width=2\columnwidth]{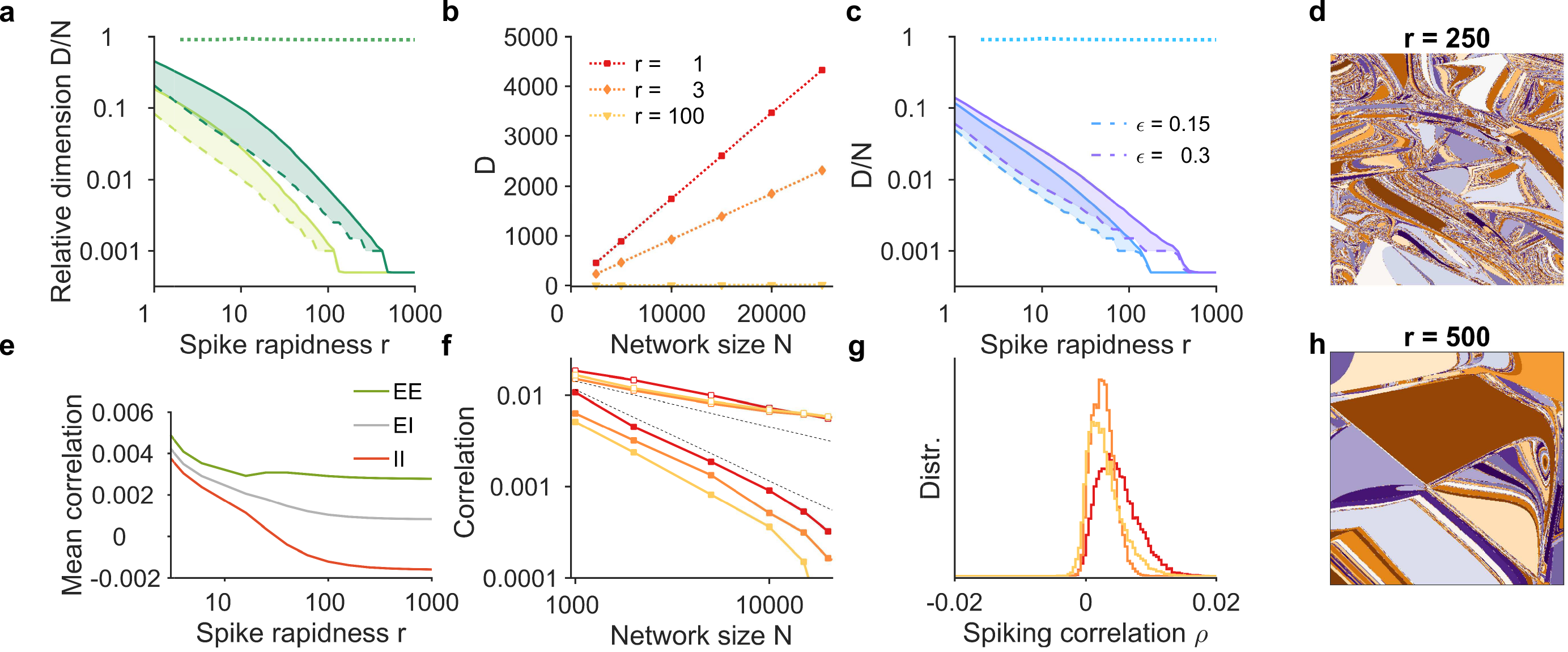}\vspace{-0.5em}
	\caption{\label{fig:fig3}\textbf{Reduction of attractor dimensionality in
			the asynchronous state despite low pairwise spike count correlations.}
		\textbf{a} Attractor dimension for different mean firing rates and
		varying spike rapidness $r$, dotted line: dimensionality estimate
		based on principal components of pairwise spike count correlations
		matrix (Supplementary Information), solid line: Kaplan-Yorke (KY)
		attractor dimension, dashed line: lower bound on attractor dimension
		(fraction of positive exponents) \textbf{b} KY Attractor dimension
		grows linearly with network size $N$ \textbf{c} same as \textbf{a}
		for different scaling $\epsilon$ of the excitatory couplings\textbf{
			d,~h} cross sections of basis on attraction in a plane perpendicular
		to the trajectory for $r=250,\:500$. Colors indicate basins of attraction
		of different trajectories ($N=200$,~$K=100$~$r_{c}\approx203$)
		\textbf{e} Mean pairwise spike count correlations for different values
		of rapidness $r$ between excitatory (E) and inhibitory (I) neurons,
		excitatory-excitatory pairs (EE) in green, inhibitory-inhibitory pairs
		(II) in red, mixed pairs in yellow \textbf{f} Mean pairwise spike
		count correlations decay $\propto1/N$, their standard deviations
		decays $\propto1/\sqrt{N}$, (color code as in\textbf{ b}) \textbf{g}
		histograms of spike count correlations for different rapidness $r$
		(EE-pairs)(color code as in\textbf{ b})(parameters as in Fig.~\ref{fig:fig2},
		spike count window $20$~ms)}\vspace{-1em}
\end{figure*}

Interestingly, while the dynamical stability changes drastically for
different values of $r$, the spike train statistics remain almost
unchanged. Both the distribution of firing rates and the coefficients
of variation are insensitive to changes in $r$ (Fig.~\ref{fig:fig2}b).
The mean pairwise Pearson correlation $\bar{\rho}$ of the spike counts
is weak and approaches zero for large networks, scaling as $\bar{\rho}\propto1/N$,
while the standard deviation $\bar{\sigma}$ of the pairwise correlations
decreases as $\bar{\sigma}\propto1/\sqrt{N}$ (Fig.~\ref{fig:fig3}e,~f,~g).
This confirms theoretical predictions of broadly distributed, but
weak pairwise correlations in the balanced state \citep{key-Renart}. 
\vspace{-2em}
\subsection*{Transition from \textit{Dense} to \textit{Sparse Chaos} with Increasing Spike Onset
Rapidness}
\vspace{-1em}
To further characterize the effect of spike-onset rapidness $r$ on
network dynamics, we compare two networks with different values of
rapidness but similar largest Lyapunov exponent: one at low rapidness
($r=1.33$) representing \textit{dense chaos}, and one at high rapidness
($r=31.6$) representing \textit{sparse chaos} (Fig.~\ref{fig:fig2-NEW}).

At higher values of spike onset rapidness $r$, we observe a transition
in the nature of chaos within the network, revealing two distinct
regimes: \textit{dense chaos} and \textit{sparse chaos}, characterized
by a peak in the largest Lyapunov exponent and the localization of
the first covariant Lyapunov vector (Fig. \ref{fig:fig2-NEW}). In
the \textit{dense chaos regime} at low rapidness (Fig.~\ref{fig:fig2-NEW}a-c),
neurons launch into action potentials gradually and spend a considerable
time in the unstable voltage range. This extended duration increases
the probability that postsynaptic neurons are in the unstable voltage
range when they receive input spikes, leading to an increased probability
of postsynaptic neurons contributing to the chaotic dynamics at any
given moment. The participation ratio $P(t)$, which quantifies the
number of neurons actively contributing to the first Lyapunov vector,
increases with network size $N$ in the \textit{dense chaos} regime (Fig.~\ref{fig:fig2-NEW}g).
The first local Lyapunov exponent $\lambda_{1}^{\textrm{local}}(t)$
exhibits temporal fluctuations with frequent small, positive peaks
(Fig.~\ref{fig:fig2-NEW}b), reflecting that instability is widespread
across the network and frequent in time.

As $r$ increases, neurons initiate action potentials more rapidly,
spending less time in the unstable voltage range. This reduction leads
to fewer synaptic inputs during the critical upswing phase. Therefore,
the central limit theorem underlying the diffusion approximation no
longer applies, resulting in the breakdown of the diffusion approximation.
This breakdown coincides with the peak of the largest Lyapunov exponent
at $r_{\mathrm{peak}}$ (Fig.~\ref{fig:fig2-NEW}h) and triggers
the localization of the first covariant Lyapunov vector, marking the
transition from \textit{dense} to \textit{sparse chaos} (Fig. \ref{fig:fig2-NEW}d-f).
In the \textit{sparse chaos} regime, the time a neuron spends in the
unstable upswing toward the action potential becomes so brief that
it is unlikely for a postsynaptic neuron to receive an input spike
during this period. Consequently, the chaotic dynamics become \textit{sparse}
both temporally and across the neuronal population: instability occurs
infrequently and is concentrated in brief events when a neuron in
the upstroke phase receives an input spike. During these instability
events, the local Lyapunov exponent $\lambda_{1}^{\textrm{local}}(t)$
exhibits large peaks (Fig.~\ref{fig:fig2-NEW}e), and the participation
ratio $P(t)$ drops sharply, approaching values close to 1 (Fig.~\ref{fig:fig2-NEW}f).
This indicates that only a few neurons dominate the tangent space
dynamics during those moments. The average participation ratio $\bar{P}$
becomes independent of network size $N$ in the \textit{sparse chaos}
regime, providing clear evidence of the localization of the first
covariant Lyapunov vector (Fig.~\ref{fig:fig2-NEW}g). This qualitative
change in network dynamics occurs concurrently with the peak of the
largest Lyapunov exponent and the breakdown of the diffusion approximation
at high $r$. As neurons spend less time in the unstable voltage range,
the average number of synaptic inputs received during the critical
upswing phase decreases. When this number becomes very low, individual
synaptic events (shot noise) become significant, and the central limit
theorem underlying the diffusion approximation no longer applies.
Analytical calculations of the frequency response reveal a different
frequency response at large $r$, confirming the breakdown of
the diffusion approximation (see Supplementary Information). 

Interestingly, the rapidness $r$ at which this breakdown occurs scales
similarly to $r_{{\rm peak}}$, suggesting a deeper connection between
the localization of the Lyapunov vector, the peak in the largest Lyapunov
exponent, and the breakdown of the diffusion approximation.

A visualization of the local Lyapunov exponents further illustrates
the effect of increasing rapidness $r$. At low $r$, a larger portion
of the Poincaré section displays slightly unstable dynamics, indicating
that instability is broadly distributed across the phase space (Fig.~\ref{fig:fig2-NEW}j).
In contrast, at high $r$, the unstable regions of the Poincaré section
become smaller but exhibit much higher local instability, as quantified
by the local Lyapunov exponent (Fig.~\ref{fig:fig2-NEW}k). This
indicates that instability is concentrated in specific regions of
the neural state space at high $r$.

In summary, increasing the spike onset rapidness $r$ beyond $r_{\mathrm{peak}}$
leads to a transition from \textit{dense }to \textit{sparse chaos},
coinciding with the breakdown of the diffusion approximation. This
transition is marked by the peak in the largest Lyapunov exponent
and the localization of the first covariant Lyapunov vector, highlighting
the profound impact of single-neuron properties on collective network
dynamics. 
\vspace{-2em}
\subsection*{Attractor dimensionality}
\vspace{-1em}
How is the drastic change in collective dynamics reflected in the
attractor dimension and the structure of pairwise correlations? We
find that increasing spike onset rapidness reduces the attractor dimension,
including both its upper and lower bounds by orders of magnitude (Fig.~\ref{fig:fig3}a).
We find that this reduction is independent of network size and also
occurs in mixed excitatory-inhibitory networks (Fig.~\ref{fig:fig3}b,~c).
The dimensionality of neural activity is often measured by the number
of principal components required to explain a fixed fraction of the
variance \citep{key-MachensPCA,key-LaurentPCA,key-ShenoyPCA,key-Ganguli}.
In this case, a dimensionality estimate based on pairwise statistics
vastly overestimates the attractor dimension (dotted lines in \ref{fig:fig3}a
and c). This implies that network dynamics exhibit strongly 'entangled'
statistics, which are obscured when inspecting only pairwise correlations
(Fig.~\ref{fig:fig3}e,~f,~g). Thus, the twisted, low-dimensional
strange attractor is interlaced in a high-dimensional phase space
\citep{key-Bialek}. In the extreme case of very high rapidness beyond
the critical rapidness $r_{\mathrm{crit}}$ the network dynamics become stable, and the basins of attraction can be visualized by random cross sections of the phase space along two random $N$-dimensional vectors
(Fig.~3d,~h). Adjacent initial conditions that converge to the same
trajectory are assigned the same color. As $r$ approaches $r_{\mathrm{crit}}$
from above, the basins of attraction become smaller and more curved,
and at $r_{\mathrm{crit}}$ they vanish (see Supplementary Information).
\vspace{-2em}
\subsection*{Cortical circuit models}
\vspace{-1em}
In the previous sections, we studied the dynamics of random (Erdős–Rényi)
balanced networks, which serve as canonical, idealized models of neocortical
networks. Since cortical tissue has a distinct architecture with a
non-random microscopic motif structure and a layered organization
where each layer has different wiring probabilities and distinct thalamic
inputs, we next investigate whether the previous findings are robust
under a more realistic microscopic and macroscopic structure. To analyze
the effect of a multilayered topology, we took experimental data of
wiring probability and synapse count from \citep{key-Potjans} to
build a full cortical column with 77,169 neurons, approximately 285
million synapses, and four layers, each containing excitatory and
inhibitory neuron populations.
\begin{figure*}[htbp!]
\includegraphics[clip,width=2\columnwidth]{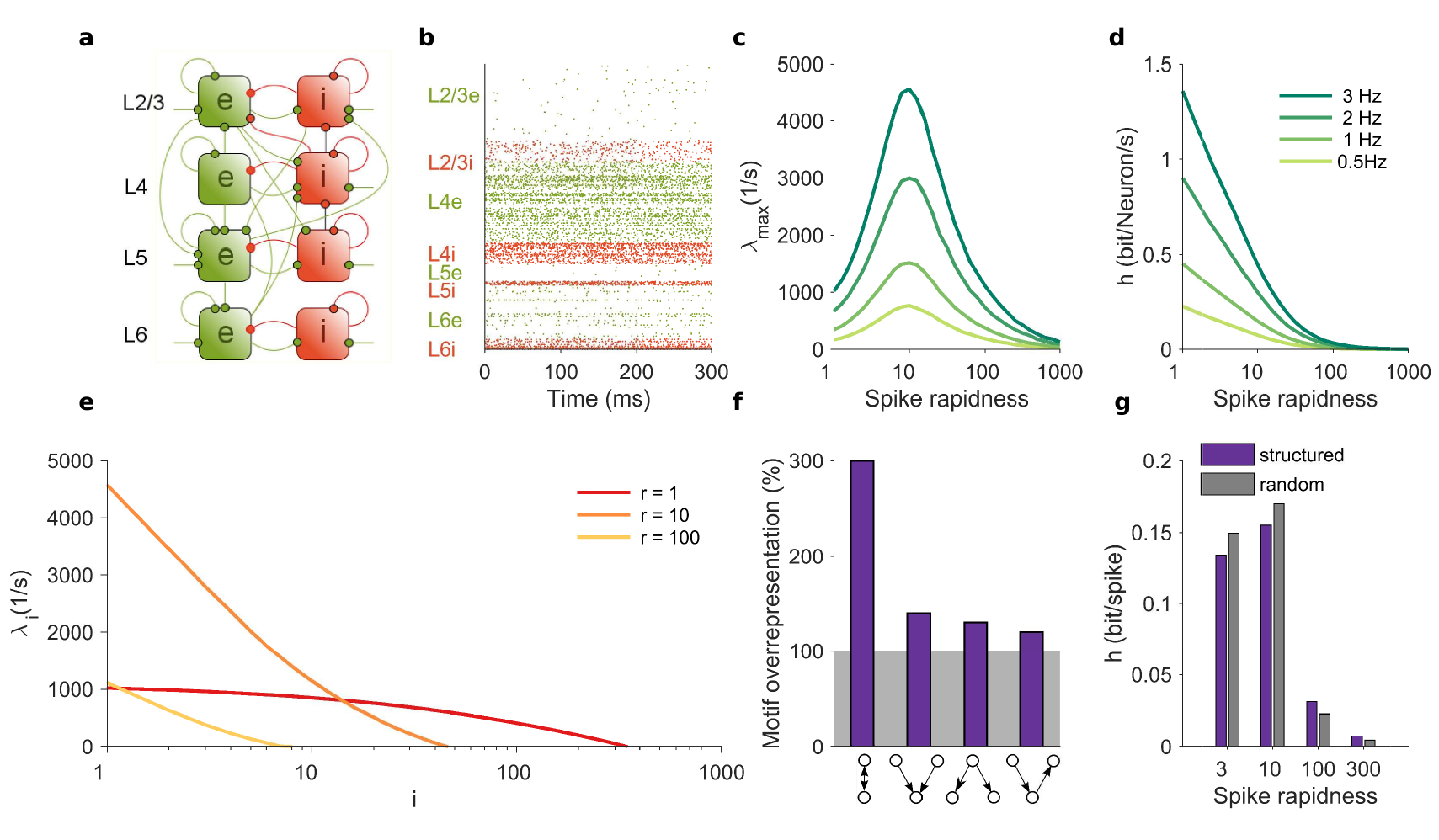}\vspace{-1em}
\caption{\label{fig:fig4}\textbf{High spike onset rapidness $r$ reduces chaos
and entropy rate in cortical circuit models:} \textbf{a} Multilayered
cortical column network model with layer- and cell type specific connection
probabilities, 77,169 neurons, $\sim$285 Million synapses \textbf{b}
spike raster illustrating layer-specific firing rates \textbf{c} largest
Lyapunov exponent vs. spike onset rapidness $r$ \textbf{d} entropy
rate $h$ vs. spike onset rapidness $r$, the gray area indicate a
physiologically plausible regime of spike onset rapidness $r$, \textbf{e}
positive Lyapunov exponents of multilayered model.\textbf{ f} Second
order network motif overrepresentation estimated from experiments
\textbf{g} dynamical entropy rate for random and realistic second
order motif structure at different values of spike onset rapidness
$r$. }\vspace{-1em}
\end{figure*}
This model produced spiking dynamics with layer-specific firing rates
(Fig.~\ref{fig:fig4}a,~b). We calculated Lyapunov spectra in large,
more realistic networks using an efficient, massively parallelized
implementation (Fig.~\ref{fig:fig4}e). We also constructed large
mixed excitatory-inhibitory circuits that incorporate experimentally
measured motif frequencies (Fig.~\ref{fig:fig4}f) \citep{key-Song}.

In these more realistic network structures, our findings from idealized
random cortex models were confirmed (Fig.~\ref{fig:fig4}). In the
multilayered network, the largest Lyapunov exponent behaves similarly
to that in the random network. It initially increases with rapidness
before decreasing (Fig.~\ref{fig:fig4}c).The entropy rate declines
as rapidness increases (Fig.~\ref{fig:fig4}d). When comparing a
random topology to a network with experimentally measured motif frequencies,
we observe a similarly strong reduction in dynamical entropy rate
with increasing spike onset rapidness (Fig.~\ref{fig:fig4}g). Thus,
the drastic reduction in chaos and dynamical entropy rate due to high
spike onset rapidness is independent of network structure. This finding
justifies analyzing more idealized random networks.
\vspace{-2em}
\subsection*{Conclusion and summary}
\vspace{-1em}
Recent theoretical and experimental studies have shown that cortical
neurons possess a surprisingly broad encoding bandwidth, which depends
on details of the spike initiation mechanism. Here, we investigated
the effect of this on the collective recurrent dynamics of neocortical
spiking circuits. We found that canonical measures of the collective
dynamics are not universal and insensitive to single-cell properties;
rather, they show a strong dependence on spike onset rapidness. Here,
we show that increasing spike onset rapidness transforms the network’s
chaos from a \textit{dense} regime, in which instability is continuously
fueled by many neurons, to a \textit{sparse} regime, characterized by sporadic,
sharply localized instability events. This transition defies the intuitive
expectation that higher single-cell instability would simply amplify
chaos. Instead, it reduces the overall chaotic intensity and complexity
of the collective state. This also holds in more realistic network
structures, including multilayered and second-order motif networks.
The effect of spike onset on chaotic entropy rate is orders of magnitude
greater than its effect on information encoding bandwidth. The importance
of single cell dynamics limiting the encoding bandwidth in a feedforward
architecture is thus not washed out by the collective network dynamics.

The dimensionality of collective states in neural circuits is a fundamental
measure. Different dimensionality metrics may be necessary to characterize
the complex dynamics. Using concepts from ergodic theory, we show
that attractor dimension decreases drastically with increasing spike
onset rapidness, a change hidden from conventional dimensionality
estimates based on activity correlations. This implies that neuron
states have strong statistical dependencies. How such dependencies
might be leveraged for computations in neocortical circuits is a question
for future research.

Since the spike threshold acts as an unstable fixed point for single-cell
dynamics, one might expect that high single-cell instability (i.e.,
high spike onset rapidness $r$) would increase network chaos. Surprisingly,
we find the opposite: large single-cell instability stabilizes collective
dynamics. Chaotic dynamics may be useful in computation for amplifying
small differences in initial conditions. If such a mechanism is used
by cortical circuits, spike onset rapidness would be an important
parameter to regulate this. Certainly, the dynamical entropy rate
contributes to noise entropy and can therefore impair coding capacity.

The cortex relies on intricate communication between layered circuits. Given that controlling highly chaotic networks with spike trains is likely difficult, we conjecture that high spike onset rapidness facilitates state control and thus enhances information transmission between circuits.

The ability of one ciruits's spike train to control dynamics in subsequent circuit is key for encoding information. Controlling highly chaotic networks
with input spike trains is presumably more challenging. We therefore
conjecture that high spike onset rapidness facilitates network state
control and enhances information transmission.

The use of ergodic theory to understand neural computation is only
just beginning. By applying these concepts to large-scale neural circuits,
we have laid the foundations for further investigation. Until now,
computational ergodic theory of spiking networks has been the only
approach to measure information-theoretic quantities in large recurrent
circuits. An important challenge is to extend this approach to other
quantities, such as transfer entropy and mutual information rate.
\vspace{-2em}
\subsection*{Methods}
\vspace{-1em}
The governing piecewise differential equation for the single-neuron
dynamics is

\begin{equation}
\tau_{m}\dot{V}_{i}=\begin{cases}
a_{U}(V_{i}-V_{G})^{2}+I_{i}(t), & \quad V>V_{G}\\
a_{S}(V_{i}-V_{G})^{2}+I_{i}^ {}(t), & \quad V\le V_{G}
\end{cases}
\end{equation}
where $\tau_{m}$ is the membrane time constant, $V_{G}=\frac{1}{2}\frac{r-1}{r+1}$
is the glue point, $a_{S}=\frac{r+1}{2r}$ and $a_{U}=r^{2}a_{S}$
are the curvatures, and the synaptic input current is 
\begin{equation}
I_{i}(t)=-I_{T}+I_{ext}+\tau_{m}\sum_{j\in\textrm{pre}(i)}J\delta(t-t_{j}^{(s)}))\label{eq:current}
\end{equation}
with $I_{T}=\frac{1}{2}\frac{r}{r+1}$, $J=J_{0}/\sqrt{K}$. $I_{ext}$
is adapted to achieve the desired target firing rate $\bar{\nu}$.
The elements of the Jacobian of the flow of the dynamics are

\begin{equation}
D_{ij}(t_{s})=\begin{cases}
1+Z^{\prime}(\phi_{i^{*}}(t_{s+1}^{-})), & \mathrm{for}\;i=j\in\textrm{post}(j^{*})\\
-\frac{\omega_{i^{*}}}{\omega_{j^{*}}}Z^{\prime}(\phi_{i^{*}}(t_{s+1}^{-})), & \mathrm{for}\;i\in\textrm{post}(j^{*})\textrm{ and }j=j^{*}\\
\delta_{ij} & \mathrm{otherwise}
\end{cases}\label{eq:current-2}
\end{equation}
where $Z$ is the phase response curve, $\omega=\frac{2}{\tau_{\mathrm{m}}}\sqrt{I_{ext}/a_{S}}$
is the phase velocity, and stars indicate the neuron spiking at $t_{s+1}$.
The Kaplan-Yorke attractor dimension is calculated from the interpolated
number of Lyapunov exponents that sum to zero:

$D=k+\dfrac{\sum_{i=1}^{k}\lambda_{i}}{\lambda_{k+1}}\quad\text{where}\quad k=\max\limits _{n}\left\{ \sum\limits _{i=1}^{n}\lambda_{i}\ge0\right\} .$
\vspace{-2em}
\subsection*{Acknowledgements}
\vspace{-1em}
We thank L.F. Abbott, F. Farkhooi, S. Goedeke, D. Hansel, G. Lajoie, L. Logiaco, R.-M. Memmesheimer, A. Neef, A. Palmigiano, M. Puelma Touzel, A. Renart, M. Schottdorf and C. v. Vreeswijk for fruitful discussions. This work was supported by the Deutsche Forschungsgemeinschaft (DFG, German Research Foundation) 436260547 in relation to NeuroNex (National Science Foundation 2015276) \& under Germany’s Excellence Strategy - EXC 2067/1- 390729940; by the DFG - Project-ID 317475864 - SFB 1286; by the DFG as part of the SPP 2205 - Project-ID 430156276. This work was further supported by the Leibniz Association (project K265/2019); and by the Niedersächsisches Vorab of the VolkswagenStiftung through the Göttingen Campus Institute for Dynamics of Biological Networks and by the Kavli Foundation and by the Gatsby Charitable Foundation (GAT3708). RE is a Kavli Fellow.

\onecolumngrid
\cleardoublepage 
\tableofcontents{}	
\section{The rapid theta neuron model}
To examine the impact of the action potential (AP) onset rapidness
on the collective dynamics of cortical networks, we constructed a
new neuron model with variable AP onset rapidness, called the rapid
theta neuron model (see\textbf{ Fig.~1} in the main paper). This
model is similar to the exponential integrate-and-fire neuron (see
Section \ref{sec:Experimentally-measured-action}), but it is much
more tractable for high-precision calculations. The rapid theta neuron
model combines the advantages of the theta neuron model for the analytical
derivation of the phase-response curve with a modifiable AP onset
rapidness, denoted as $r$. For $r=1$, the rapid theta neuron model
is equivalent to the standard theta neuron model \citep{Skey-thetaneuron}.
Increasing $r$ decreases the time constant at the unstable fixed
point $V_{\mathrm{U}}$ (voltage threshold), leading to greater instability
and \emph{sharper} AP initiation. The membrane time constant $\tau_{\mathrm{m}}$,
which is the time constant at the stable fixed point $V_{\mathrm{S}}$
(resting potential), remains unchanged. This is achieved by smoothly
joining two parabolas at $V_{\mathrm{G}}$. In the dimensionless voltage
representation, the resulting rapid theta neuron model is described
by the following piecewise ordinary differential equation:
\begin{equation}
\tau_{\mathrm{m}}\frac{\mathrm{d}V}{\mathrm{d}t}=\begin{cases}
a_{\mathrm{S}}(V-V_{\mathrm{G}})^{2}-I_{\mathrm{T}}+I(t) & \text{for }V\le V_{\mathrm{G}}\\
a_{\mathrm{U}}(V-V_{\mathrm{G}})^{2}-I_{\mathrm{T}}+I(t) & \text{for }V>V_{\mathrm{G}}.
\end{cases}\label{eq:rapid-theta-eq-V}
\end{equation}
In this equation, $I_{\mathrm{T}}$ denotes the rheobase current,
and $I(t)$ is the synaptic input current. The curvatures $a_{\mathrm{U,S}}$
depend on the AP onset rapidness $r$ and, together with $V_{\mathrm{G}}$
and $I_{\mathrm{T}}$, define the positions of the two branches of
the parabolas. The glue point, denoted $V_{\mathrm{G}}$, where the
two branches are continuously and smoothly joined, divides the single-neuron
phase space into two parts: $V\le V_{\mathrm{G}}$ and $V>V_{\mathrm{G}}$
. At the stable fixed point $V_{\mathrm{S}}$, the slope of the subthreshold
parabola is set to $-1/\tau_{\mathrm{m}}$, and at the unstable fixed
point $V_{\mathrm{U}}$, the slope is $r/\tau_{\mathrm{m}}$. This
leads to the following expressions
\begin{eqnarray*}
\frac{\partial\dot{V}(V_{S})}{\partial V}=-1 & = & 2a_{S}(V_{S}-V_{G})\\
a_{S} & = & \frac{1}{2}\frac{1}{(V_{G}-V_{S})}\\
\frac{\partial\dot{V}(V_{U})}{\partial V}=r & = & 2a_{U}(V_{U}-V_{G})\\
a_{U} & = & \frac{r}{2}\frac{1}{(V_{U}-V_{G})}.
\end{eqnarray*}
The derivative of the voltage vanishes at the two fixed points for
zero synaptic inputs ($I(t)\equiv0$). This defines the glueing point
$V_{G}$ and the rheobase current $I_{T}$:
\begin{eqnarray*}
\dot{V}(V_{S})=0 & = & a_{S}(V_{S}-V_{G})^{2}-I_{T}\\
I_{T} & = & \frac{V_{G}-V_{S}}{2}\\
\dot{V}(V_{U})=0 & = & a_{U}(V_{U}-V_{G})^{2}-I_{T}\\
 & = & \frac{rV_{U}-V_{G}(r+1)+V_{S}}{2}\\
V_{G} & = & \frac{rV_{U}+V_{S}}{r+1}.
\end{eqnarray*}

Without loss of generality, we set the stable and unstable fixed points
to $V_{\mathrm{S}}=-0.5$ and $V_{\mathrm{U}}=+0.5$, respectively,
yielding: 
\begin{eqnarray}
V_{\mathrm{G}} & = & \frac{1}{2}\frac{r-1}{r+1}\label{eq:rapid-VG}\\
I_{\mathrm{T}} & = & \frac{1}{2}\frac{r}{r+1}\label{eq:rapid-Ir}\\
a_{\mathrm{S}} & = & \frac{r+1}{2r}\label{eq:rapid-as}\\
a_{\mathrm{U}} & = & \frac{r(r+1)}{2}=r^{2}a_{\mathrm{S}}.\label{eq:rapid-au}
\end{eqnarray}
With Eq.~\eqref{eq:rapid-VG}-\eqref{eq:rapid-au} the governing
equation of the rapid theta neuron model \eqref{eq:rapid-theta-eq-V}
becomes

\begin{equation}
\tau_{\mathrm{m}}\frac{\textrm{d}V}{\textrm{d}t}=\begin{cases}
\frac{r+1}{2r}\left(V-\frac{1}{2}\frac{r-1}{r+1}\right)^{2}-I_{T}+I(t) & V\le\frac{1}{2}\frac{r-1}{r+1}\\
\frac{r(r+1)}{2}\left(V-\frac{1}{2}\frac{r-1}{r+1}\right)^{2}-I_{T}+I(t) & V>\frac{1}{2}\frac{r-1}{r+1}.
\end{cases}\label{eq:rapid-DiffEq-V}
\end{equation}

\section{Stationary firing rate of the rapid theta neuron}

The stationary firing rate of the rapid theta neuron with constant
input current follows directly from solving equation \ref{eq:rapid-DiffEq-V}.
The firing rate is determined by the inverse of the time taken for
the membrane potential to move from reset ( $V=-\infty$) to threshold
($V=\infty$):

\begin{equation}
\nu(I^{\mathrm{ext}})=\frac{\sqrt{I^{\mathrm{ext}}}}{\pi\tau_{\textnormal{\ensuremath{\mathrm{m}}}}}\sqrt{\frac{2r}{r+1}}\label{eq:nu-Iext}
\end{equation}
where the rheobase current $I_{T}$ is absoved in the constant $I^{\mathrm{ext}}=I+I_{T}$.
Cortical neurons are driven by a dense stream of input spikes. The
resulting compound spike train can be modeled as a Poisson process,
if the input spike trains are uncorrelated and random. However, the
superposition of many uncorrelated, non-Poissonian spike trains generally
deviates from a Poisson process \citep{key-Lindner2006,key-Cateau}\citep{key-Lindner2006,key-Cateau}.
When many weak and uncorrelated spikes arrive at a neuron, treating
the random component of the synaptic currents as Gaussian white noise
is justified \citep{key-Ricciardi1979,key-Lansky}. Using this diffusion
approximation, we obtained the mean firing rate by solving the stationary
Fokker-Planck equation with additive Gaussian white noise input current.
We start by writing a piecewise Langevin equation for the rapid theta
model:

\begin{equation}
\tau_{\mathrm{m}}\frac{\textrm{d}V}{\textrm{d}t}=\begin{cases}
a_{S}(V-V_{G})^{2}+\mu(t)+\sigma\sqrt{2\tau_{m}}\xi(t) & V\le V_{G}\\
a_{U}(V-V_{G})^{2}+\mu(t)+\sigma\sqrt{2\tau_{m}}\xi(t) & V>V_{G}
\end{cases}\label{eq:rapid-DiffEq-Langevin}
\end{equation}
where $\xi(t)$ is Gaussian white noise with unit variance, $\sigma\sqrt{2\tau_{\mathrm{m}}}$
is the noise intensity of the input and $\mu$ is a constant input
comprising the constant external input, the rheobase current and the
mean recurrent input. This results in a piecewise Fokker-Planck equation:

\begin{equation}
\frac{\partial P(V,\,t)}{\partial t}=\begin{cases}
\frac{\sigma^{2}}{\tau_{\mathrm{m}}}\frac{\partial^{2}P}{\partial V^{2}}+\frac{\partial}{\partial V}\left(\frac{-\mu(t)-a_{\mathrm{S}}(V-V_{\mathrm{G}})^{2}}{\tau_{\mathrm{m}}}P\right) & V\le V_{\mathrm{G}}\\
\frac{\sigma^{2}}{\tau_{\mathrm{m}}}\frac{\partial^{2}P}{\partial V^{2}}+\frac{\partial}{\partial V}\left(\frac{-\mu(t)-a_{\mathrm{U}}(V-V_{\mathrm{G}})^{2}}{\tau_{\mathrm{m}}}P\right) & V>V_{\mathrm{G}}
\end{cases}\label{eq:rapid-Fokker-Planck}
\end{equation}
where $P(V,\,t)$ is the time-dependent probability density of finding
a neuron at voltage $V$ at time $t$. The stationary (time-independent)
Fokker-Planck equation can be solved numerically using the efficient
threshold integration method proposed in~\citep{key-Richardson-2007}.
Briefly, the Fokker-Planck equation is set to zero and rewritten as
two first-order equations for probability flux and probability density
in $V$: 

\begin{equation}
\tau_{\textnormal{\textnormal{m}}}\frac{\partial J_{0}}{\partial V}=\nu_{0}\delta(V-V_{\textnormal{th}})-\nu_{0}\delta(V-V_{\textnormal{re}})\label{eq:rapid-stationary-FP-flux}
\end{equation}

\begin{equation}
-\frac{\partial P_{0}}{\partial V}=\begin{cases}
\frac{\tau_{\mathrm{m}}}{\sigma_{0}^{2}}\left(\frac{-\mu_{0}-a_{\mathrm{S}}(V-V_{\mathrm{G}})^{2}}{\tau_{\mathrm{m}}}P_{0}+J_{0}\right) & V\le V_{\mathrm{G}}\\
\frac{\tau_{\mathrm{m}}}{\sigma_{0}^{2}}\left(\frac{-\mu_{0}-a_{\mathrm{U}}(V-V_{\mathrm{G}})^{2}}{\tau_{\mathrm{m}}}P_{0}+J_{0}\right) & V>V_{\mathrm{G}}.
\end{cases}\label{eq:rapid-stationary-FP-density}
\end{equation}
Here $P_{0}(V)$ is the stationary probability distribution of membrane
potentials and $J_{0}(V)$ is the probability flux. As we know the
boundary conditions $\lim_{V\to\infty}P_{0}(V)=0$ and $\lim_{V\to\infty}J_{0}(V)=\nu_{0}$,
we can simultaneously integrate $P_{0}$ and $J_{0}$ from threshold
to some lower bound $V_{\mathrm{lb}}$. The rate $\nu_{0}$ is initially unknown but can be scaled out as
$p_{0}=P_{0}/\nu_{0}$. The normalization $\int_{V_{\mathrm{lb}}}^{V_{\mathrm{th}}}P_{0}dV=1$
then yields the firing rate

\begin{equation}
\nu_{0}=\left(\int_{V_{\mathrm{lb}}}^{V_{\mathrm{th}}}p_{0}dV\right)^{-1}\label{eq:rate-stationary-FP}
\end{equation}

\paragraph{Convergence of the numerical integration of the Fokker-Planck solution:}

\textbf{Fig.}~\textbf{\ref{fig:FPA_rate_nu_sigma_convergence}c}
and \textbf{d} displays the steady-state firing rate of the stationary
solution of the Fokker-Planck approach. As the rapid theta model has
neither finite threshold nor finite reset in the voltage representation,
the threshold, reset and lower bound of the numerical threshold integration
scheme have to be chosen sufficiently far away from zero such that
the results do not change (\textbf{Fig.}~\textbf{\ref{fig:FPA_rate_nu_sigma_convergence}c}).
The integration step size $\Delta V$ has to be chosen sufficiently
small (\textbf{Fig.}~\textbf{\ref{fig:FPA_rate_nu_sigma_convergence}f}).
For higher rapidness, smaller step sizes are necessary to get the
same precision. This is because there is a drastic change of the dynamics
at $V_{\mathrm{G}}$ therefore close to $V_{\mathrm{G}}$ the voltage
integration steps have to be small. To increase the numerical accuracy
at high rapidness, we chose $\Delta V$, such that both $V_{\mathrm{re}}$
and $V_{\mathrm{G}}$ fall on a lattice point of the integration scheme.
Code for Julia and MATLAB$\textsuperscript{\textregistered}$/Octave
is available upon request. We found that setting $V_{\mathrm{th}}=1000$
and $\Delta V=10^{-3}$ is sufficient for a relative precision of
the firing rate, $\frac{\Delta\nu_{0}}{\nu_{0}}<10^{-2}$.

\paragraph{Diffusion approximation and shot noise:}

\begin{figure}[tp]
\includegraphics[width=1\textwidth]{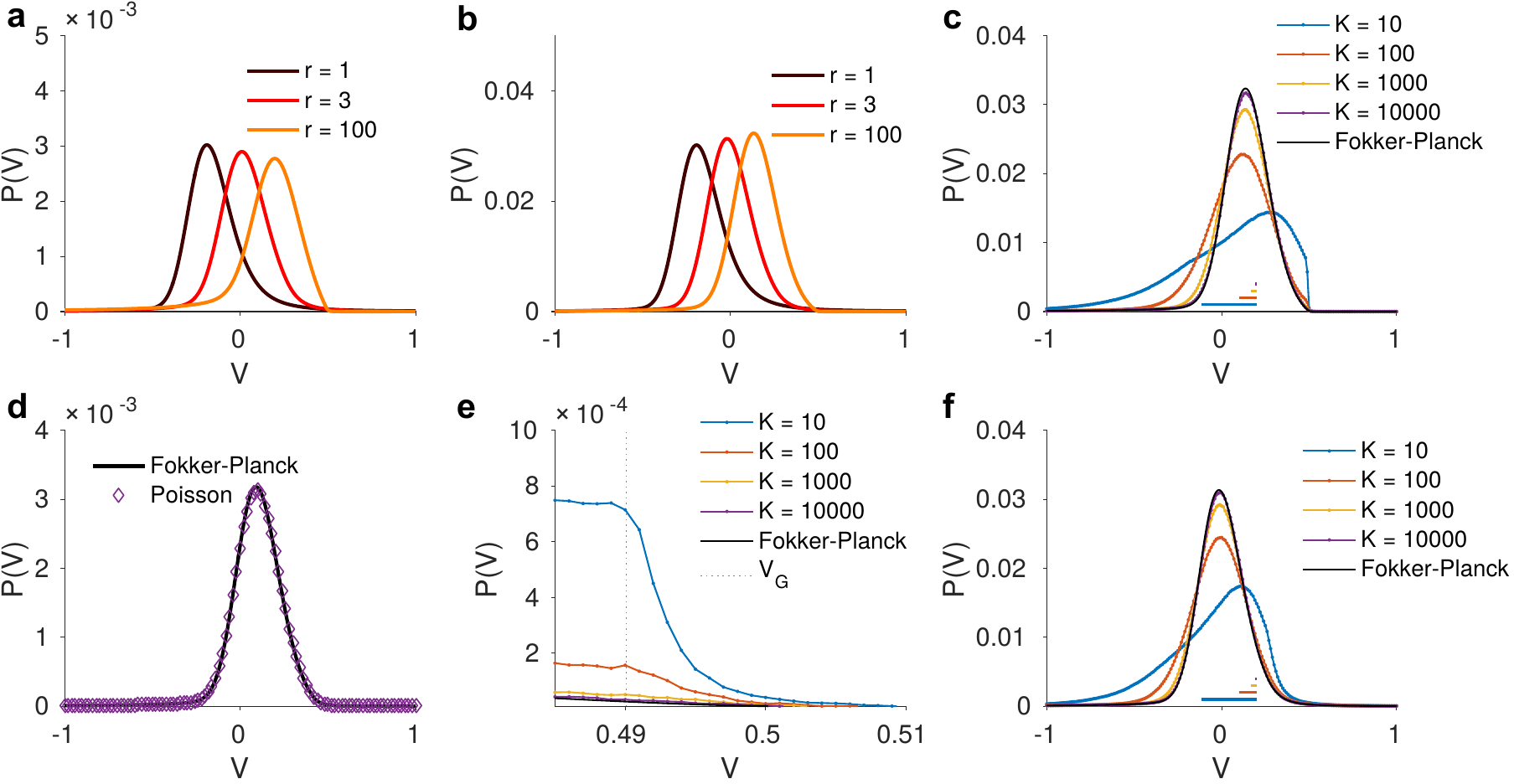}

\caption[Phase-transition curve, phase-response and infinitesimal phase-response
of the rapid theta neuron model with AP onset rapidness.]{\textbf{\label{fig:FPA_DNS_P_V}Action potential onset rapidness
$r$ shapes stationary voltage distribution}. \textbf{a)} Voltage
distribution for different rapidness for fixed external input obtained
from the solution of the stationary Fokker Planck equation Eq.~\ref{eq:rapid-stationary-FP-flux}
and Eq.~\ref{eq:rapid-stationary-FP-density} \textbf{b)} same as
\textbf{a}, but for fixed mean firing rate. \textbf{c)} Voltage distribution
obtained with inhibitory Poisson input spikes trains. For small K,
the shot noise nature of the Poisson input becomes important. (\textbf{$r=100$}
) \textbf{d)~}Comparison of Poisson input and Fokker Planck solution
for \textbf{$K=10000$} and \textbf{$r=10$}. \textbf{e)}~same as
\textbf{c)}~zoomed in for \textbf{$r=10$}. The probability density
drops at \textbf{$V_{\mathrm{G}}.$} \textbf{f)}~same as \textbf{c)}~for
\textbf{$r=3$} (parameters: $\nu_{0}=1\,\mathrm{Hz}$, $J_{0}=1$,
$\tau_{\mathrm{m}}=10\,\mathrm{ms}$).}
\end{figure}

\begin{figure}[tp]
\includegraphics[width=1\textwidth]{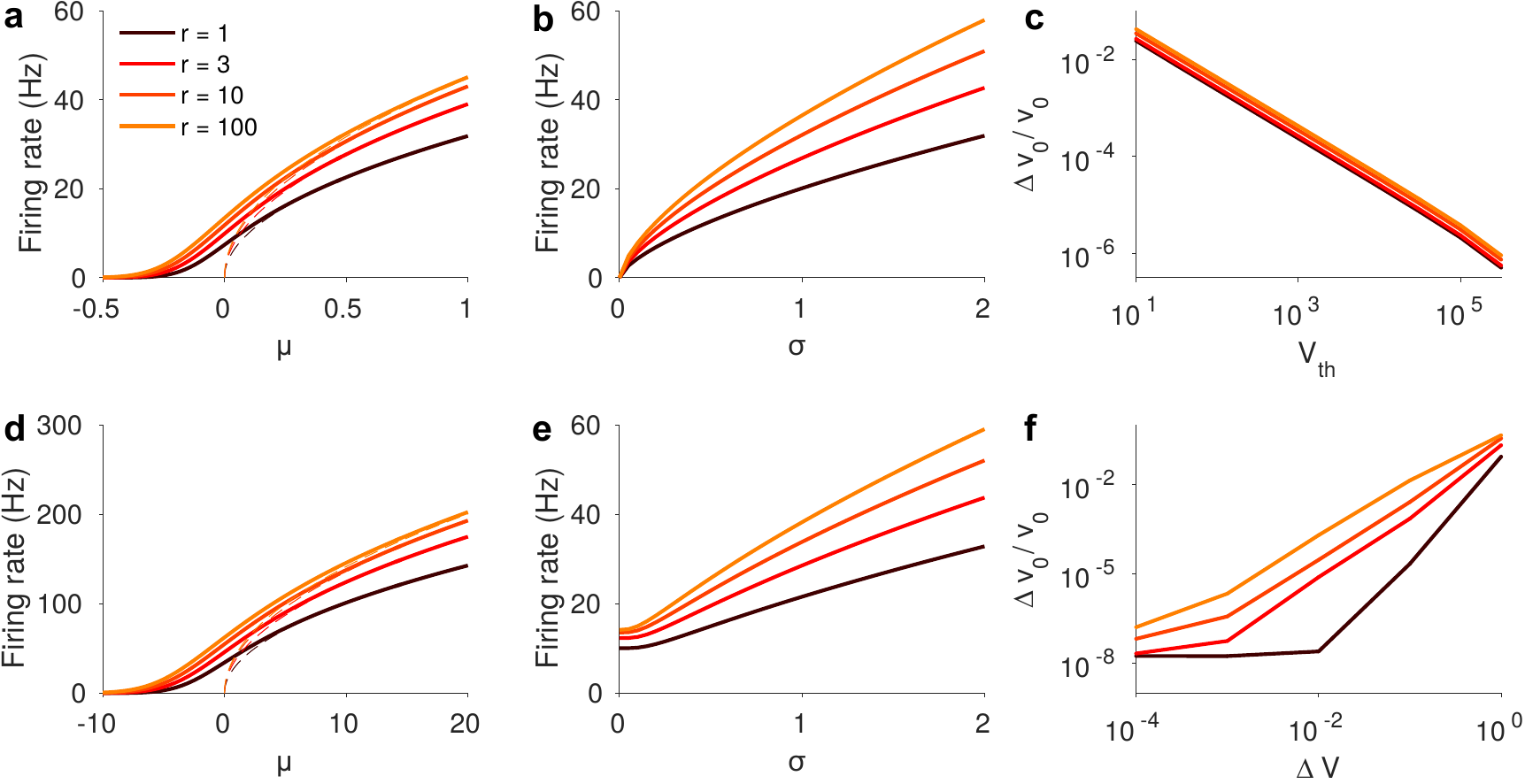}

\caption[Phase-transition curve, phase-response and infinitesimal phase-response
of the rapid theta neuron model with AP onset rapidness.]{\textbf{\label{fig:FPA_rate_nu_sigma_convergence}Spike onset rapidness
$r$ mildly affects steady-state firing rate}. \textbf{a)} Firing
rate as a function of constant input \textbf{$\mu$} for different
values of rapidness obtained from stationary Fokker-Planck solution
for weak noise (\textbf{$\sigma=\sqrt{0.05}$}). Dashed line is analytical
noise-free result (Eq.~\ref{eq:nu-Iext}). \textbf{b)} same as \textbf{a}
for varying noise strength \textbf{$\sigma$} at \textbf{$\mu=I_{T}$}.
\textbf{c)} rate deviation for different threshold levels of numerical
integration scheme. \textbf{d)} same as \textbf{a)} for strong noise
\textbf{($\sigma=\sqrt{0.5}$)} \textbf{e)} same as \textbf{b)} for
strong constant input \textbf{$\mu=0.1+I_{T}$} \textbf{f)} rate deviation
for different integration step sizes in the numerical integration
scheme (parameters: $\tau_{\mathrm{m}}=10\,\mathrm{ms}$, $V_{\mathrm{re}}=-V_{\mathrm{th}}$,
$\Delta V=10^{-3}$, $V_{\mathrm{th}}=10^{3}$).}
\end{figure}

The Fokker-Planck approach approximates the synaptic input as Gaussian
white noise. This is justified in the limit of uncorrelated input
and large rate of infinitesimally strong received postsynaptic currents
per neurons. To investigate the impact of finite postsynaptic potentials,
we replaced the Gaussian white noise with inhibitory Poisson pulses
of rate $\nu_{\mathrm{p}}=\nu_{0}K$ and strength $J_{\mathrm{p}}=-J_{0}/\sqrt{K}$,
keeping the variance $\sigma_{\mathrm{p}}^{2}=\nu_{\mathrm{p}}J_{\mathrm{\mathrm{p}}}^{2}=\nu_{0}J_{0}^{2}$
fixed and varied $K$. The mean firing rate $\nu_{0}$ was kept constant
for different values of rapidness $r$ and $K$ by adapting the constant
input current $\mu$. For large $K$, the diffusion approximation
is valid, whereas for small $K$, the shot noise nature of the Poisson
input becomes relevant (\textbf{Fig.}~\textbf{\ref{fig:FPA_DNS_P_V}c+f}).
This effect is particularly strong for large rapidness $r$. For small
$K$, the density just below $V_{\mathrm{G}}$ obtained from direct
numerical simulation is larger than the probability density obtained
from the Fokker-Planck solution. The numerical density for small $K$
drops at $V_{\mathrm{G}}$ (\textbf{Fig.}~\textbf{\ref{fig:FPA_DNS_P_V}e}).
This will help later in understanding the dependence of the largest
Lyapunov exponent on $K$ and $r$. In the next paragraph, we provide
additional analysis of the shot noise regime.

\section{Stationary characterization of rapid theta neurons driven by inhibitory shot noise}

We start by writing a piecewise stochastic differential equation for the rapid theta model:
\begin{equation}
F(V)=\tau_{\mathrm{m}}\frac{\textrm{d}V}{\textrm{d}t}=\begin{cases}
a_{S}(V-V_{G})^{2}+I_{ext} + s(t) & V\le V_{G}\\
a_{U}(V-V_{G})^{2}+I_{ext} + s(t) & V>V_{G}
\end{cases}\label{eq:rapid-DiffEq-Langevin-1}
\end{equation}
where $s(t)$ represents inhibitory input spikes arriving at rate $R$ with amplitude $\frac{-J_0}{\sqrt{K}}$. The inhibitory input $s(t)$ can be represented
as a sum of inhibitory spikes:
$s(t)=\sum_{k}\frac{-J_0}{\sqrt{K}}\delta(t-t_{k})$ where $t_{k}$ are the Poisson-distributed
spike arrival times with rate $R=K\nu$. Each input spike causes a decrement
of $\frac{-J_0}{\sqrt{K}}$ in the membrane potential.
Thus, the neuron's stochastic differential equation (SDE) is:
\[dV(t)=F(V)dt+ds(t)
\]
or explicitly including the jumps:
\[dV(t)=F(V)dt-J\,dN(t)\]
where $N(t)$ is a Poisson process with rate $R$, and $dN(t)$ represents the number of spikes in the interval $[t,t+dt)$.
We consider a population of neurons described by the probability density
$P(V,t)$ of the membrane potential $V$ at time $t$. The continuity
equation for $P(V,t)$ is:
\[\frac{\partial P(V,t)}{\partial t}+\frac{\partial J(V,t)}{\partial V}=0\]
where $J(V,t)$ is the probability flux. The total flux $J(V,t)$ consists of two components: The drift flux
$J_{\text{drift}}(V,t)$, due to the deterministic dynamics $f(V)$
and the jump flux $J_{\text{jump}}(V,t)$ due to the inhibitory Poisson input.
\[J(V,t)=J_{\text{drift}}(V,t)+J_{\text{jump}}(V,t).\]
We next describe $\frac{dV}{dt}=F(v)+s(v,t).$ with driving force
\[F(v)=\begin{cases}
\frac{1}{\tau}(a_{S}(V-V_{G})^{2}+\mu) & V\le V_{G}\\
\frac{1}{\tau}(a_{U}(V-V_{G})^{2}+\mu) & V>V_{G}
\end{cases}.\]
The voltage trajectories of a population of neurons, each obeying the voltage dynamics with an uncorrelated but statistically identical realization of the stochastic drive in can be described by a probability density $P(V,t)$. The stochastic single neuronal dynamics allow for the construction of a master equation that describes the deterministic dynamics of the ensemble at the population
level. As well as the probability density, it is convenient to consider the probability flux $J(V,t)$. This describes the flow-rate of trajectories passing a particular voltage. Note that the flux at threshold $J(V_{th})$
is equal to the instantaneous spike-rate $\nu(t)$ of the population with the flow then reinserted at the reset $V_{re}$.
These quantities are connected by a continuity equation 

\[\frac{\partial P}{\partial t}+\frac{\partial J}{\partial V}=\nu(t)\left(\delta(V-V_{re})-\delta(V-V_{th})\right)\]

\begin{figure}[tp]
\includegraphics[width=1\textwidth]{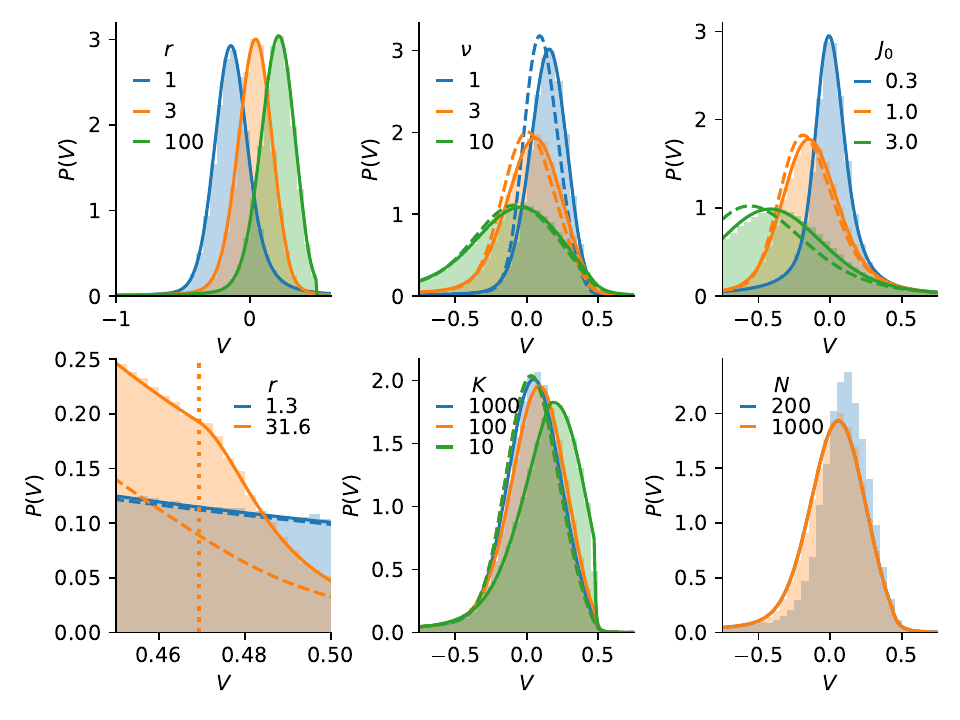}

\caption[Phase-transition curve, phase-response and infinitesimal phase-response
of the rapid theta neuron model with AP onset rapidness.]{\textbf{\label{fig:FPA_rate_nu_sigma_convergence-1}Shot noise affects stationary voltage distribution}. \textbf{a)} Voltage distribution
for different rapidness for fixed external input obtained analytically
from solving continuity equation with shot noise, transparent histogram
shows network simulation\textbf{ b)} same as\textbf{ a} for different
firing rates $\mu$, dashed lines is solution to Fokker-Planck equation.
\textbf{c)} Voltage distribution obtained with inhibitory Poisson
input spikes trains. For small K, the shot noise nature of the Poisson
input becomes important. (\textbf{$r=100$} ) \textbf{d)~}Comparison
of Poisson input and Fokker Planck solution for \textbf{$K=10000$}
and \textbf{$r=10$}. \textbf{e)}~same as \textbf{c)}~zoomed in
for \textbf{$r=10$}. The probability density drops at \textbf{$V_{\mathrm{G}}.$}
\textbf{f)}~same as \textbf{c)}~for \textbf{$r=3$} (parameters:
$\nu_{0}=1\,\mathrm{Hz}$, $J_{0}=1$, $\tau_{\mathrm{m}}=10\,\mathrm{ms}$).}\label{fig:stationary-shot-noise-figure}
\end{figure}

The stationary continuity equation that takes the form of a delay differential equations (DDE) \cite{key-Manz2019}:
\[\frac{d}{dV}[P(V)F(V)]=\bar\nu\left[P(V-C)-h(V)P(V)\right]
+\rho\left[\delta(V-V_{re})-\delta(V-V_{th})\right]\]
with the average firing rate $\bar\nu$.  We solve this delay differential equations by the method of steps using higher-order solvers with high accuracy by setting up a threshold integration analogously to the established threshold-integration method proposed in~\citep{key-Richardson-2007} as proposed for the DDE in \cite{key-Manz2019}. Again, the unknown rate $\bar \nu$ can be scaled out as $\bar p=\bar P/\bar \nu$. The normalization $\int_{V_{\mathrm{lb}}}^{V_{\mathrm{th}}}\bar PdV=1$
then yields the 
$\bar\nu=\left(\int_{V_{\mathrm{lb}}}^{V_{\mathrm{th}}}\bar p dV\right)^{-1}\label{eq:rate-stationary-DDE}
$ with the initial condition $\bar p(V_{th})=1 / F(V_{th})$.
We find excellent agreement between the voltage distribution obtained by solving the delay differential equations with network simulations in Figure~\ref{fig:stationary-shot-noise-figure}, where we usually use bisectioning to adapt the external input current to achieve a target firing rate $\bar\nu$. We note that for large rapidness $r$, we find systematic deviations of  $P(V)$ from the Fokker-Planck approach close to the unstable fixed point $V_U$, that are captured well by the DDE. As expected, for large values of $K$. We further note that there exist deviations for finite network size $N$ even for the DDE solution, that probably step from the fact that the DDE approach ignores pairwise correlations and population fluctuations.

\section{Linear response of an ensemble of rapid theta neurons}

\paragraph*{Figure 1 e of main paper.}

\begin{figure*}
\noindent\begin{minipage}[t]{1\columnwidth}%
\includegraphics[clip,width=0.49\columnwidth]{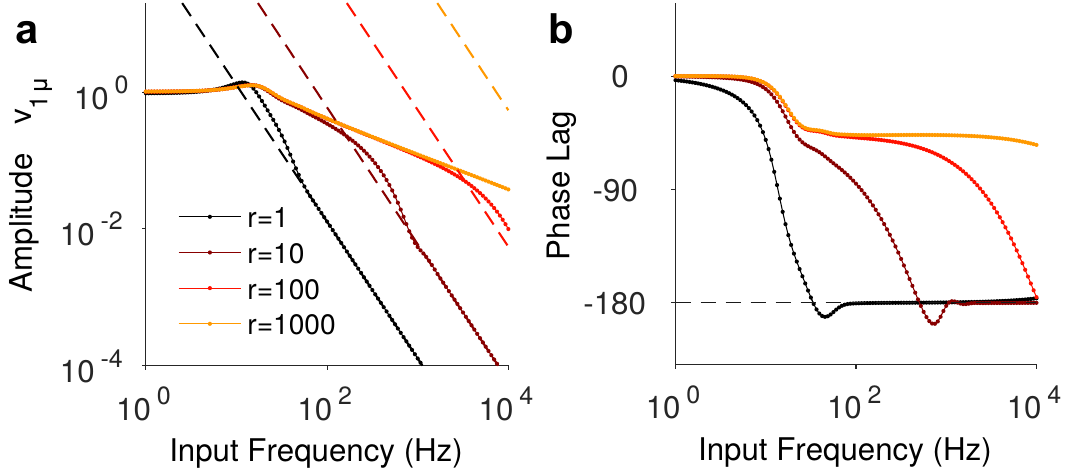}\includegraphics[clip,width=0.49\columnwidth]{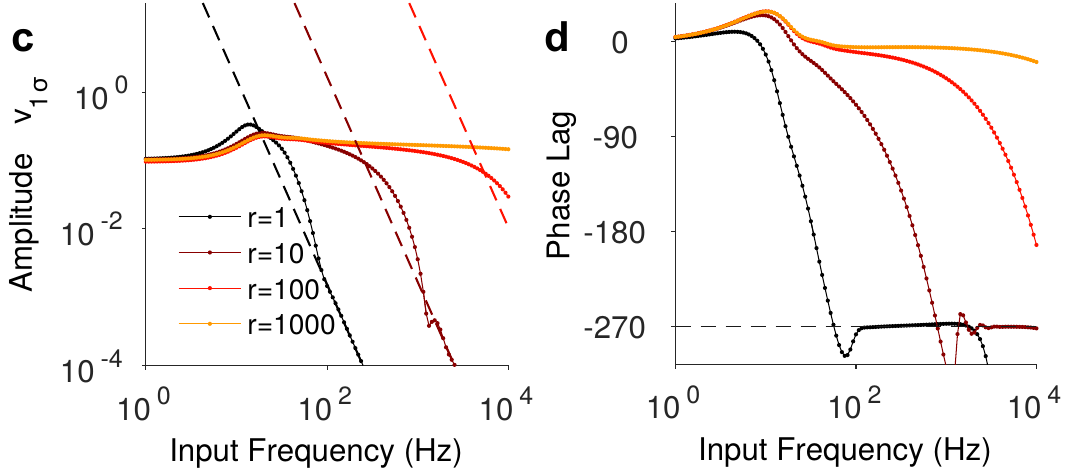}

\includegraphics[clip,width=0.49\columnwidth]{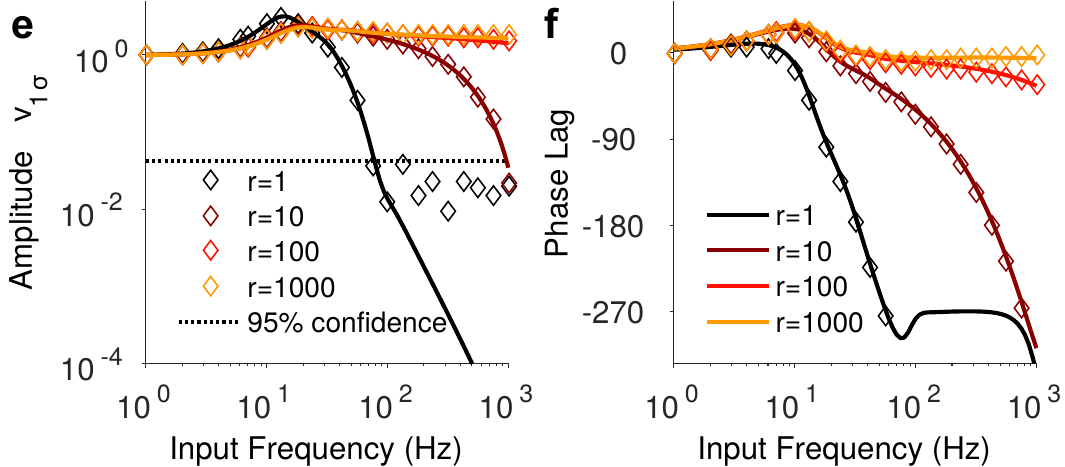}\includegraphics[clip,width=0.49\columnwidth]{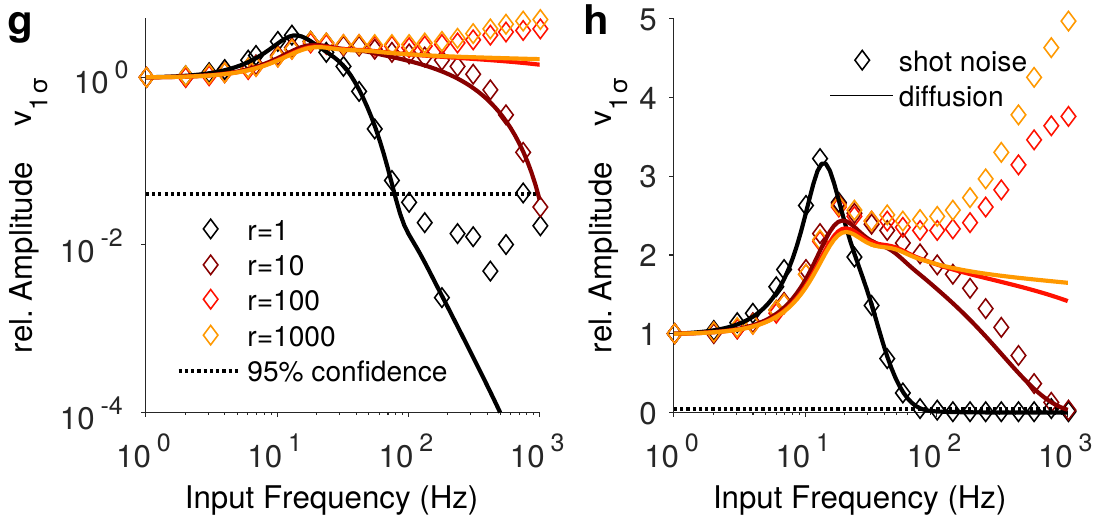}
\end{minipage}

\caption{\label{fig:FPA_DNS_linear_response}\textbf{Action potential onset
rapidness $r$ drastically shapes linear firing rate response}:\textbf{
Fokker-Planck approach (FPA) vs. direct numerical simulations (DNS)
vs. high-frequency analytical solution (HFA)} \textbf{a)} Linear response
for mean modulation of input current. Increasing rapidness leads to
a higher cutoff frequency. Full lines are FPA, dashed lines correspond
to HFA (Eq.~\ref{eq:rapid-frequency-response-mean}). \textbf{b)}
Phase lag for different input frequencies. \textbf{c) }and\textbf{
d)} the same as \textbf{a)}+\textbf{b)} for modulation of variance
of the input current. \textbf{e)} Comparison of DNS (diamonds) and
FPA (full line) for variance modulation with $K=1000$. Dashed line
indicates 95\% significance threshold for DNS. \textbf{f)} Same for
phase lag. \textbf{g)} Impact of shot noise: Same as \textbf{e)} for
$K=30$. For large values rapidness and large input frequencies, the
shot noise becomes noticeable. \textbf{h)} Same as \textbf{g)} with
linear amplitude axis (parameters: $\nu_{0}=10\,$Hz, $K=1000$, $\tau_{\mathrm{m}}=10\,$ms.)}
\end{figure*}

We calculated the linear response of ensembles of rapid theta neurons
using three different approaches: first, by solving the time-dependent
Fokker-Planck-equation; second, by direct numerical simulation using
Poisson input; and third, by direct numerical simulations using band-limited
white noise. The same efficient numerical threshold integration method
used for the stationary response can be adapted to obtain the linear
response to a sinusoidal modulation of the input current. A modulation
of the mean input current, $\mu(t)=\mu_{0}+\mu_{1}e^{i\omega t}$,
results in a firing rate modulation, which in linear response is $\nu(t)=\nu_{0}+\hat{\nu}_{\mu}e^{i\omega t}$.
The absolute value $|\hat{\nu}_{\mu}(\omega)|$ represents the rate
modulation strength, and the phase of $\hat{\nu}(\omega)$ gives the
phase lag between input and output modulation. The expansion of the
flux in first order therefore satisfies:

\begin{equation}
\tau_{\mathrm{m}}\frac{\partial\hat{J}_{\mu}}{\partial V}=i\omega\hat{P}_{\mu}+\hat{\nu}_{\mu}\delta(V-V_{th})-\hat{\nu}_{\mu}\delta(V-V_{\mathrm{re}})\label{eq:meanmod-FP-flux}
\end{equation}

\begin{equation}
-\frac{\partial\hat{P}_{\mu}}{\partial V}=\begin{cases}
\frac{\tau_{\mathrm{m}}}{\sigma_{0}^{2}}\left(\frac{-\mu_{0}-a_{\mathrm{S}}(V-V_{\mathrm{G}})^{2}}{\tau_{\mathrm{m}}}\hat{P}_{\mu}+\hat{J}_{\mu}+F_{\mu}\right) & V\le V_{\mathrm{G}}\\
\frac{\tau_{\mathrm{m}}}{\sigma_{0}^{2}}\left(\frac{-\mu_{0}-a_{\mathrm{U}}(V-V_{\mathrm{G}})^{2}}{\tau_{\mathrm{m}}}\hat{P}_{\mu}+\hat{J}_{\mu}+F_{\mu}\right) & V>V_{\mathrm{G}}.
\end{cases}\label{eq:meanmod-FP-density}
\end{equation}
$F_{\mu}$ describes the inhomogeneous term given by $F_{\mu}=-\frac{\partial J}{\partial\mu}P_{0}=-\frac{\mu_{1}P_{0}}{\tau_{\mathrm{m}}}$.
The boundary conditions are $\lim_{V\to\infty}\hat{P}_{\mu}(V)=0$
and $\lim_{V\to\infty}\hat{J}_{\mu}(V)=\hat{\nu}_{\mu}$. The solution
can be separated into two parts, one proportional to $\hat{\nu}_{\mu}$
and a second proportional to $\mu_{1}$. 
\begin{equation}
\hat{P}_{\mu}=\hat{\nu}_{\mu}\hat{p}_{\nu}+\mu_{1}\hat{p}_{\mu}\text{ and }\hat{J}_{\mu}=\hat{\nu}_{\mu}\hat{j}_{\nu}+\mu_{1}\hat{j}_{\mu}\label{eq:varmod-FP-flux-1}
\end{equation}
The resulting pair of equations can be numerically integrated yielding
the linear response rate amplitude and phase lag. 

Similarly, for variance modulations of the input current, the expansion
of the flux satisfies in first order:

\begin{equation}
\tau_{\mathrm{m}}\frac{\partial\hat{J}_{\sigma^{2}}}{\partial V}=i\omega\hat{P}_{\sigma^{2}}+\hat{\nu}_{\sigma^{2}}\delta(V-V_{\mathrm{th}})-\hat{\nu}_{\sigma^{2}}\delta(V-V_{\mathrm{re}})\label{eq:varmod-FP-flux}
\end{equation}

\begin{equation}
-\frac{\partial\hat{P}_{\sigma^{2}}}{\partial V}=\begin{cases}
\frac{\tau_{\mathrm{m}}}{\sigma_{0}^{2}}\left(\frac{-\mu_{0}-a_{\mathrm{S}}(V-V_{\mathrm{G}})^{2}}{\tau_{\mathrm{m}}}\hat{P}_{\sigma^{2}}+\hat{J}_{\sigma^{2}}+F_{\sigma^{2}}\right) & V\le V_{\mathrm{G}}\\
\frac{\tau_{\mathrm{m}}}{\sigma_{0}^{2}}\left(\frac{-\mu_{0}-a_{\mathrm{U}}(V-V_{\mathrm{G}})^{2}}{\tau_{\mathrm{m}}}\hat{P}_{\sigma^{2}}+\hat{J}_{\sigma^{2}}+F_{\sigma^{2}}\right) & V>V_{\mathrm{G}}.
\end{cases}\label{eq:varmod-FP-density}
\end{equation}
The inhomogeneous term $F_{\sigma^{2}}$ is defined by $F_{\sigma^{2}}=-\frac{\partial J}{\partial\sigma^{2}}P_{0}$.
It is given by 

\begin{equation}
F_{\sigma^{2}}=\begin{cases}
\frac{\sigma_{1}^{2}}{\sigma^{2}}(-\mu-a_{\mathrm{S}}(V-V_{\mathrm{G}})^{2}-V_{\mathrm{T}}) & V\le V_{\mathrm{G}}\\
\frac{\sigma_{1}^{2}}{\sigma^{2}}(-\mu-a_{\mathrm{U}}(V-V_{\mathrm{G}})^{2}-V_{\mathrm{T}}) & V>V_{\mathrm{G}}.
\end{cases}\label{eq:varmod-FP-density-1}
\end{equation}
The linear response of the Fokker Planck ansatz (FPA) is obtained with the same threshold integration
scheme as above.  We solve the linear response for the delay differential equation (DDE) similarly as above by the method of steps using higher-order solvers with high accuracy by setting up a threshold integration analogously to the established threshold-integration method proposed in~\citep{key-Richardson-2007}, extending the previous DDE approach in \cite{key-Manz2019} from the stationary case to the time-dependent case. A more detailed account of this method will be given elsewhere. We find excellent agreement between the DDE approach and direct numerical simulations using forward Euler in \textbf{Fig.}~\textbf{\ref{fig:FPA_DNS_linear_response-1-1}}. We note as described in the main text and in more detail below, that we observe a mismatch between the predictions of the FPA and direct numerical simulations especially for large $r$, and large $J_0$ in a regime of intermediate to high frequencies, which we explain by a breakdown of the diffusion approximation. This breakdown of the diffusion approximation is related but different from the breakdown of the diffusion approximation described previously \cite{key-Richardson2018}. As we are concerned about the inhibition-dominated balanced regime on recurrent spiking networks, we focus here on the effect of inhibitory shot-noise, in contrast to \cite{key-Richardson2018} that focused on excitatory shot noise in EIF neurons with exponentially distributed synaptic weights and also found  an effect of shot noise for small $\delta_T/a_s$, corresponding to large rapidness or large synaptic amplitude. A more recent paper extended that to conductance-based integrate-and-fire neurons and  considered both excitation and inhibition, but did not report the breakdown of the diffusion approximation for large rapidness $r$ discussed here \cite{key-Richardson2024}.
We finally note that the Fokker-Planck ansatz has the identical asymptotic limit for large frequencies, but that shot noise effects become relevant in an intermediate frequency range when $r$ and the synaptic amplitude are large. This is also different from the previously described excitatory shot-noise effect that changes also the asymptotic exponent of the frequency-response, due to the model choice (EIF) and the choice of an exponential synaptic weight distribution \cite{key-Richardson2018}.
\paragraph*{High-frequency limit:}

The high-frequency response is obtained by expanding $\hat{P}(t)$
in terms of $\frac{1}{\omega},$ similarly to \citep{key-Fourcaud-Trocm=0000E92003,key-Richardson-2007}.
The response to modulations of the mean is:

\begin{equation}
\hat{\nu}_{\mu}(\omega)=\nu_{0}\mu_{1}r(r+1)\frac{1}{(i\omega\tau_{\mathrm{m}})^{2}}\label{eq:rapid-frequency-response-mean}
\end{equation}

The high-frequency response for modulations of the variance is:

\begin{equation}
\hat{\nu}_{\sigma}(\omega)=3\nu_{0}\sigma_{1}^{2}r^{2}(r+1)^{2}\frac{1}{(i\omega\tau_{\mathrm{m}})^{3}}\label{eq:rapid-frequency-response-variance}
\end{equation}

The excellent agreement between analytical high-frequency response
and the Fokker-Planck approach is shown in \textbf{Fig.}~\textbf{\ref{fig:FPA_DNS_linear_response}}.
Building on previous work, the functional form of the voltage dependence
of the activation variable of fast sodium currents determines how
well the spiking of a fluctuation-driven neuron can reflect high frequency
input \citep{key-Fourcaud-Trocm=0000E92003,key-Fourcaud-Trocm=0000E92005,key-Naundorf2005,key-Naundorf2006,key-wei-2011}.
High spike onset rapidness allows neurons to precisely position their
spikes in time. Neurons with low spike onset rapidness are susceptible
to input fluctuations after crossing the unstable fixed point $V_{\mathrm{U}}$.
High-frequency components in the input are then washed out and only
weakly reflected in the spiking. Therefore, in the presence of noise,
the spikes of neurons with high action potential onset rapidness can
convey more information about high-frequency input currents. To support
this claim quantitatively, we calculated a lower bound on the mutual
information rate between input current and spike trains (Section \ref{sec:Mutual-information-rate}).

\paragraph*{Analytical high-frequency response for high rapidness limit:}

The high-frequency response in the limit of large rapidness is also
obtained by expanding $\hat{P}(t)$ in $\frac{1}{\omega}$ for $r=\infty$.
In this case $V_{\mathrm{G}}=V_{\mathrm{U}}$ acts as a hard threshold
and neurons crossing it spike instantaneously. The high-frequency
response to modulations of the mean is similar to that of the leaky
integrate-and-fire neuron:

\begin{equation}
\hat{\nu}_{\mu}=\frac{\nu_{0}\mu_{1}}{\sqrt{i\omega\tau_{\mathrm{m}}}}\label{eq:large-rapid-frequency-response-mean}
\end{equation}

\section{High-frequency behavior of rapid theta neurons with inhibitory shot noise}
\begin{figure*}
\noindent\begin{minipage}[t]{1\columnwidth}%
\includegraphics[clip,width=0.49\columnwidth]{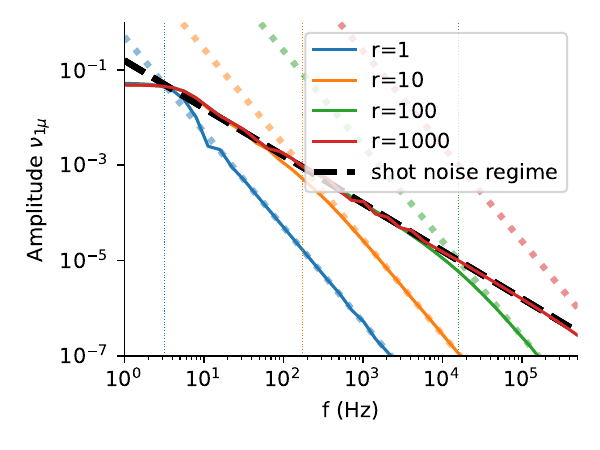}\includegraphics[clip,width=0.49\columnwidth]{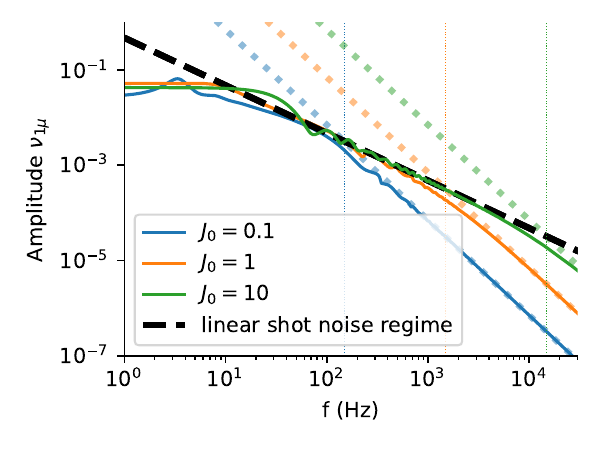}%
\end{minipage}

\caption{\label{fig:FPA_DNS_linear_response-1}\textbf{Qualitative change in
population response of shot-noise-driven rapid theta neurons to synaptic
rate modulations}: \textbf{a)} Linear response for mean modulation
of synaptic input rate for varying spike onset rapidness $r$. Dotted
lines represent the high-frequency asymptotics (Eq.~\ref{eq:quadratic-asymptotic-regime}),
while the black dashed line corresponds to the linear shot-noise limit
(Eq.~\ref{eq:linear-shot-noise-asymptotics}). For smaller $r$,
the frequency response transitions to a decay proportional to $\frac{1}{f^{2}}$
at much lower frequencies compared to larger values of $r$. \textbf{b)}
Same as \textbf{a)} for varying coupling strength $J_{0}.$ Weak coupling
($J_{0}=0.1)$ begins to fall of proportional to $\frac{1}{f^{2}}$
at lower frequencies compared to larger values of $r$ ($J_{0}=10)$.
(parameters: $\nu_{0}=1\,$Hz, $K=100$, $\tau_{\mathrm{m}}=10\,$ms.)}
\end{figure*}

\begin{figure*}
\noindent\begin{minipage}[t]{1\columnwidth}%
\includegraphics[clip,width=0.49\columnwidth]{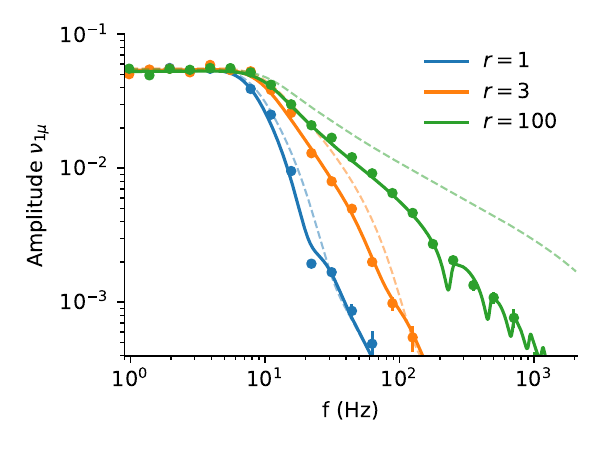}\includegraphics[clip,width=0.49\columnwidth]{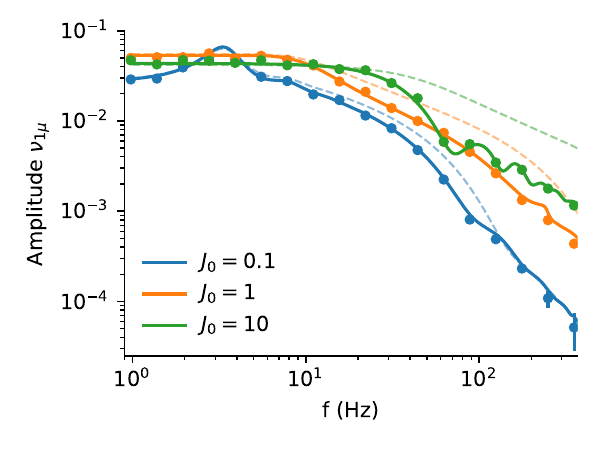}%
\end{minipage}

\caption{\label{fig:FPA_DNS_linear_response-1-1}\textbf{Frequency response
of rapid theta neurons driven by inhibitory shot-noise, described
by Fokker-Planck ansatz}: \textbf{a)} Linear response for mean modulation
of synaptic input rate for varying spike onset rapidness $r$. Full
lines are the shot-noise theory, dashed lines are the Fokker-Planck
theory dots are direct numerical simulations using Forward Euler.
Note that for large rapidness, the direct numerical simulations are
not well-captured by the Fokker-Planck theory but agree well with
the preditions by the shot-noise theory. This confirms that the diffusion
approximation inherent to the Fokker-Planck ansatz breaks down for
large $r$. \textbf{b)} Same as \textbf{a)} for varying coupling strength
$J_{0}.$ (parameters: $\nu_{0}=3\,$Hz, $K=100$, $\tau_{\mathrm{m}}=10\,$ms.
a) $J_{0}=1$, b) $r=10$)}
\end{figure*}

The rapid theta neuron model in rescaled variables is given by

\begin{equation}
F(V)\coloneqq\frac{dV}{dt}=\begin{cases}
(I+a_{s}V^{2})/\tau & V<0\\
(I+a_{u}V^{2})/\tau & V\geq0
\end{cases}\label{eq:rescaled-rapid-theta-1}
\end{equation}
where $V$ represents the rescaled voltage, $I$ is the constant input
current between two incoming spikes, $a_{s}$ and $a_{u}$ are parameters
controlling the voltage dynamics below and above $V=0$, respectively,
and $\tau$ is the time constant. We consider the regime $I>0$, ensuring
that $F(V)>0$.

We consider rapid theta neurons subject to inhibitory Poisson input
with a time-dependent rate $R(t)=\bar{R}+\hat{R}\mathrm{e}^{i\omega t}$
and a negative amplitude $W<0$. $\bar{R}$ denotes the average input
rate and $\hat{R}$ is the modulation amplitude. Similarly, the resulting
firing rate is by $\nu(t)=\bar{\nu}+\hat{\nu}\mathrm{e}^{i\omega t}$.

For large values of $V$, the flux $J(t)$ approximates the firing
rate $\nu(t)$, and the stationary voltage probability density $P(V)$
is given by 
\[
\bar{P}(V)\approx\frac{\bar{\nu}}{F(V)}\approx\frac{\bar{\nu}\tau}{a_{u}V^{2}}.
\]
where we ignored $I$ in the denominator because $a_{u}V^{2}$dominates
for large $V$. The stationary shot noise flux $\bar{J}_{{\rm s}}(V)$,
representing voltage jumps induced by incoming spikes, is approximated
as 
\begin{align*}
\bar{J}_{{\rm s}}(V) & \approx\bar{R}\int_{V}^{V+W}\bar{P}(u)\,du\\
 & \approx\bar{R}\int_{V}^{V+W}\,\frac{\bar{\nu}\tau}{a_{u}y^{2}}dy\\
 & =\frac{\bar{\nu}\tau W\bar{R}}{a_{u}V(V+W)}
\end{align*}

The high-frequency response $\hat{r}(\omega)$ is expressed as the
Laplace transform of the steady-state shot-noise flux: 
\[
\hat{r}(\omega)=\frac{\hat{R}}{\bar{R}}\int_{0}^{\infty}\mathrm{e}^{-\lambda}\bar{J_{s}}(\lambda,\omega)\,d\lambda,
\]
 where $\lambda(V,\omega)\coloneqq i\omega T(V)$. The escape time
$T(V)$ from voltage $V$ to spike time $V=\infty$ is approximated
as: 
\[
T(V)=\int_{V}^{\infty}\frac{dy}{F(y)}\approx\int_{V}^{\infty}\frac{\tau}{a_{u}y^{2}}\,dy=\frac{\tau}{a_{u}V}.
\]
Thus, 

\[
\lambda=i\omega T=\frac{i\omega\tau}{a_{u}V},\quad\text{and hence}\quad V=\frac{i\omega\tau}{a_{u}\lambda}.
\]

Plugging this into $\bar{J_{s}}(V)$, we get:
\[
\bar{J_{s}}(\lambda,\omega)=\frac{\bar{\nu}\tau W\bar{R}}{a_{u}\frac{i\omega\tau}{a_{u}\lambda}(\frac{i\omega\tau}{a_{u}\lambda}+W)}=\frac{\bar{\nu}\tau\bar{R}\lambda^{2}}{i\omega\tau(\frac{i\omega\tau}{a_{u}W}+\lambda)}.
\]

The high-frequency response $\hat{\nu}(\omega)$ becomes
\begin{align}
\hat{\nu}(\omega) & =\frac{\hat{R}}{\bar{R}}\int_{0}^{\infty}\mathrm{e}^{-\lambda}\bar{J_{s}}(\lambda,\omega)\,d\lambda\nonumber \\
 & =\hat{R}\bar{\nu}\tau\int_{0}^{\infty}\frac{\lambda^{2}\mathrm{e}^{-\lambda}}{i\omega\tau(\frac{i\omega\tau}{a_{u}W}+\lambda)}d\lambda\label{eq:high-frequency-full}
\end{align}

Let $\ensuremath{x=\frac{i\omega\tau}{a_{u}W}}$ and $t=\lambda+x$,
so $\lambda=t-x$ and $d\lambda=dt$. The limits of integration become
$x$ to $\infty$. Then the integral becomes

\begin{align*}
\hat{\nu}(\omega) & =\frac{\hat{R}\bar{\nu}e^{x}}{i\omega}\left[\int_{x}^{\infty}t\mathrm{e}^{-t}dt-2b\int_{x}^{\infty}\mathrm{e}^{-t}dt+x^{2}\int_{x}^{\infty}\frac{\mathrm{e}^{-t}}{t}dt\right]\\
 & =\hat{R}\bar{\nu}\left[\frac{2}{i\omega}-\frac{2\tau}{a_{u}W}-\frac{\omega\tau^{2}}{(a_{u}^{2}W^{2}}e^{\frac{i\omega\tau}{a_{u}W}}E_{1}\left(\frac{i\omega\tau}{a_{u}W}\right)\right]
\end{align*}
where $E_{1}(x)$ is the exponential integral function, $E_{1}(z)=\int_{z}^{\infty}\frac{\mathrm{e}^{-t}}{t}dt$.

The expression Eq.~\ref{eq:high-frequency-full} contains two terms
in the denominator, and depending on which term dominates, different
behaviors of $\hat{r}$ emerge. We will now analyze all possible regimes
and develop analytical transition conditions between these regimes.

\paragraph{Quadratic Asymptotic Regime (large $\omega$):}

For very large $\omega$ (i.e., $\omega\tau\gg1,a_{u},W$), the term
involving $\omega^{2}$ dominates the denominator. The denominator
simplifies to:

\[
D_{1}(\lambda)\approx-\dfrac{\omega^{2}\tau^{2}}{a_{u}W}.
\]

Thus, the response becomes:

\begin{align}
\hat{\nu}(a_{u},\omega,W) & \approx-\hat{R}\bar{\nu}\int_{0}^{\infty}d\lambda\,e^{-\lambda}\dfrac{a_{u}\lambda^{2}}{\omega^{2}\tau^{2}}\nonumber \\
 & =-\hat{R}\bar{\nu}\tau\dfrac{a_{u}}{(\omega\tau)^{2}}\int_{0}^{\infty}d\lambda\,e^{-\lambda}\lambda^{2}\nonumber \\
 & =-\dfrac{2a_{u}\hat{R}\bar{\nu}}{\omega^{2}\tau}.\label{eq:quadratic-asymptotic-regime}
\end{align}

Here, we used $\int_{0}^{\infty}d\lambda\,e^{-\lambda}\lambda^{2}=2$.

\paragraph{Linear Shot Noise Regime (large $W$, large $r$)}

When $a_{u}$ and $W$ are large and $\omega^{2}/(a_{u}\lambda^{2})$
is not dominant, the term $\dfrac{2i\omega W}{\lambda}$ dominates
the denominator:

\[
D_{2}(\lambda)\approx\dfrac{2i\omega W}{\lambda}.
\]

The response becomes:

\begin{equation}
\hat{\nu}(a_{u},\omega,W)\approx W\hat{R}\bar{\nu}\int_{0}^{\infty}d\lambda\,e^{-\lambda}\dfrac{\lambda}{2i\omega W}=\dfrac{\hat{R}\bar{\nu}}{2i\omega}\int_{0}^{\infty}d\lambda\,e^{-\lambda}\lambda=\dfrac{\hat{R}\bar{\nu}}{2i\omega}.\label{eq:linear-shot-noise-asymptotics}
\end{equation}

We used $\int_{0}^{\infty}d\lambda\,e^{-\lambda}\lambda=1$.

\paragraph{Constant Regime}

For low frequencies $\omega$, the linear response $\hat{r}$ can
be estimated using an adiabatic approximation, we perform a linear
expansion of the steady-state firing rate $\bar{\nu}$ with respect
to small perturbations in the input rate $\hat{R}$:

\[
\hat{\nu}_{\text{adiabatic}}(a_{u},\omega,W)\approx\dfrac{d\bar{\nu}}{d\bar{R}}\hat{R}
\]

\[
\hat{\nu}_{\text{adiabatic}}(a_{u},\omega,W)\approx-\dfrac{1}{2}\dfrac{W\tau\hat{R}\sqrt{\nu}}{\sqrt{\bar{R}}}
\]

\paragraph{Transition Between Quadratic and Linear Regime}

Set the quadratic frequency term equal to the linear frequency term:

\[
\left|-\dfrac{\omega^{2}\tau^{2}}{a_{u}\lambda^{2}}\right|\approx\left|\dfrac{2i\omega\tau W}{\lambda}\right|
\]

Assuming the most significant contributions come from $\lambda\approx1$:

\[
\omega_{\textnormal{transition}}\approx\frac{2a_{u}W}{\tau}
\]
This transition happens at a value of $|\hat{\nu}|$ of 
\[
|\hat{\nu}_{\textnormal{\textnormal{transition}}}|=W\hat{R}\bar{\nu}\dfrac{2a_{u}}{\omega_{\textnormal{\textnormal{transition}}}^{2}\tau^{2}}=W\hat{R}\bar{\nu}\dfrac{2a_{u}}{(2a_{u}W)^{2}}=\dfrac{\hat{R}\bar{\nu\tau}}{2Wa_{u}}
\]

\paragraph{Linking shot noise to recurrent network dynamics}

To convert back to the variables used in recurrent neural network,
we have to first go back to the original rapid theta model without
rescaled variables and then replace $R$ by compound input spike rate
$K\bar{\nu}$, where $K$ is the number of synapses per neuron and
$\bar{\nu}$ is the average firing rate. Moreover, we set $W=J_{0}/\sqrt{K}$,
which is the balanced scaling of the negative coupling strength that
maintains a finite variance of the input for large $K$. Moreover,
we remember that $a_{u}=r(r+1)/2$.

This gives for the 'quadratic regime' (large $\omega$ limit): 
\[
\hat{\nu}_{\text{I}}(a_{u},\omega,W)=-\dfrac{2W\hat{R}\bar{\nu}a_{u}}{\omega^{2}\tau^{2}}.
\]
For the linear 'shot noise regime' (large $W$ and large $r$): 
\[
\hat{\nu}_{\text{II}}(a_{u},\omega,W)=\dfrac{\hat{R}\bar{\nu}\tau}{2i\omega\tau}.
\]
Thus, the intersection of the linear and quadratic regime is at: 
\[
\omega_{\textnormal{transition}}\approx\frac{2a_{u}W}{\tau}
\]
which corresponds to a value of $|\hat{\nu}|$ of 
\[
|\hat{\nu}_{\textnormal{critical}}|=W\hat{R}\bar{\nu\tau}\dfrac{2a_{u}}{\omega_{\textnormal{transition}}^{2}\tau^{2}}=W\hat{R}\bar{\nu}\tau\dfrac{2a_{u}}{(2a_{u}W)^{2}}=\dfrac{\hat{R}\bar{\nu}\tau}{2Wa_{u}}
\]
To determine if the linear shot noise regime contributes at all, we
have to determine if the transition between quadratic and linear regime
happens before or after the constant regime dominates with $|\hat{\nu}_{\textnormal{constant}}|=\frac{\hat{R}}{\sqrt{K}}$

\[
|\hat{\nu}_{\textnormal{transition}}|=W\hat{R}\bar{\nu}\tau\dfrac{2a_{u}}{\omega_{\textnormal{transition}}^{2}\tau^{2}}=W\hat{R}\bar{\nu}\tau\dfrac{2a_{u}}{(2a_{u}W)^{2}}=\dfrac{\hat{R}\bar{\nu}\tau}{2Wa_{u}}=\frac{\hat{R}}{\sqrt{K}}
\]
Plugging in $a_{u}=\frac{r(r+1)}{2}$ and $W=\frac{J_{0}}{\sqrt{K}}$
gives: 
\[
|\hat{\nu}_{\textnormal{transition}}|=\dfrac{\hat{R}\bar{\nu}\tau}{J_{0}r(r+1)}=\frac{\hat{R}}{\sqrt{K}}
\]
Solving by $r$ gives: 
\[
r=\frac{\sqrt{1+4\frac{K\bar{\nu}\tau}{J_{0}}}-1}{2}
\]
For large $r$ this is 
\[
r\approx\sqrt{\frac{K\bar{\nu}\tau}{J_{0}}}
\]
Compare to $r_{\mathrm{localization}}$ 
\[
r_{\mathrm{localization}}\propto\sqrt{\frac{K\nu_{0}\tau_{\mathrm{m}}}{J_{0}}}.
\]
and 
\[
r_{\mathrm{peak}}\propto\sqrt{\frac{K\nu_{0}\tau_{\mathrm{m}}}{J_{0}}}
\]

\section{Mutual information rate and action potential onset rapidness\label{sec:Mutual-information-rate}}

\paragraph{Figure 1 h of main paper.}

\paragraph{Lower bound on mutual information rate:}

\begin{figure*}
\noindent\begin{minipage}[t]{1\columnwidth}%
\includegraphics[clip,width=1\columnwidth]{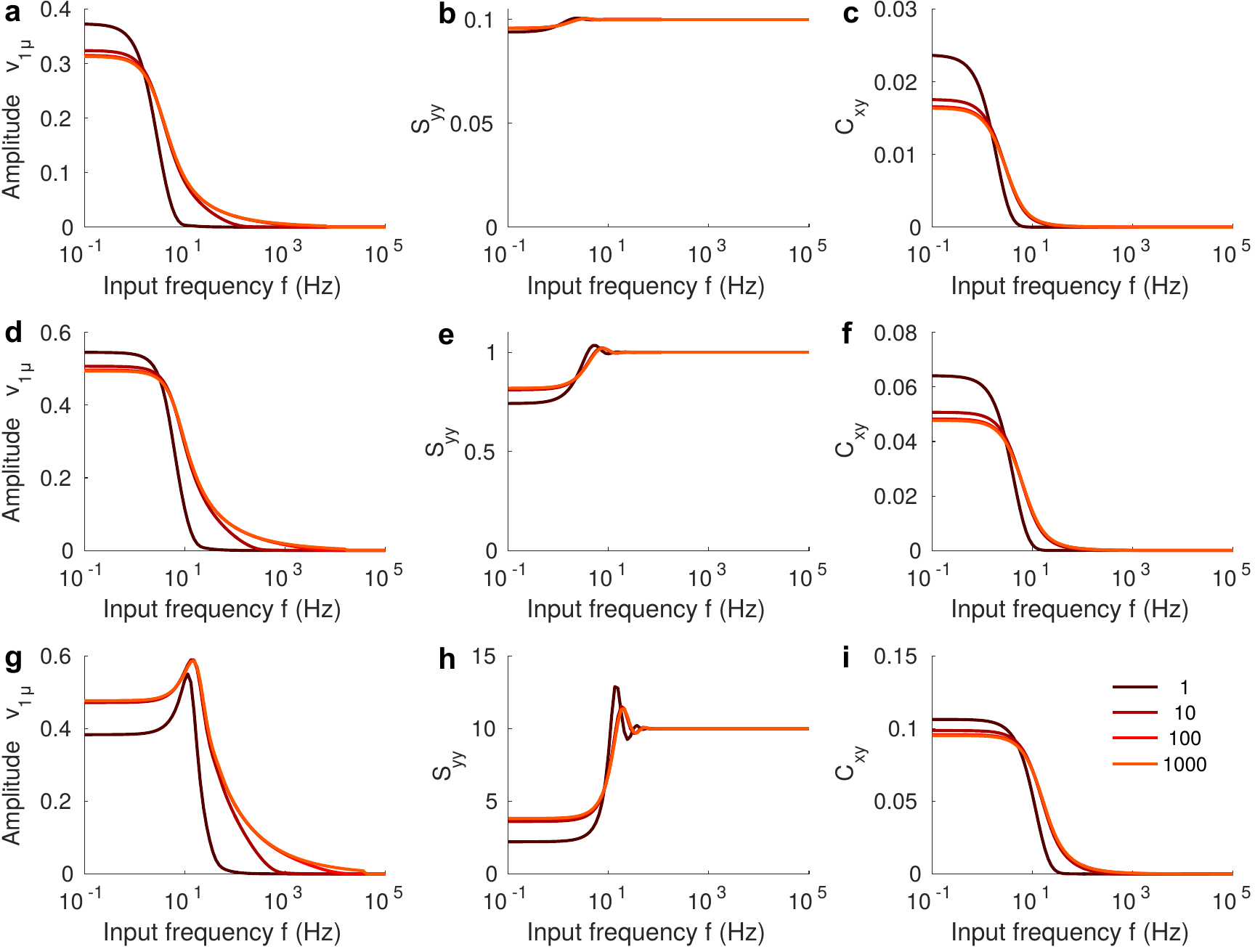}%
\end{minipage}

\caption{\label{fig:FPA_Syy_Sxy_coherence}\textbf{Action potential onset rapidness
}$r$\textbf{ and }$\nu_{0}$\textbf{ shape frequency-response, spike
power spectra and spectral coherence:} \textbf{a)} Linear response
for mean modulation of input current for $\nu_{0}=0.1\,$Hz, \textbf{b)}
Spike power spectra for different $r$\textbf{ }for $\nu_{0}=0.1\,$Hz.
\textbf{c)} spectral coherence for different $r$ for $\nu_{0}=0.1\,$Hz.
\textbf{d)}, \textbf{e)} and \textbf{f)} same as \textbf{a}, \textbf{b}
and \textbf{c} for $\nu_{0}=1\,$Hz. At low rates, neurons are in
the fluctuation-driven regime, therefore $S_{yy}$ is flat. At high
rates (mean-driven regime), the response amplitude $\nu_{1\mu}$ has
a resonance close to $\nu_{0}$ (parameters: $\nu_{0}=10\,$Hz, $\mu_{1}=0.01$,
$\tau_{\mathrm{m}}=10\,$ms, $\Delta V=10^{-4}$, $V_{\mathrm{th}}=10^{4}=-V_{\mathrm{r}}$).}
\end{figure*}

Using information theory, we can treat a neuron as a noisy communication
channel, transforming a signal embedded in a noisy current into a
spiking response sent to its postsynaptic partners. The mutual information
rate measures how much the uncertainty about the input $x(t)$ is
reduced given the spiking output $y(t)$ per unit time:

\begin{equation}
R(X,Y)=h(Y)-h(Y|X)=h(X)-h(X|Y)=\lim_{T\to\infty}\frac{1}{T}\intop_{X}\intop_{Y}p(x,y)\log{}_{2}\left(\frac{p(x,y)}{p(x)p(y)}\right)\label{eq:mutual-info}
\end{equation}
where $p(x,y)$ is the joint probability density function of $x(t)$
and $y(t)$, $h(X)$ and $h(Y)$ are their entropy rates, and $h(X|Y)$
is the conditional entropy rate of $x(t)$ given $y(t)$. The continuous
Gaussian channel provides a lower bound on the mutual information
rate for a Gaussian input signal~\citep{key-Bialek1991,key-Rieke}:
\begin{equation}
R(X,Y)=h(X)-h(X|Y)\geq h(X)-h_{\mathrm{Gaussian}}(X|Y)=R_{\mathrm{lb}}(X,Y)\label{eq:mutual-info-lb}
\end{equation}
The inequality results from the property that a Gaussian process has
the maximum entropy of all processes with fixed variance. Recently,
it was shown that for moderate input modulation in a fluctuation-driven
regime, this lower bound is very close to the mutual information rate
estimated from direct methods \citep{key-Bernardi}. This is convenient
because estimating the mutual information from empirical data is computationally
costly, requiring the sample size to be much larger than the size
of the alphabet \citep{key-Strong-1998}. Continuous processes usually
have to be discretized resulting in very large alphabets. In our case,
a Gaussian channel approximation to the mutual information rate between
input current and output spike train was estimated based on the spectral
coherence between input current and output spike train, which relies
purely on second-order statistics. 

\begin{equation}
R_{\mathrm{lb}}(X,Y)=-\int_{0}^{f_{\mathrm{cutoff}}}df\log_{2}\left(1-C_{xy}(f)\right),\label{eq:mutual-info-lb-coherence}
\end{equation}
where $C_{xy}(f)$ denotes the magnitude squared spectral coherence.
The spectral coherence is the frequency-domain analog of correlation,
measuring the linear relationship between frequency components of
input and output signal. Its magnitude squared is 

\begin{equation}
C_{xy}(f)=\frac{|S_{xy}(f)|^{2}}{|S_{xx}(f)||S_{yy}(f)|}\label{eq:spectral-coherence}
\end{equation}
$S_{xx}(f)$ is the power spectrum of band-limited Gaussian white
noise:

\begin{equation}
S_{xx}(f)=\begin{cases}
\sigma^{2} & f\le f_{\mathrm{cutoff}}\\
0 & \mathrm{else}.
\end{cases}\label{eq:bandlimited-noise-psd}
\end{equation}
$S_{yy}(f)$ is the power spectrum of the spike train: $S_{yy}(f)=\lim_{T\to\infty}\frac{1}{T}\left\langle \tilde{y}\tilde{y}^{*}\right\rangle $,
where $\tilde{y}(f)$ is the Fourier transform of the spike train.
If and only if $x(t)$ and $y(t)$ are linearly scaled copies of each
other, then $C_{xy}(f)$=1. If $x(t)$ and $y(t)$ are independent,
then $C_{xy}(f)$=0; however, the reverse generally does not hold.
Nonlinear effects and noise reduce the coherence. We calculated the
spectral coherence in two independent ways: first, based on the numerical
solution of the time-dependent Fokker-Planck equation, and second,
based on direct numerical simulations with band-limited white noise.
The time-dependent Fokker-Planck solution using the threshold integration
scheme above yields a linear response approximation of $S_{xy}(f)$
directly from the linear response. Assuming weak modulations of the
input, the stationary power spectrum of the spike train is sufficient
and can be obtained using again threshold integration again~\citep{key-Richardson-2007}.
The dependences of $S_{yy}(f)$, $S_{xy}(f)$ and $C_{xy}(f)$ on
rapidness $r$ and mean rate $\nu_{0}$ are depicted in \textbf{Fig.~\ref{fig:FPA_Syy_Sxy_coherence}}.
At low rates, due to our balanced scaling (see Section~\ref{sec:Setup-of-network}),
neurons are in the fluctuation driven regime, therefore, $S_{yy}$
is flat. At high firing rates, corresponding to a more mean-driven
regime, the response amplitude $\nu_{1\mu}$ has a resonance close
to $\nu_{0}$.

\begin{figure*}
\noindent\begin{minipage}[t]{1\columnwidth}%
\includegraphics[clip,width=1\columnwidth]{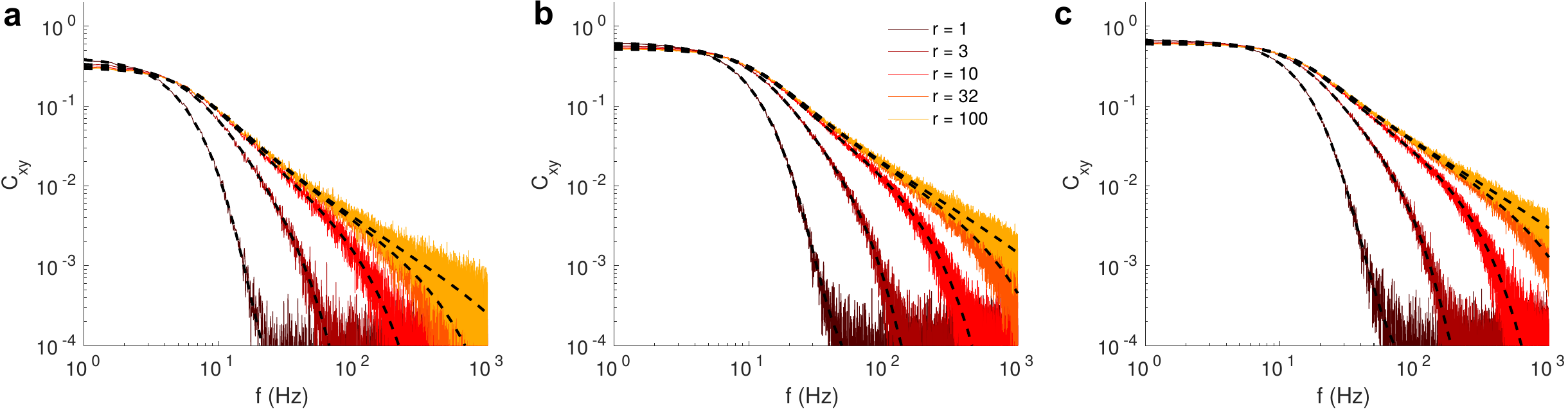}%
\end{minipage}

\caption{\label{fig:FPA_mi_coherence_DNS-varyRap-varyRate}\textbf{Action potential
onset rapidness $r$} \textbf{and $\nu_{0}$ shape spectral coherence}:
\textbf{(Fokker-Planck ansatz vs. direct numerical simulations).}
\textbf{a)} \textbf{$\nu_{0}=1\,$}Hz \textbf{b)} \textbf{$\nu_{0}=5\,$}Hz
\textbf{c)} \textbf{$\nu_{0}=10\,$}Hz. \textbf{$C_{xy}$} normalized
by input modulation \textbf{$\mu_{1}^{2}$} (parameters: $\mu_{1}=0.01$,
$\tau_{\mathrm{m}}=10\,$ms, $\Delta V=10^{-3}$, $V_{\mathrm{th}}=10^{3}=-V_{\mathrm{r}}$)}
\end{figure*}

The Fokker-Planck ansatz and direct numerical simulations with band-limited
white noise are in excellent agreement (\textbf{Fig.}~\textbf{\ref{fig:FPA_mi_coherence_DNS-varyRap-varyRate}}).

\begin{figure*}
\noindent\begin{minipage}[t]{1\columnwidth}%
\includegraphics[clip,width=1\columnwidth]{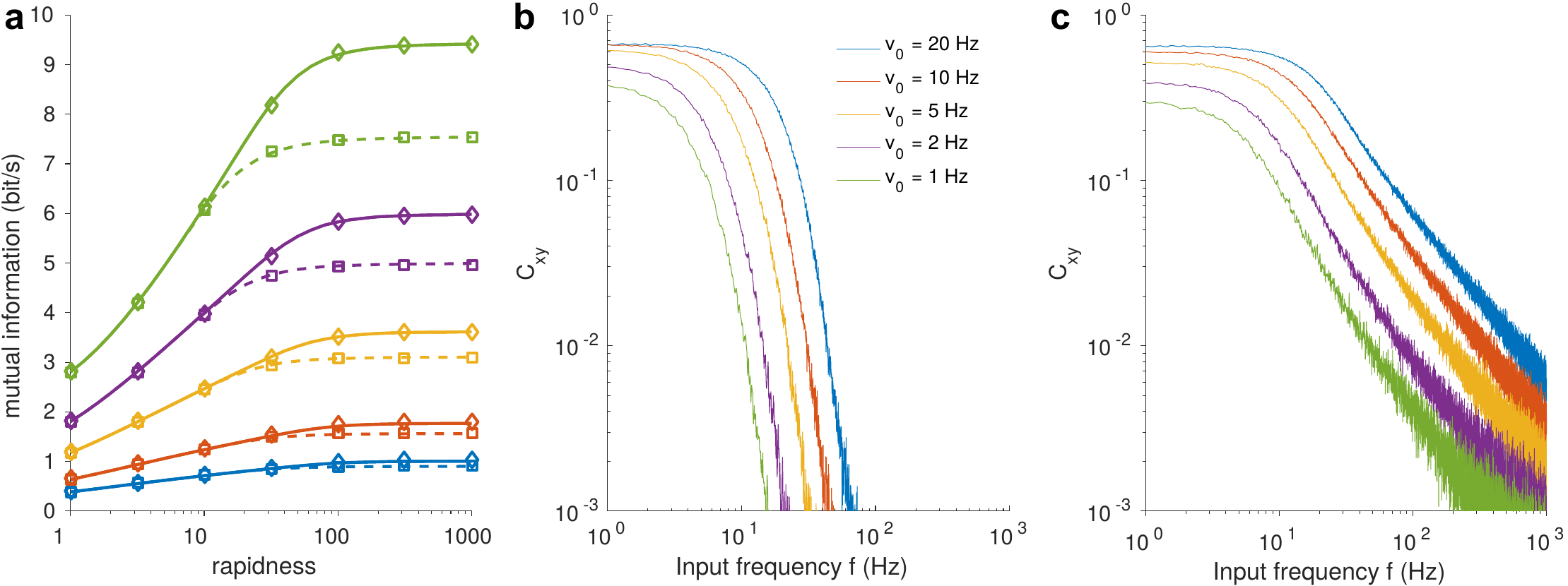}%
\end{minipage}

\caption{\label{fig:FPA_mi_coherence_DNS_varyRate}\textbf{Action potential
onset rapidness $r$ and $\nu_{0}$ limit mutual information rate
and spectral coherence}: \textbf{a)} Mutual information rate for two
band limits of the incoming white noise: dashed lines for \textbf{$f_{\mathrm{bandlimit}}=500\,$}Hz
and full lines for \textbf{$f_{\mathrm{bandlimit}}=2000\,$}Hz both
from Fokker-Planck approach. Diamonds and squares are the corresponding
direct numerical simulations with band-limited white noise input.
\textbf{b)} Spectral coherence for different firing rates from direct
numerical simulations \textbf{$r=1$}. \textbf{c)} same as \textbf{b)}
for \textbf{$r=1000$} (parameters: $\nu_{0}=10\,$Hz, $\mu_{1}=0.01$,
$\tau_{\mathrm{m}}=10\,$ms, $\Delta V=10^{-3}$, $V_{\mathrm{th}}=10^{3}=-V_{\mathrm{r}}$).}
\end{figure*}

 The mutual information rate scales approximately logarithmically
with $r$. For large rapidness, the mutual information rate saturates,
with the saturation level is determined by the band limit of the incoming
white noise (\textbf{Fig.~\ref{fig:FPA_mi_coherence_DNS_varyRate}}). 

\paragraph{Dependence of mutual information rate on spike rapidness}

We know $S_{xy}(f)$ in both in the high-frequency limit from the
analytical high-frequency response and the low-frequency limit because,
for very low frequency input modulations ($f\tau_{\mathrm{m}}\lll1)$,
the spike rate follows the input adiabatically: 
\begin{equation}
\lim_{f\to0}S_{xy}(f)=\mu_{1}\frac{\partial\nu_{0}}{\partial\mu}\label{eq:cross-spectral-density-low}
\end{equation}
This is the slope of the $\nu_{0}-\mu$ curve depicted in \textbf{Fig.~\ref{fig:FPA_rate_nu_sigma_convergence}}.
For renewal processes, $S_{yy}$ is given, for high input frequencies,
by the firing rate 

\begin{equation}
\lim_{f\to\infty}S_{yy}(f)=\nu_{0}\label{eq:autospectrum-low}
\end{equation}
and for low input frequencies 

\begin{equation}
\lim_{f\to0}S_{yy}(f)=\nu_{0}cv{}^{2},\label{eq:autospectrum-high}
\end{equation}
where $\mathrm{cv}$ is the coefficient of variation of the interspike
intervals $T$: $\mathrm{cv}=\sqrt{\langle T^{2}\rangle-\langle T\rangle^{2}}/\langle T\rangle$.
For intermediate frequencies at intermediate to high onset rapidness,
an approximation of the frequency-response of the rapid theta neuron
can be obtained from the analytical high-frequency response for high
rapidness (Eq.~\ref{eq:large-rapid-frequency-response-mean}). The
reason is that for intermediate input frequencies, the glue point
$V_{\mathrm{G}}$ acts like a hard threshold, if the time from threshold
$V_{\mathrm{U}}$ to spike is smaller than the inverse frequency.
Plugging these estimates into Eq.~\ref{eq:spectral-coherence}, we
obtain an analytical approximation of the spectral coherence for different
frequency regimes:

\begin{equation}
C_{xy}^{\mathrm{low}}(f)=\frac{|S_{xy}(f)|^{2}}{|S_{xx}(f)||S_{yy}(f)|}=\frac{|\mu_{1}\frac{\partial\nu_{0}}{\partial\mu}|^{2}}{\sigma^{2}\nu_{0}cv^{2}}=\mu_{1}^{2}\frac{\left(\frac{\partial\nu_{0}}{\partial\mu}\right)^{2}}{\sigma^{2}\nu_{0}cv^{2}}\label{eq:coherence-analytics-low}
\end{equation}

\begin{equation}
C_{xy}^{\mathrm{mid}}(f)=\frac{|S_{xy}(f)|^{2}}{|S_{xx}(f)||S_{yy}(f)|}==\frac{|\frac{\nu_{0}\mu_{1}}{\sqrt{i\omega\tau_{\mathrm{m}}}}|^{2}\sigma^{4}}{\sigma^{2}\nu_{0}}=\frac{\nu_{0}\mu_{1}^{2}\sigma^{2}}{2\pi f\tau_{\mathrm{m}}}\label{eq:coherence-analytics-mid}
\end{equation}
\begin{equation}
C_{xy}^{\mathrm{high}}(f)=\frac{|S_{xy}(f)|^{2}}{|S_{xx}(f)||S_{yy}(f)|}=\frac{|\frac{\nu_{0}\mu_{1}r(r+1)}{(i\omega\tau_{\mathrm{m}})^{2}}|^{2}\sigma^{4}}{\sigma^{2}\nu_{0}}=\frac{\nu_{0}\mu_{1}^{2}r^{2}(r+1)^{2}\sigma^{2}}{(2\pi f\tau_{\mathrm{m}})^{4}}\label{eq:coherence-analytics-high}
\end{equation}

\begin{figure*}
\noindent\begin{minipage}[t]{1\columnwidth}%
\includegraphics[clip,width=1\columnwidth]{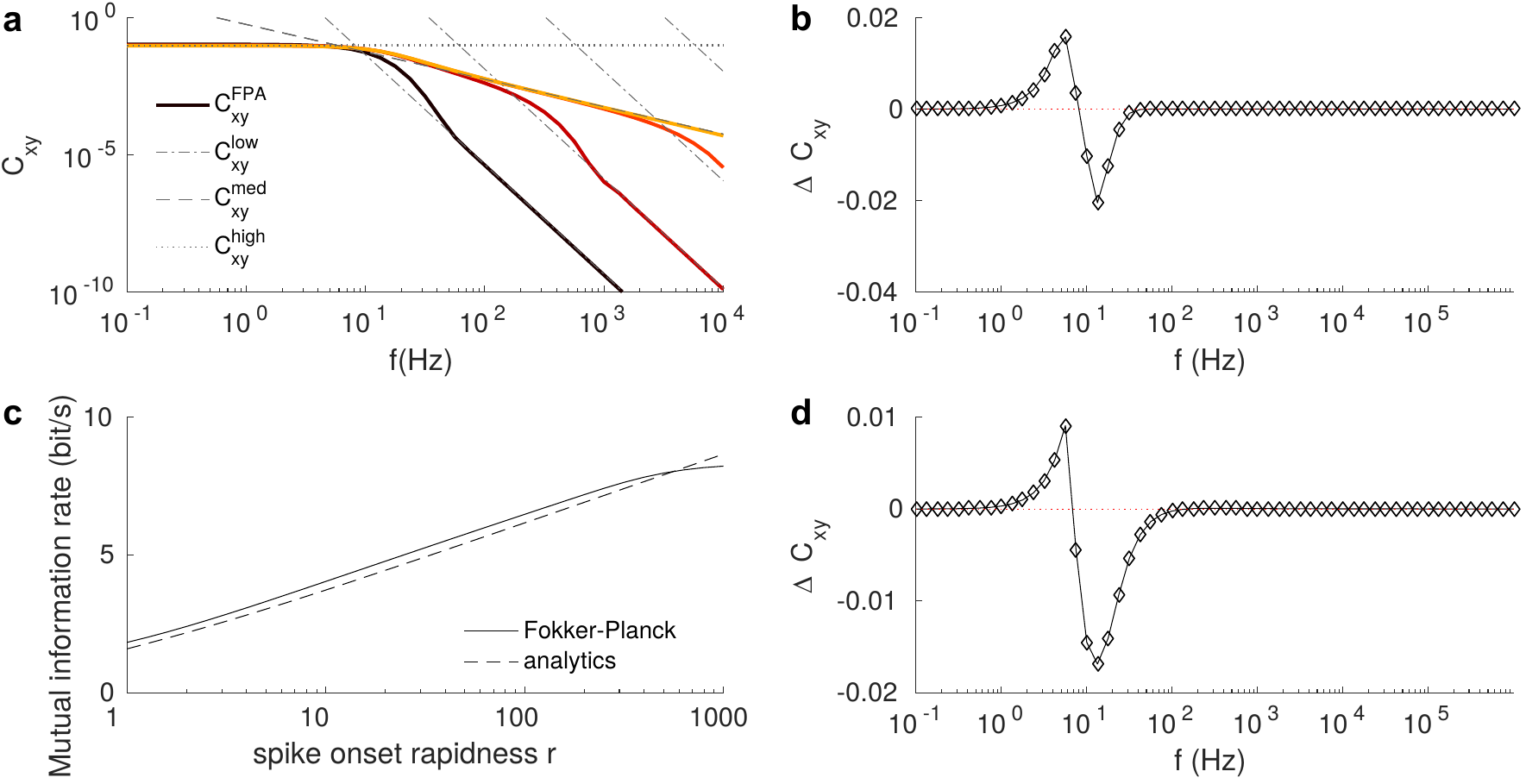}%
\end{minipage}

\caption{\label{fig:FPA_mi_coherence_FPA-analytic10Hz}\textbf{Analytical estimate
of the mutual information rate and spectral coherence explains logarithmic
scaling with action potential onset rapidness }$r$\textbf{ }$R_{\mathrm{lb}}(r)\propto\log\left(r\right)$\textbf{:}
\textbf{a)} Spectral coherence obtained from Fokker-Planck ansatz
vs analytical estimates for low mid and high frequency regime. Dashed
line: high-rapidness scaling, dash-dotted line: high-frequency scaling,
dotted line: low frequency estimate. \textbf{b)} Deviation between
Fokker-Planck ansatz and analytical estimates for \textbf{$r=1$}.
\textbf{c)} comparison of mutual information rates obtained from Fokker-Planck
ansatz vs. analytical estimate. \textbf{d)} Same as \textbf{b} for
\textbf{$r=1000$} (parameters: $\nu_{0}=10\,$Hz, $\mu_{1}=0.01$,
$\tau_{\mathrm{m}}=10\,$ms, $\Delta V=10^{-3}$, $V_{\mathrm{th}}=10^{3}=-V_{\mathrm{r}}$).}
\end{figure*}

These approximations are precise in their respective limits of low
and high frequencies. However, in the intermediate transition regimes,
deviations occur (\textbf{Fig.~\ref{fig:FPA_mi_coherence_FPA-analytic10Hz}}). 

\begin{figure*}
\noindent\begin{minipage}[t]{1\columnwidth}%
\includegraphics[clip,width=1\columnwidth]{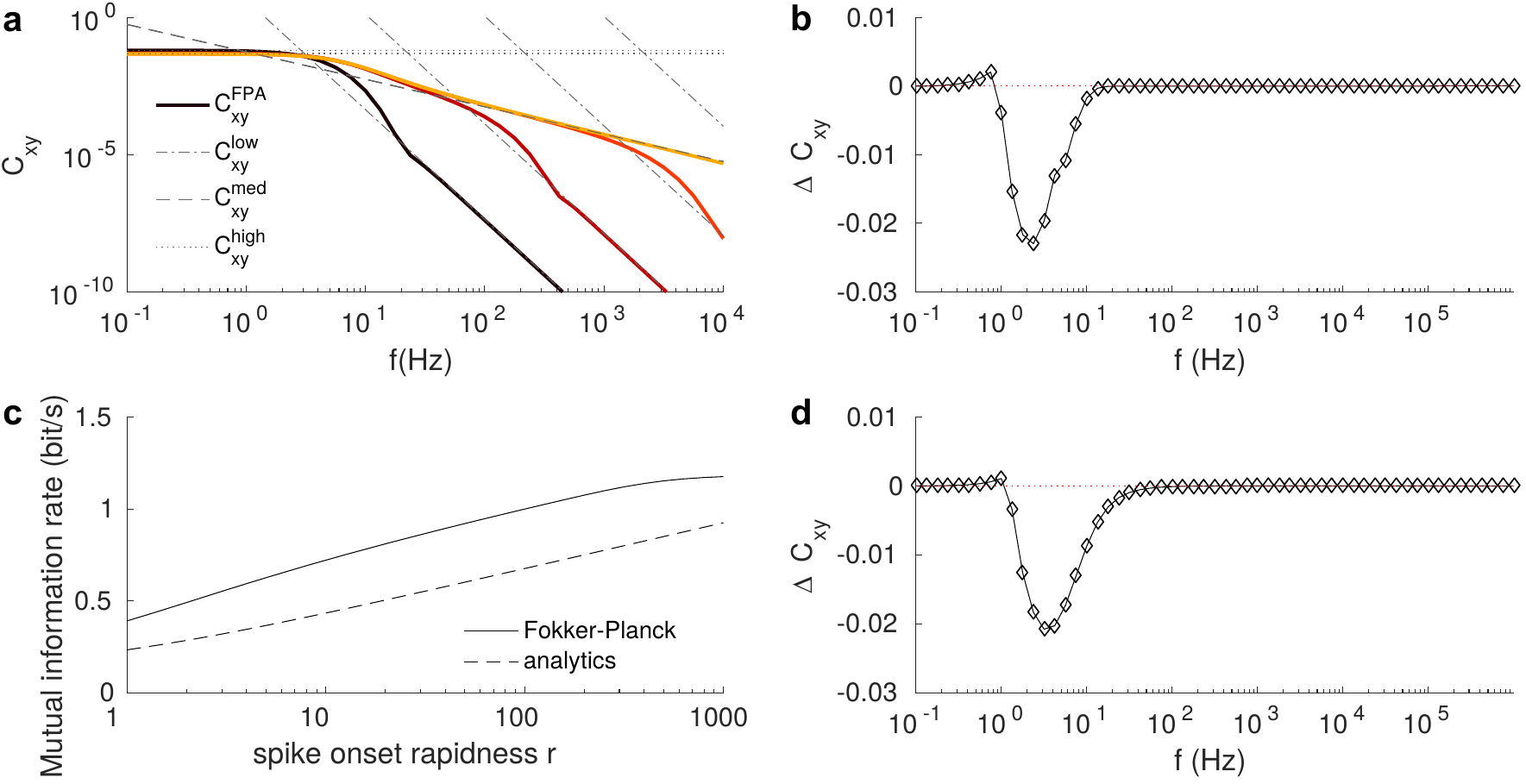}%
\end{minipage}

\caption{\label{fig:FPA_mi_coherence_FPA-analytic1Hz}\textbf{Analytical estimate
of the mutual information rate and spectral coherence with action
potential onset rapidness $r$ for low firing rate:} \textbf{$\nu_{0}=1$}
(parameters same as \textbf{Fig.~\ref{fig:FPA_mi_coherence_FPA-analytic1Hz}}).}
\end{figure*}

For intermediate mean firing rates, the analytical low coherence approximation
overestimates the coherence, while at the transition zone, the analytical
intermediate approximation underestimates the coherence. These two
errors approximately compensate each other (\textbf{Fig.~\ref{fig:FPA_mi_coherence_FPA-analytic10Hz}b}).
At low rates, the overestimation in the analytical low coherence regime
is smaller; therefore, the analytically obtained approach underestimates
the mutual information rate (\textbf{Fig.~\ref{fig:FPA_mi_coherence_FPA-analytic1Hz}}).

\begin{figure*}
\noindent\begin{minipage}[t]{1\columnwidth}%
\includegraphics[clip,width=1\columnwidth]{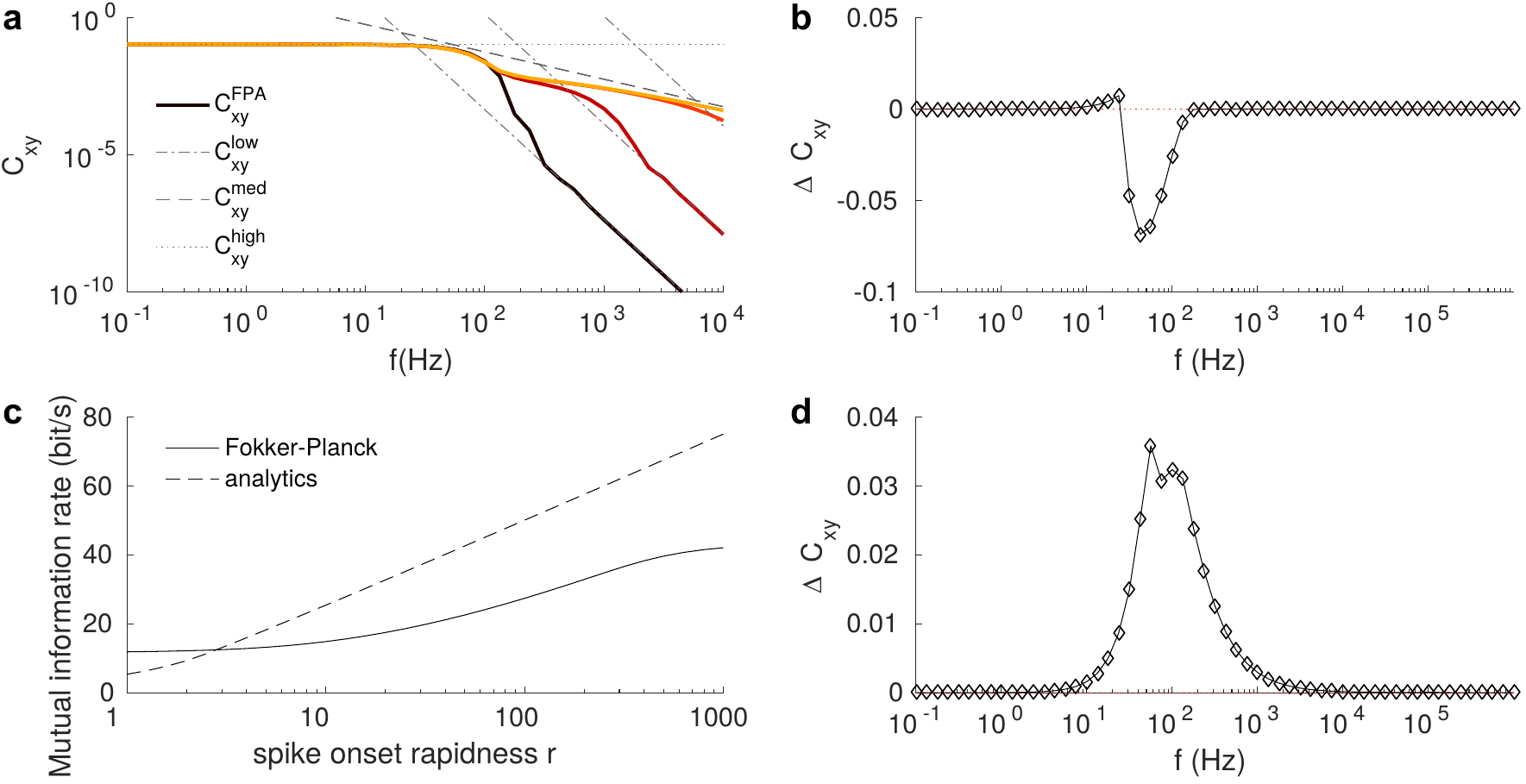}%
\end{minipage}

\caption{\label{fig:FPA_mi_coherence_FPA-analytic100Hz}\textbf{Analytical
estimate of the mutual information rate and spectral coherence for
different values of action potential onset rapidness $r$ for high
firing rate} (\textbf{$\nu_{0}=100$}, other parameters same as \textbf{Fig.~\ref{fig:FPA_mi_coherence_FPA-analytic1Hz}}).}
\end{figure*}

For high firing rates, the analytical coherence approximation strongly
underestimates the coherence, therefore the analytical approach overestimates
the mutual information rate (\textbf{Fig.~\ref{fig:FPA_mi_coherence_FPA-analytic100Hz}}).
The estimation errors are largely insensitive to rapidness, therefore
the slopes are still in good agreement with the Fokker-Planck approach.

The transition point between the low and intermediate regimes is obtained
by calculating the crossings between $C_{xy}^{\mathrm{low}}(f)$ and
$C_{xy}^{\mathrm{mid}}(f)$ as follows:

\begin{equation}
f^{\mathrm{lm}}(r,\nu_{0})=\frac{\nu_{0}^{2}\sigma^{4}cv^{2}}{\left(\frac{\partial\nu_{0}}{\partial\mu}\right)^{2}\tau_{\mathrm{m}}}\label{eq:coherence-crossing-low-middle}
\end{equation}
For weak noise input, we obtain from Eq. \ref{eq:nu-Iext} that $\frac{\partial\nu_{0}}{\partial\mu}=\frac{1}{\pi\tau_{\mathrm{m}}}\sqrt{\frac{r}{2r(r+1)I_{\mathrm{ext}}}}$.
The same is done for the second crossing with $C_{xy}^{\mathrm{mid}}(f)$
and $C_{xy}^{\mathrm{high}}(f)$:

\begin{equation}
f^{\mathrm{mh}}(r,\nu_{0})=\left[r(r+1)\sqrt{(\nu_{0}\tau_{\mathrm{m}}}\right]^{2/3}\label{eq:coherence-crossing-middle-high}
\end{equation}
To obtain an analytical approximation of the mutual information rate,
we stitched together analytical high-, middle-, and low-frequency
approximations at the transition points. Plugging these into Eq.~\ref{eq:mutual-info-lb-coherence}
gives us:
\begin{eqnarray}
R_{\mathrm{lb}} & = & R_{\mathrm{lb}}+R_{\mathrm{lb}}+R_{\mathrm{lb}}\nonumber \\
 & \thickapprox & -\int_{0}^{f_{\mathrm{lm}}}df\log_{2}\left(1-C_{xy}^{\mathrm{low}}(f)\right)-\int_{f_{\mathrm{lm}}}^{f_{\mathrm{mh}}}df\log_{2}\left(1-C_{xy}^{\mathrm{mid}}(f)\right)-\int_{f_{\mathrm{mh}}}^{f_{\mathrm{cutoff}}}df\log_{2}\left(1-C_{xy}^{\mathrm{high}}(f)\right)\nonumber \\
\label{eq:mutual-information-analytics}
\end{eqnarray}
Only the contribution of the intermediate frequency term changes strongly
as a function of rapidness in the relevant parameter regime. To obtain
a simple analytical approximation for the dependence on rapidness,
we need to evaluate only the second integral.
\begin{eqnarray}
R_{\mathrm{lb}} & \thickapprox & -\int_{f_{\mathrm{lm}}}^{f_{\mathrm{mh}}}df\log_{2}\left(1-C_{xy}^{\mathrm{mid}}(f)\right)\nonumber \\
 & = & -\int_{f_{\mathrm{lm}}}^{f_{\mathrm{mh}}}df\log_{2}\left(1-\frac{c_{1}}{f}\right)\nonumber \\
 & = & \left[c_{1}\log_{2}(f-c_{1})-f\log_{2}(1-\frac{c_{1}}{f})\right]_{f_{\mathrm{lm}}}^{f_{\mathrm{mh}}}\label{eq:mutual-information-analytics-1}\\
 & = & c_{1}\log_{2}(f_{\mathrm{mh}}-c_{1})-f_{\mathrm{lm}}\log_{2}(f_{\mathrm{lm}})+(f_{\mathrm{lm}}-c_{1})\log_{2}(f_{\mathrm{lm}}-c_{1})-f_{\mathrm{mh}}\log_{2}\left(1-\frac{c_{1}}{f_{\mathrm{mh}}}\right)\nonumber \\
\nonumber 
\end{eqnarray}
As short-hand notation, we defined $c_{1}=\frac{\nu_{0}\mu_{1}^{2}\sigma^{2}}{2\pi\tau_{\mathrm{m}}}$.
Only the first term changes strongly as a function of rapidness, therefore

\begin{eqnarray}
R_{\mathrm{lb}}(r) & \overset{}{\thickapprox} & c_{1}\log_{2}(f_{\mathrm{mh}}-c_{1})+const\nonumber \\
 & = & c_{1}\log_{2}\left(\left[r(r+1)\sqrt{(\nu_{0}\tau_{\mathrm{m}}}\right]^{2/3}-c_{1}\right)+const\nonumber \\
 & \overset{r\gg c_{1}}{\thickapprox} & c_{1}\log_{2}\left(\left[r(r+1)\sqrt{(\nu_{0}\tau_{\mathrm{m}}}\right]^{2/3}\right)+const\label{eq:mutual-information-analytics-1-1}\\
 & = & c_{1}\frac{2}{3}\log_{2}\left(r(r+1)\sqrt{(\nu_{0}\tau_{\mathrm{m}}}\right)+const\nonumber \\
 & = & c_{1}\frac{2}{3}\log_{2}\left(r(r+1)\right)+const\nonumber \\
 & \overset{r\gg1}{\thickapprox} & c_{1}\frac{4}{3}\log_{2}\left(r\right)+const\nonumber \\
 & = & \frac{2\nu_{0}\mu_{1}^{2}\sigma^{2}}{3\pi\tau_{\mathrm{m}}}\log_{2}\left(r\right)+const\nonumber 
\end{eqnarray}

We conclude that the mutual information rate scales approximately
logarithmically with action potential onset rapidness for sufficiently
large rapidness, weak input, and high band limit. Where does the logarithmical
scaling of the mutual information rate with rapidness arises from?
The reason is that the rapidness determines the cutoff frequency,
up to which the rapid theta neuron transmits information approximately
like a hard-threshold neuron (e.g., a leaky integrate-and-fire neuron).
Contributions from frequencies beyond this cutoff are negligible to
the first order. As a hard threshold always causes an asymptotic linear
response $\propto1/\sqrt{f}$, which becomes proportional to $1/f$
in the squared coherence, the integral that yields the lower bound
of the mutual information (Eq.~\ref{eq:mutual-info-lb-coherence})
always scales approximately proportional to $\log\left(f_{\mathrm{cutoff}}\right)$.
Therefore $R_{\mathrm{lb}}(r)\propto\log\left(r\right)$. This conclusion
is not restricted to the rapid theta model; a similar line of argumentation
holds also, e.g., for the exponential integrate-and-fire model and
the rapid $\tau$ model \citep{key-Fourcaud-Trocm=0000E92003,key-wei-2011}.
For the exponential integrate-and-fire neuron model, the mutual information
rate also scales approximately logarithmically with the action potential
onset rapidness parameter $1/\Delta_{T}$ despite the different asymptotic
linear response, which scales $\hat{\nu}_{\mu}\propto\frac{1}{f}$
\citep{key-Fourcaud-Trocm=0000E92003}. In the case of the rapid $\tau$
model, the same holds, despite the asymptotic linear response, which
scales $\hat{\nu}_{\mu}\propto\frac{1}{\sqrt{f}}\exp(\frac{\omega}{r})$
\citep{key-wei-2011}.

\section{Phase representation of the rapid theta neuron}

A phase representation of the rapid theta neuron model, similar to
the classical theta neuron model, is obtained with the transformation
$\tan\frac{\theta}{2}=V-V_{\mathrm{G}}$ and $\theta\in[-\pi,\pi)$,
yielding 
\begin{equation}
\tau_{\mathrm{m}}\frac{\mathrm{d}\theta}{\mathrm{d}t}=\begin{cases}
\frac{r+1}{2r}\big(1-\cos\theta\big)+\big(I(t)-I_{\mathrm{T}}\big)\big(1+\cos\theta\big) & \theta\le0\\
\frac{r(r+1)}{2}\big(1-\cos\theta\big)+\big(I(t)-I_{\mathrm{T}}\big)\big(1+\cos\theta\big) & \theta>0.
\end{cases}\label{eq:rapid-DiffEq-theta}
\end{equation}
For $r=1$, the theta neuron model \citep{Skey-thetaneuron} is recovered.

The exact solutions of the rapid theta neuron model for constant positive
external currents and $\delta$ pulse coupling allow us to derive
a phase representation with constant phase velocity. This phase representation
is convenient both for efficient, numerically exact, event-based simulation
and also for analytical tractability. 

The solution of the governing differential equation in the dimensionless
voltage representation \eqref{eq:rapid-theta-eq-V} for constant input
currents $I(t)\equiv I_{\mathrm{T}}+I$ is
\begin{eqnarray}
\frac{1}{I}\frac{\mathrm{d}V}{1+\left(\dfrac{V-V_{\mathrm{G}}}{\sqrt{I/a_{\mathrm{S,U}}}}\right)^{2}} & = & \frac{1}{\tau_{\mathrm{m}}}\mathrm{d}t\nonumber \\
\frac{1}{I}\sqrt{I/a_{\mathrm{S,U}}}\left[\arctan\left(\frac{V-V_{\mathrm{G}}}{\sqrt{I/a_{\mathrm{S,U}}}}\right)\right]_{V_{1}}^{V_{2}} & = & \frac{t_{2}-t_{1}}{\tau_{\mathrm{m}}}\nonumber \\
\arctan\left(\frac{V_{2}-V_{\mathrm{G}}}{\sqrt{I/a_{\mathrm{S,U}}}}\right) & = & \arctan\left(\frac{V_{1}-V_{\mathrm{G}}}{\sqrt{I/a_{\textnormal{\ensuremath{\mathrm{S,U}}}}}}\right)+\sqrt{I\,a_{\mathrm{S,U}}}\frac{t_{2}-t_{1}}{\tau_{\textnormal{\ensuremath{\mathrm{m}}}}}.\label{eq:rapid-Vevolution}
\end{eqnarray}
This equation represents the solution for both branches of Eq.~\eqref{eq:rapid-theta-eq-V},
separated by $V_{\mathrm{G}}$ as before. For the subthreshold part
($V\le V_{\mathrm{G}}$), the curvature is $a_{\mathrm{S}}=\frac{r+1}{2r},$
and for the suprathreshold part ($V>V_{\mathrm{G}}$), the curvature
is $_{\mathrm{U}}=\frac{r(r+1)}{2}$. In the phase representation
with phase $\phi\in[-\pi,\pi)$ and constant phase velocity, the phase
evolution is given by 
\begin{equation}
\phi_{2}=\phi_{1}+\omega\frac{t_{2}-t_{1}}{\tau_{\mathrm{m}}}.\label{eq:rapid-PHIevolution}
\end{equation}

Identifying Eq.~\eqref{eq:rapid-Vevolution} and \eqref{eq:rapid-PHIevolution}
enables us to derive the constant phase velocity $\omega$ and the
gluing point $\phi_{\mathrm{G}}$ to define the transformation between
the two representations
\begin{equation}
\frac{\phi-\phi_{\mathrm{G}}}{\omega}=\arctan\left(\frac{V-V_{\mathrm{G}}}{\sqrt{I/a_{\mathrm{S,U}}}}\right)\frac{1}{\sqrt{I\,a_{\mathrm{S,U}}}}.\label{eq:rapid-trafo}
\end{equation}
During one complete cycle, the time $T_{\mathrm{S}}$ spent in the
subthreshold part ($V_{2}=V_{\mathrm{G}}$ and $V_{1}\to-\infty$)
and the time $T_{\mathrm{U}}$ spent in the suprathreshold part ($V_{2}\to\infty$
and $V_{1}\to V_{\mathrm{G}}$) are obtained from Eq.~\eqref{eq:rapid-Vevolution}:
\[
T_{\mathrm{S}}=\frac{\pi\tau_{\textnormal{\ensuremath{\mathrm{m}}}}}{\sqrt{2I(r+1)/r}}\quad\textrm{and}\quad T_{\mathrm{U}}=\frac{\pi\tau_{\mathrm{m}}}{\sqrt{2I(r+1)r}}.
\]
The time spent in the subthreshold part is thus $T_{\mathrm{S}}/T_{\mathrm{U}}=r$
times as long as that in the suprathreshold part. The total cycle
length, or unperturbed interspike interval, is
\begin{eqnarray}
T^{\mathrm{free}} & = & (r+1)T_{\mathrm{U}}\nonumber \\
 & = & \frac{\pi\tau_{\mathrm{m}}}{\sqrt{I}}\sqrt{\frac{r+1}{2r}}.\label{eq:rapid-isi}
\end{eqnarray}
Its inverse gives the firing rate for constant external input. 

The constant phase velocity is then 
\begin{eqnarray}
\omega & = & \frac{2\pi}{T^{\mathrm{free}}}\nonumber \\
 & = & \frac{2\sqrt{I}}{\tau_{\mathrm{m}}}\sqrt{\frac{2r}{r+1}}=\frac{2}{\tau_{\mathrm{m}}}\sqrt{I/a_{\mathrm{S}}}.\label{eq:rapid-phasevelocity}
\end{eqnarray}
The phase corresponding to the gluing point is
\begin{eqnarray}
\phi_{\mathrm{G}} & = & -\pi+\omega T_{\mathrm{S}}\nonumber \\
 & = & \pi\frac{r-1}{r+1}.\label{eq:rapid-phasegluepoint}
\end{eqnarray}
The constant phase velocity \eqref{eq:rapid-phasevelocity} and the
gluing point \eqref{eq:rapid-phasegluepoint} define the transformation
\eqref{eq:rapid-trafo} between the voltage representation and the
phase representation:
\begin{eqnarray}
\phi & = & \phi_{\mathrm{G}}+\begin{cases}
\tfrac{2}{a_{\textnormal{\textnormal{S}}}}\arctan\left(\frac{V-V_{\mathrm{G}}}{\sqrt{I/a_{\mathrm{S}}}}\right) & V\le V_{\mathrm{G}}\\
\tfrac{2}{ra_{\mathrm{S}}}\arctan\left(r\frac{V-V_{\mathrm{G}}}{\sqrt{I/a_{\mathrm{S}}}}\right) & V>V_{\mathrm{G}}
\end{cases}\label{eq:rapid-trafo1}\\
V & = & V_{\mathrm{G}}+\begin{cases}
\sqrt{I/a_{\mathrm{S}}}\tan\left(a_{\mathrm{S}}\frac{\phi-\phi_{\mathrm{G}}}{2}\right) & \phi\le\phi_{\mathrm{G}}\\
\sqrt{I/r^{2}a_{\mathrm{S}}}\tan\left(ra_{\mathrm{S}}\frac{\phi-\phi_{\mathrm{G}}}{2}\right) & \phi>\phi_{\mathrm{G}}.
\end{cases}\label{eq:rapid-trafo2}
\end{eqnarray}

This transformation between the two equivalent representations is
now used to calculate the phase-transition curve $g(\phi)$ and the
phase-response curve $Z(\phi)$. Receiving a $\delta$ pulse of strength
$J$ leads to a step like change of the neuron's voltage $V^{+}=V^{-}+J$.
If this change does not lead to a change from the subthreshold to
the suprathreshold part or vice versa, the calculation of the phase-transition
curve is straightforward. However, some care is needed if the $\delta$
pulse does lead to such a change.

An inhibitory pulse $J<0$ can lead to a change from the suprathreshold
to the subthreshold part. This happens if the neuron's phase is between
$\phi_{\mathrm{G}}$ and $\phi_{-}$. The phase-transition curve for
inhibitory $\delta$ pulses of strength $J$ and constant external
currents $I$ with the effective coupling $C=J/\sqrt{I}$ and $\phi_{-}=\phi_{\mathrm{G}}+\frac{2}{ra_{\mathrm{S}}}\arctan\big(r(V_{\mathrm{G}}-J-V_{\mathrm{G}})/\sqrt{I/a_{\mathrm{S}}}\big)=\phi_{\mathrm{G}}-\frac{2}{ra_{\mathrm{S}}}\arctan\big(r\sqrt{a_{\mathrm{S}}}C\big)$
is 
\begin{eqnarray}
\mathrm{g}_{-}(\phi) & = & \phi_{\mathrm{G}}+\begin{cases}
\frac{2}{a_{\mathrm{S}}}\arctan\left(\tan\left(a_{\mathrm{S}}\frac{\phi-\phi_{\mathrm{G}}}{2}\right)+\sqrt{a_{\mathrm{S}}}C\right) & -\pi<\phi\le\phi_{\mathrm{G}}\\
\frac{2}{a_{\mathrm{S}}}\arctan\left(\frac{1}{r}\tan\left(ra_{\mathrm{S}}\frac{\phi-\phi_{\mathrm{G}}}{2}\right)+\sqrt{a_{\mathrm{S}}}C\right) & \phi_{\mathrm{G}}<\phi<\phi_{-}\\
\frac{2}{ra_{\mathrm{S}}}\arctan\left(\tan\left(ra_{\mathrm{S}}\frac{\phi-\phi_{\mathrm{G}}}{2}\right)+r\sqrt{a_{\mathrm{S}}}C\right) & \phi_{-}\le\phi<\pi.
\end{cases}\label{eq:rapid-PTC-inh}
\end{eqnarray}

For excitatory $\delta$ pulses of strength $J>0$, the phase can
change from the subthreshold to the suprathreshold part if the phase
is between $\phi_{+}$ and $\phi_{\mathrm{G}}$. The phase-transition
curve for excitatory $\delta$ pulses of strength $J$ and constant
external currents $I$ with the effective coupling $C=J/\sqrt{I}$
and $\phi_{+}=\phi_{\mathrm{G}}-\frac{2}{a_{\mathrm{S}}}\arctan(\sqrt{a_{\mathrm{S}}}C)$
(displayed in \textbf{Fig.~\ref{fig:rapid-PRC-varyR}}) is
\begin{eqnarray}
g_{+}(\phi) & = & \phi_{\mathrm{G}}+\begin{cases}
\frac{2}{a_{\mathrm{S}}}\arctan\left(\tan\left(a_{\mathrm{S}}\frac{\phi-\phi_{\textnormal{\ensuremath{\mathrm{G}}}}}{2}\right)+\sqrt{a_{\mathrm{S}}}C\right) & -\pi<\phi\le\phi_{+}\\
\frac{2}{ra_{\mathrm{S}}}\arctan\left(r\tan\left(a_{\mathrm{S}}\frac{\phi-\phi_{\mathrm{G}}}{2}\right)+r\sqrt{a_{\mathrm{S}}}C\right) & \phi_{+}<\phi<\phi_{\mathrm{G}}\\
\frac{2}{ra_{\mathrm{S}}}\arctan\left(\tan\left(ra_{\mathrm{S}}\frac{\phi-\phi_{\mathrm{G}}}{2}\right)+r\sqrt{a_{\mathrm{S}}}C\right) & \phi_{\mathrm{G}}\le\phi<\pi.
\end{cases}\label{eq:rapid-PTC-exc}
\end{eqnarray}

The phase-response curve is $Z_{\pm}(\phi)=g_{\pm}(\phi)-\phi$. Thus,
the infinitesimal phase-response curve is the same for both excitatory
and inhibitory pulses, since $\phi_{\pm}\to\phi_{\mathrm{G}}$ for
$C\to0$: 
\begin{equation}
Z(\phi)\stackrel{C\to0}{\simeq}C\begin{cases}
\frac{2\sqrt{a_{\mathrm{S}}}}{a_{\mathrm{S}}}\frac{1}{1+\tan\left(a_{\mathrm{S}}\frac{\phi-\phi_{\mathrm{G}}}{2}\right)^{2}}=\dfrac{1+\cos\left(a_{\mathrm{S}}(\phi-\phi_{\mathrm{G}})\right)}{\sqrt{a_{\mathrm{S}}}} & -\pi<\phi\le\phi_{\mathrm{G}}\\
\frac{2r\sqrt{a_{\mathrm{S}}}}{ra_{\mathrm{S}}}\frac{1}{1+\tan\left(ra_{\mathrm{S}}\frac{\phi-\phi_{\mathrm{G}}}{2}\right)^{2}}=\dfrac{1+\cos\left(ra_{\mathrm{S}}(\phi-\phi_{\mathrm{G}})\right)}{\sqrt{a_{\mathrm{S}}}} & \phi_{\mathrm{G}}\le\phi<\pi.
\end{cases}\label{eq:rapid-iPRC}
\end{equation}

\section{Single spike Jacobian of the rapid theta neuron network}

The analytical expression of the derivative of the evolution map,
called the single spike Jacobian, is necessary for calculating the
full Lyapunov spectrum with high precision. The single spike Jacobian
describes the linear evolution of infinitesimal perturbations of the
neuron\textquoteright s states and will be used to numerically calculate
the Lyapunov spectra. Since infinitesimal perturbations are considered
here, the spike-order in the networks is preserved, provided that
there are no exactly synchronous spike events. Such exactly synchronous
events generally should which generally should not occur in the asynchronous
network states considered here. In a phase representation, the iterative
map, which maps the state of the network at one spike time to the
state at the next spike in the network, reads
\begin{equation}
\phi_{i}(t_{s+1})=\phi_{i}(t_{s})+\omega_{i}(t_{s+1}-t_{s})+Z\big(\phi_{i}(t_{s})+\omega_{i}(t_{s+1}-t_{s})\big)\delta_{i\in\textrm{post}(j^{*})},\label{eq:theta-phasemapPRC}
\end{equation}
where $\delta_{i\in\textrm{post}(j^{*})}$ is one if $i$ is a postsynaptic
neuron of the spiking neuron $j^{*}$ and zero otherwise and $Z(\phi_{i})$
is the phase-response curve.

Consequently, the single spike Jacobian reads 
\begin{equation}
D_{ij}(t_{s})=\frac{\mathrm{d}\phi_{i}(t_{s+1})}{\mathrm{d}\phi_{j}(t_{s})}=\begin{cases}
1+Z^{\prime}(\phi_{i^{*}}(t_{s+1}^{-})) & \mathrm{for}\;i=j=i^{*}\\
-\frac{\omega_{i^{*}}}{\omega_{j^{*}}}Z^{\prime}(\phi_{i^{*}}(t_{s+1}^{-})) & \mathrm{for}\;i=i^{*}\;\mathrm{and}\;j=j^{*}\\
\delta_{ij} & \mathrm{otherwise},
\end{cases}\label{eq:theta-jacobianPRC}
\end{equation}
where $j^{*}$ denotes the spiking neuron in the considered interval,
firing at time $t_{s+1}$, $i^{*}\in\mathrm{post(}j^{*})$ are the
spike receiving neurons and $\delta_{ij}$ is the Kronecker delta.
The derivatives of the phase-response curves $Z^{\prime}(\phi)$ are
evaluated at the phases of the spike receiving neurons $\phi_{i^{*}}(t_{s+1}^{-})=\phi_{i^{*}}(t_{s})+\omega_{i^{*}}(t_{s+1}-t_{s})$
just before spike reception. To investigate the collective dynamics
of networks of rapid theta neurons, it is necessary to obtain the
derivative $d(\phi_{i^{*}}(t_{s+1}^{-}))=1+Z^{\prime}(\phi_{i^{*}}(t_{s+1}^{-}))$
of the phase-transition curve for calculation of the single spike
Jacobians. The derivative $d(\phi)$ in the case of inhibitory pulses
is

\begin{eqnarray}
d_{-}(\phi) & = & \begin{cases}
\frac{\tan\left(a_{\mathrm{S}}\frac{\phi-\phi_{\mathrm{G}}}{2}\right)^{2}+1}{\left(\tan\left(a_{\mathrm{S}}\frac{\phi-\phi_{\mathrm{G}}}{2}\right)+\sqrt{a_{\mathrm{S}}}C\right)^{2}+1} & -\pi<\phi\le\phi_{\mathrm{G}}\\
\frac{\tan\left(ra_{\mathrm{S}}\frac{\phi-\phi_{\mathrm{G}}}{2}\right)^{2}+1}{\left(\frac{1}{r}\tan\left(ra_{\mathrm{S}}\frac{\phi-\phi_{\mathrm{G}}}{2}\right)+\sqrt{a_{\mathrm{S}}}C\right)^{2}+1} & \phi_{\mathrm{G}}<\phi<\phi_{-}\\
\frac{\tan\left(ra_{\mathrm{S}}\frac{\phi-\phi_{\mathrm{G}}}{2}\right)^{2}+1}{\left(\tan\left(ra_{\mathrm{S}}\frac{\phi-\phi_{\mathrm{G}}}{2}\right)+r\sqrt{a_{\mathrm{S}}}C\right)^{2}+1} & \phi_{-}\le\phi<\pi.
\end{cases}\label{eq:rapid-d-inh}
\end{eqnarray}
\begin{figure}[tp]
\includegraphics[width=1\textwidth]{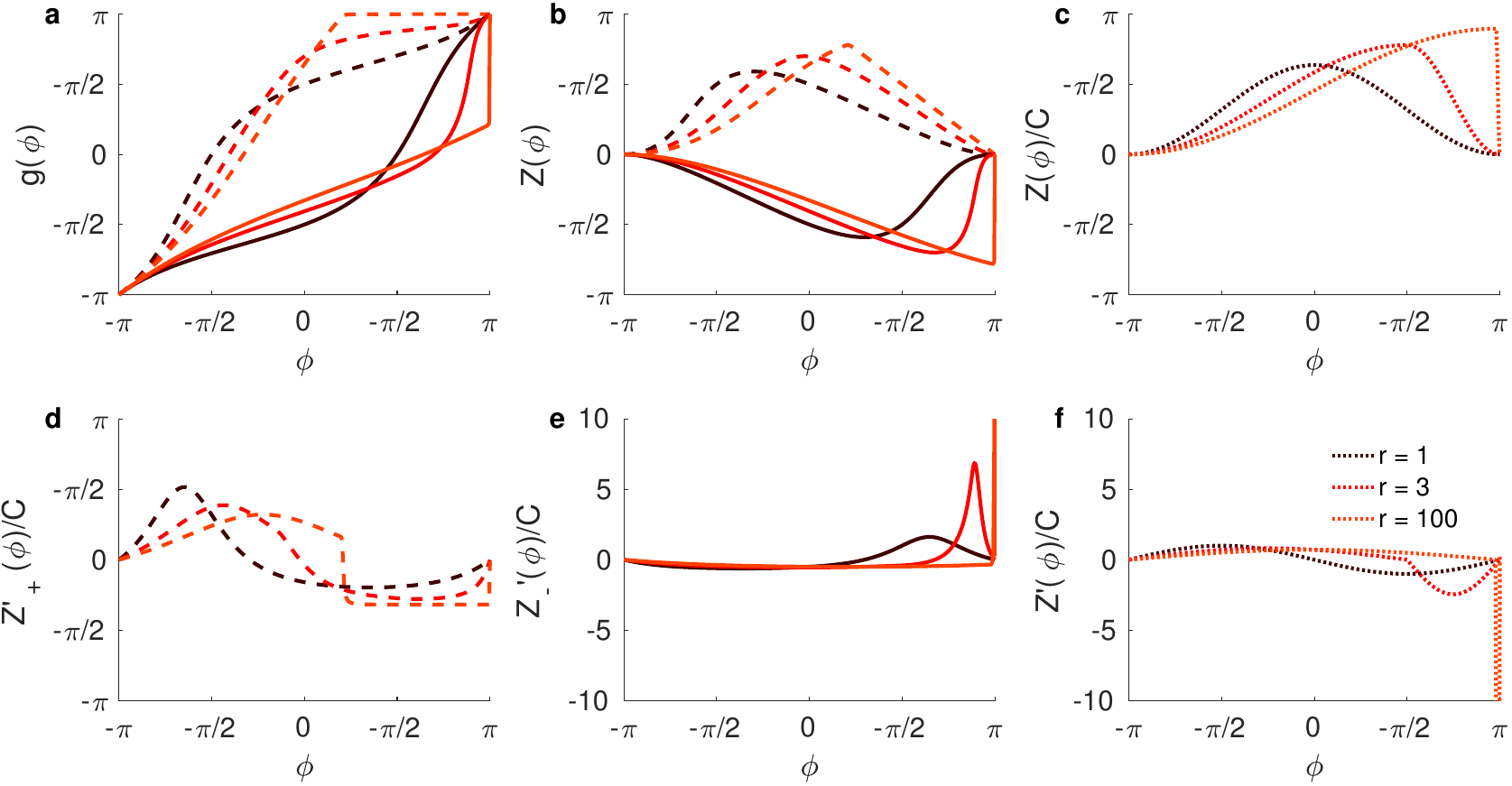}

\caption[Phase-transition curve, phase-response and infinitesimal phase-response
of the rapid theta neuron model with AP onset rapidness.]{\textbf{\label{fig:rapid-PRC-varyR}Phase-transition curve (PTC),
phase-response (PRC) and infinitesimal phase-response (iPRC) explain
reduction of chaos for high AP onset rapidness $r$}. \textbf{a)}
The phase-transition curve (PTC) $g(\phi)$ shown with inhibitory
coupling $C=-1$ (full lines, Eq.~\eqref{eq:rapid-PTC-inh}) and
excitatory coupling $C=+1$ (dashed lines, Eq.~\eqref{eq:rapid-PTC-exc})
for three values of spike onset rapidness $r=1,\,3,\,100$. \textbf{b)}
Same for phase response curve (PRC) $Z(\phi)=g(\phi)-\phi$. \textbf{c)}
Same for infinitesimal PRC (Eq.~\eqref{eq:rapid-iPRC}). \textbf{d)}
Derivative of PRC for excitatory coupling. \textbf{e)} Derivative
of PRC for inhibitory coupling. \textbf{f)} Derivative of derivative
of infinitesimal phase response curve (Eq.~\eqref{eq:rapid-iPRC-derivative}).
Note that in the limit $r\to\infty$ the iPRC becomes monotonically
increasing and its derivative is positive almost everywhere. }
\end{figure}
In the voltage representation, the derivative $d(V)$ is useful for
the analytical calculation of the mean Lyapunov exponent in section
\ref{sec:Random-matrix-theory}: 
\begin{eqnarray}
d_{-}(V) & = & \begin{cases}
\frac{r^{2}\left(V-V_{\mathrm{G}}\right)^{2}+\frac{I_{0}\sqrt{K}}{a_{\mathrm{S}}}}{\left(V-V_{\mathrm{G}}+C\sqrt{I}\right)^{2}+\frac{I_{0}\sqrt{K}}{a_{\mathrm{S}}}} & V\le V_{\mathrm{G}}\\
\frac{r^{2}\left(V-V_{\mathrm{G}}\right)^{2}+\frac{I_{0}\sqrt{K}}{a_{\mathrm{S}}}}{\left(V-V_{\mathrm{G}}+C\sqrt{I}\right)^{2}+\frac{I_{0}\sqrt{K}}{a_{\mathrm{S}}}} & V_{\mathrm{G}}<V<V_{-}\\
\frac{r^{2}\left(V-V_{\mathrm{G}}\right)^{2}+\frac{I_{0}\sqrt{K}}{a_{\mathrm{S}}}}{r^{2}\left(V-V_{\mathrm{G}}+C\sqrt{I}\right)^{2}+\frac{I_{0}\sqrt{K}}{a_{\mathrm{S}}}} & V_{-}\le V.
\end{cases}\label{eq:rapid-d-inh-V}
\end{eqnarray}
The derivative of the phase-transition curve in the case of excitatory
pulses is

\begin{eqnarray}
d_{+}(\phi) & = & \begin{cases}
\frac{\tan\left(a_{\mathrm{S}}\frac{\phi-\phi_{\mathrm{G}}}{2}\right)^{2}+1}{\left(\tan\left(a_{\mathrm{S}}\frac{\phi-\phi_{\mathrm{G}}}{2}\right)+\sqrt{a_{\mathrm{S}}}C\right)^{2}+1}-1 & -\pi<\phi\le\phi_{+}\\
\frac{\tan\left(a_{\mathrm{S}}\frac{\phi-\phi_{\mathrm{G}}}{2}\right)^{2}+1}{\left(r\tan\left(a_{\mathrm{S}}\frac{\phi-\phi_{\mathrm{G}}}{2}\right)+r\sqrt{a_{\mathrm{S}}}C\right)^{2}+1}-1 & \phi_{+}<\phi<\phi_{\mathrm{G}}\\
\frac{\tan\left(ra_{\mathrm{S}}\frac{\phi-\phi_{\mathrm{G}}}{2}\right)^{2}+1}{\left(\tan\left(ra_{\mathrm{S}}\frac{\phi-\phi_{\mathrm{G}}}{2}\right)+r\sqrt{a_{\mathrm{S}}}C\right)^{2}+1}-1 & \phi_{\mathrm{G}}\le\phi<\pi.
\end{cases}\label{eq:rapid-d-exc}
\end{eqnarray}
The derivative of the phase-response curve is $Z_{\pm}^{\prime}(\phi)=d_{\pm}(\phi)-1$
and the derivative of the infinitesimal phase-response curve is
\begin{equation}
Z^{\prime}(\phi)\stackrel{C\to0}{\simeq}-C\begin{cases}
\sqrt{a_{\mathrm{S}}}\sin\left(a_{\mathrm{S}}(\phi-\phi_{\mathrm{G}})\right) & -\pi<\phi\le\phi_{\mathrm{G}}\\
r\sqrt{a_{\mathrm{S}}}-\sin\left(ra_{\mathrm{S}}(\phi-\phi_{\mathrm{G}})\right) & \phi_{\mathrm{G}}\le\phi<\pi.
\end{cases}\label{eq:rapid-iPRC-derivative}
\end{equation}

\begin{figure}[tp]
\includegraphics[width=1\textwidth]{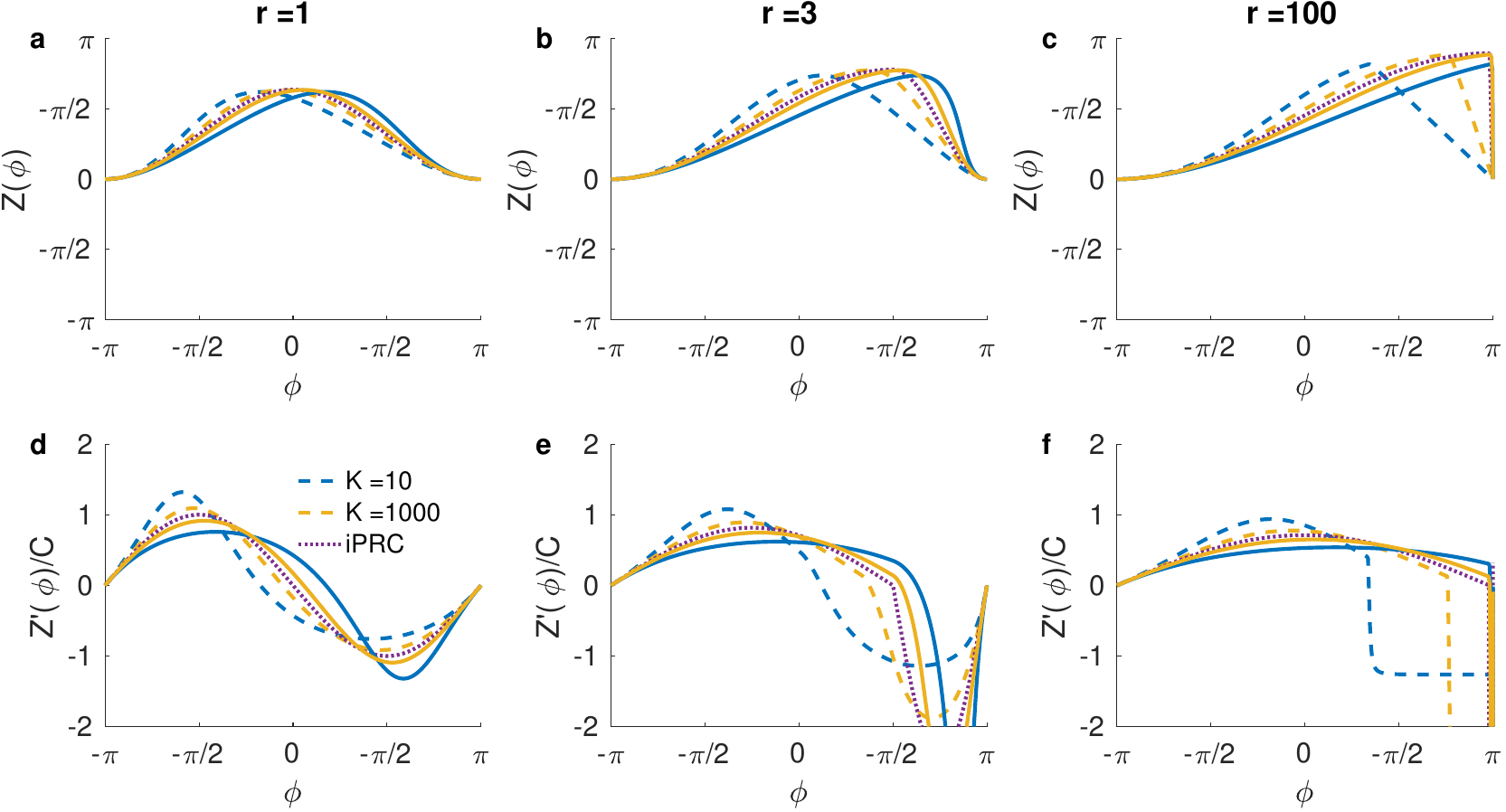}

\caption[Phase-transition curve, phase-response and infinitesimal phase-response
of the rapid theta neuron model with AP onset rapidness.]{\textbf{\label{fig:rapid-PRC-varyK}Phase-response (PRC) and infinitesimal
phase-response (iPRC) of the rapid theta neuron model for different
effective coupling strengths}. \textbf{a)} The PRC is shown for different
$K$ corresponding to different coupling strengths $C=-\frac{J_{0}}{\sqrt{K}}\sqrt{\frac{a_{\mathrm{S}}}{I}}$
for inhibitory (full lines Eq.~\eqref{eq:rapid-PTC-inh}) and excitatory
couplings (dashed lines, Eq.~\eqref{eq:rapid-PTC-exc}) for AP onset
rapidness $r=1$ . The infinitesimal PRC (Eq.~\eqref{eq:rapid-iPRC})
is also displayed for comparison (dotted lines)\textbf{ b)} Same as
\textbf{a} for AP onset rapidness $r=3$ \textbf{c)} Same as \textbf{a}
for AP onset rapidness $r=100$. \textbf{d)} Derivative of PRC normalized
by $C$ for $r=1$. The derivative of the infinitesimal PRC is shown
as dotted line (Eq.~\eqref{eq:rapid-iPRC-derivative}). \textbf{e)}
Same as \textbf{d} for $r=3$. \textbf{f)} Same as \textbf{d} for
$r=100$ (parameters: $I_{0}=1,\,J_{0}=1$).}
\end{figure}
The phase-transition curves (PTC, $g(\phi)$, Eqs.~\eqref{eq:rapid-PTC-inh}
and \eqref{eq:rapid-PTC-exc}), the phase response curves (PRC, $Z(\phi)=g(\phi)-\phi$),
and the infinitesimal phase-response curves (iPRC, Eq.~\eqref{eq:rapid-iPRC})
of the rapid theta neuron model are displayed in \textbf{Fig.}~\textbf{\ref{fig:rapid-PRC-varyR}}.
The iPRC of the theta neuron ($r=1$) is fully symmetric \citep{Skey-thetaneuron},
whereas for increasing AP onset rapidness $r$ the iPRC becomes more
and more asymmetric. In the limit $r\to\infty$ it becomes monotonically
increasing/decreasing and one might expect that this can qualitatively
change the collective network dynamics.

\section{Setup of network and event-based simulation\label{sec:Setup-of-network}}

The pattern of action potentials in cortical tissue is asynchronous
and irregular~\citep{key-irregularity}, despite the reliable response
of single neurons~\citep{key-Mainen1995}. This is commonly explained
by a balance of excitatory and inhibitory synaptic currents~\citep{key-ShadlenSoftky},
which cancels out large mean synaptic inputs. A dynamically self-organized
balance can be achieved without the fine-tuning of synaptic coupling
strength, if the connectivity is inhibition dominated~\citep{key-VreeswijkSompolinsky}.
The statistics of this state is described by a mean-field theory,
which is largely independent of neuron model. We studied large sparse
networks of $N$~rapid theta neurons arranged on a directed Erd\H{o}s\textendash Rényi
random graph of mean indegree $K$. All neurons $i=1,\dots,N$ receive
constant external currents $I_{\mathrm{ext}}$ and non-delayed $\delta$-pulses
from the presynaptic neurons $j\in\textrm{pre }(i)$. The external
currents are chosen to obtain a certain average network firing rate
$\bar{\nu}$ using a bisection method in purely inhibitory networks.
In purely inhibitory networks, the non-zero coupling strengths are
set to $J_{ij}=-J_{0}/\sqrt{K}$ and all neurons receive identical
external currents. In two-population networks, the intra-population
couplings are $J_{EE}=\frac{J_{0}}{\sqrt{K}}\eta\varepsilon$ and
$J_{II}=-\frac{J_{0}}{\sqrt{K}}\sqrt{1-\varepsilon^{2}}$ for the
excitatory (E) and inhibitory (I) population, respectively. The inter-population
coupling strengths are $J_{IE}=\frac{J_{0}}{\sqrt{K}}\varepsilon$
and $J_{EI}=-\frac{J_{0}}{\sqrt{K}}\sqrt{1-\eta^{2}\varepsilon{}^{2}}$.
At $\varepsilon=0$, all excitatory neurons are completely passive.
They only receive inputs from the inhibitory neurons but do not provide
feedback. Increasing $\varepsilon$ above zero activates the excitatory
feedback loops in the network. The specific choice of coupling strengths
preserves the temporal variance of the input currents $\sigma_{I}^{2}=J_{0}^{2}\bar{\nu}$
in both purely inhibitory networks and two-population networks, as
explained below. In mixed networks, the external currents are adapted
using the two-dimensional Newton-Raphson root-finding method, to obtain
the desired mean firing rate in the excitatory and inhibitory population.

\paragraph*{Setup of balanced network with strong couplings and nonvanishing
fluctuations.}

The coupling strengths in inhibitory and excitatory-inhibitory networks
were chosen such that the magnitudes of the input current fluctuations
are identical in all studied networks. Assuming that inputs from different
presynaptic neurons are weakly correlated, the compound input spike
train received by neuron $i$ is modeled as a Poisson process with
rate $\Omega_{i}=\sum_{j\in\textrm{pre}(i)}\nu_{j}\approx K\bar{\nu}\equiv\Omega$,
where $\bar{\nu}$ is the network-averaged firing rate and $K$ the
average number of presynaptic neurons. For inhibitory networks, the
nonzero coupling strengths were $J_{ij}=-\frac{J_{0}}{\sqrt{K}}$.
Assuming that the compound input spike train is a Poisson process,
the input current auto-correlation function is given by
\begin{eqnarray}
C(\tau) & = & \langle\delta I(t)\delta I(t+\tau)\rangle_{t}\nonumber \\
 & \approx & \left(\frac{J_{0}}{\sqrt{K}}\right)^{2}\Omega\int\delta(t-s)\delta(t+\tau-s)\mathrm{d}s\nonumber \\
 & = & \frac{J_{0}}{K}\Omega\delta(\tau)\nonumber \\
 & \approx & J_{0}^{2}\bar{\nu}\delta(\tau)\label{eq:currentcorr}
\end{eqnarray}
Thus, the fluctuations in the input currents are described as delta-correlated
white noise of magnitude
\begin{equation}
\sigma^{2}=J_{0}^{2}\bar{\nu}.\label{eq:curvar}
\end{equation}
Note that due to the scaling of the coupling strengths $J=-\frac{J_{0}}{\sqrt{K}}$
with the square root of the number of synapses $K$, the magnitude
of the fluctuations $\sigma^{2}$ is independent of the number of
synapses. Therefore, the input fluctuations do not vanish in the thermodynamic
limit, and the balanced state in sparse networks emerges robustly
\citep{key-VreeswijkSompolinsky}.

The existence of a balanced state fixed point in the large $K$-limit
can be derived from the equation of the network-averaged mean current
\[
\bar{I}\approx\sqrt{K}(I_{I}-J_{0}\bar{\nu}).
\]
In the large $K$-limit, self-consistency requires the balance of
excitation and inhibition to be $I_{I}=J_{0}\bar{\nu}$: If $\lim_{K\to\infty}(I_{I}-J_{0}\bar{\nu})>0$,
the mean current $\bar{I}$ would diverge to $\infty$ and the neurons
would fire at their maximal rate. This resulting strong inhibition
would break the inequality, leading to a contradiction. If $\lim_{K\to\infty}(I_{I}-J_{0}\bar{\nu})<0$,
the mean current $\bar{I}$ would diverge to $-\infty$, and the neurons
would be silent. This resulting lack of inhibition again breaks the
inequality. The large $K$-limit is self-consistent if 
\[
\lim_{K\to\infty}(I_{I}-J_{0}\bar{\nu})=\mathcal{O}\left(\frac{1}{\sqrt{K}}\right),
\]
such that the excitatory external drive and mean recurrent inhibitory
current cancel out each other. Note that since $I_{I}-J_{0}\bar{\nu}=\mathcal{O}(1/\sqrt{K})$,
the network mean current has a finite large $K$-limit. The average
firing rate in units of the membrane time constant $\tau_{\mathrm{m}}$
is approximately given by the balance equation (also called the van
Vreeswijk equation):
\begin{equation}
\bar{\nu}=\frac{I_{I}}{J_{0}}+\mathcal{O}\left(\frac{1}{\sqrt{K}}\right).\label{eq:balance equation}
\end{equation}

\paragraph*{Setup of mixed network}

The input current autocorrelations in balanced networks with excitatory
(E) and inhibitory (I) populations are derived analogously to Eq.~\ref{eq:currentcorr}
and also display delta-correlated white noise. Assuming the same average
indegree $K$ from each population, the magnitudes of the input fluctuations
are 
\begin{eqnarray*}
\sigma_{I}^{2} & = & J_{II}^{2}\bar{\nu}_{I}+J_{IE}^{2}\bar{\nu}_{E}\\
\sigma_{E}^{2} & = & J_{EE}^{2}\bar{\nu}_{E}+J_{EI}^{2}\bar{\nu}_{I}.
\end{eqnarray*}
Choosing the coupling weights to be

\begin{equation}
\mathbf{J}=\left(\begin{array}{cc}
J_{EE} & -J_{EI}\\
J_{IE} & -J_{II}
\end{array}\right)=\frac{J_{0}}{\sqrt{K}}\left(\begin{array}{cc}
\eta\varepsilon & -\sqrt{1-(\eta\varepsilon)^{2}}\\
\varepsilon & -\sqrt{1-\varepsilon^{2}}
\end{array}\right),\label{eq:theta-EIcoupling-1}
\end{equation}
and setting the average firing rates in both populations identical
$\bar{\nu}_{E}=\bar{\nu}_{I}\equiv\bar{\nu}$ leads to 
\[
\sigma_{E}^{2}=\sigma_{I}^{2}\equiv\sigma^{2}=J_{0}^{2}\bar{\nu}.
\]
Thus, the magnitude of fluctuations is identical in all networks considered,
independent of the excitatory feedback loop activation $\varepsilon$.
Accordingly, the statistical characteristics of the balanced state
are preserved when activating the excitatory loops. However, the
dynamical features of the collective activity, change upon activation
of the excitatory loops. Increasing the excitatory coupling enhances
chaoticity in the network dynamics. It is important to note that the
balanced state arises in an extended portion of the parameters space
in mixed networks. This region is constrained by the following inequalities,
which can be derived by self-consistency arguments similar to those
used for the purely inhibitory network: 
\begin{equation}
\frac{J_{EE}}{J_{EI}}<\frac{J_{IE}}{J_{II}}<\min\left\{ 1,\frac{\bar{\nu}_{I}}{\bar{\nu}_{E}}\right\} \,\mathrm{and}\,\frac{J_{EE}}{J_{IE}}<\frac{J_{EI}}{J_{II}}<\frac{E_{0}}{I_{0}}.\label{eq:theta-balance_inequality}
\end{equation}
The particular parametrization with $\eta$ and $\epsilon$ is used
to keep the input statistics fixed.

All simulations were run event-based, following Ref.~\citep{key-eventbased},
where an exact map is iterated from spike to spike in the $\phi$-representation
of the rapid theta neuron model with homogeneous coupling strengths
and homogeneous external currents for all neurons in each population.
The next spike time in each population is obtained by inverting Eq.~\ref{eq:rapid-PHIevolution}
\begin{equation}
t_{s}=t_{s-1}+\min_{i}\left\{ \frac{\pi-\phi_{i}(t_{s-1})}{\omega}\right\} .\label{eq:map_spiketime}
\end{equation}

The phase map $f(\vec{\phi}(t_{s-1}))=\vec{\phi}(t_{s})$, iterating
all neuron\textquoteright s phases between two successive spike events
$t_{s-1}$ and $t_{s}$ in the network, is then the concatenation
of Eq.~\ref{eq:rapid-PHIevolution} and the phase transition curve
(Eq.~\ref{eq:rapid-PTC-inh} and Eq.~\ref{eq:rapid-PTC-exc} )
\begin{equation}
f(\phi_{i}(t_{s-1}))=\begin{cases}
\phi_{i}(t_{s-1})+\omega(t_{s}-t_{s-1}) & \textrm{if }i\notin\textrm{post}(j^{*})\\
g(\phi_{i}(t_{s-1})+\omega(t_{s}-t_{s-1})) & \textrm{if }i\in\textrm{post}(j^{*}),
\end{cases}\label{eq:phasemap}
\end{equation}
where $\textrm{post}(j^{*})$ denotes the set of neurons postsynaptic
to the spiking neuron $j^{*}$ in the considered interval.

Eq.~\ref{eq:phasemap} is used for all network simulations in an
iterative event-based procedure \citep{key-eventbased}. At the beginning
of an iteration the next spike time in the network is calculated with
Eq.~\eqref{eq:map_spiketime}. Then all neuron\textquoteright s phases
are evolved until the next spike time using Eq.~\eqref{eq:phasemap}.
Since the external currents are identical for all neurons in each
population, it is sufficient to search for the neuron with the largest
phase and then calculate the corresponding next spike time.

 The average number of spikes per second of simulation time gives
the firing rate in hertz. We used the interspike intervals $T_{i}$
between subsequent spike times to calculate the coefficients of variation,
$\mathrm{cv}_{i}=\sqrt{\langle T_{i}^{2}\rangle-\langle T_{i}\rangle^{2}}/\langle T_{i}\rangle$,
where $\langle\ldots\rangle$ denotes the temporal average. The neurons'
firing rate and coefficient of variation distributions are shown in
\textbf{Fig.~2} of the main manuscript.

\section{Convergence of the Lyapunov spectra\label{sec:Numerical-convergence}}

The single spike Jacobians Eq.~\ref{eq:theta-jacobianPRC} is evaluated
using Eq.~\ref{eq:rapid-d-inh} and Eq.~\ref{eq:rapid-d-exc} with
the exact phases of the neurons before spike reception, t. These were
used to numerically calculate all Lyapunov exponents with the standard
procedure \citep{key-Benettin1980}. Following a warmup of the network
dynamics of typically $100$ spikes per neuron on average, we started
with a random $N$-dimensional orthonormal system that was evolved
in each iteration using the single spike Jacobian. Subsequently, after
a short warmup of the orthonormal system of about one spike per neuron,
these norms were used to calculate the $N$ Lyapunov exponents, $\lambda_{i}=\lim_{p\to\infty}\frac{1}{t_{p}}\sum_{s=1}^{p}\log g_{i}(t_{s})$.
The evolved vectors were reorthonormalized every $\mathcal{O}\left(N/K\right)$
network spikes, yielding the norms of the orthogonalized vectors $g_{i}(t_{s})$
and the orthonormal system to be used in the following iterations. 

All calculations were performed using custom code written in Julia
and C++ with double precision. The GNU Scientific Library (GSL) was
used for the random number generator (Mersenne-Twister), the Automatically
Tuned Linear Algebra Software (ATLAS) for matrix multiplications in
the Gram\textendash Schmidt procedure and the Message Passing Interface
(MPI) for the parallel implementation of the simulations. The sparseness
of the networks was used for efficient storage of the coupling matrices,
the updates of the postsynaptic neurons, and the matrix multiplications
of the orthonormal system with the sparse single spike Jacobians.
For very large networks ($N>10^{6}$), the topology was not stored
as a sparse matrix but generated on the fly during the simulation
using the index of the spiking neuron as a seed for the random number
generator used to generate the indices of the postsynaptic neurons
\citep{key-Rosenbaum}. For the reorthonormalization, we chose a
parallel recursive blocked version of the Gram\textendash Schmidt
procedure \citep{key-Gram-Schmidt} in our C++ implementation and
the default LAPACK (Linear Algebra PACKage) routines for our Julia
implementation which is using Householder reflections.

\begin{figure}[tp]
\begin{centering}
\includegraphics[clip,width=1\textwidth]{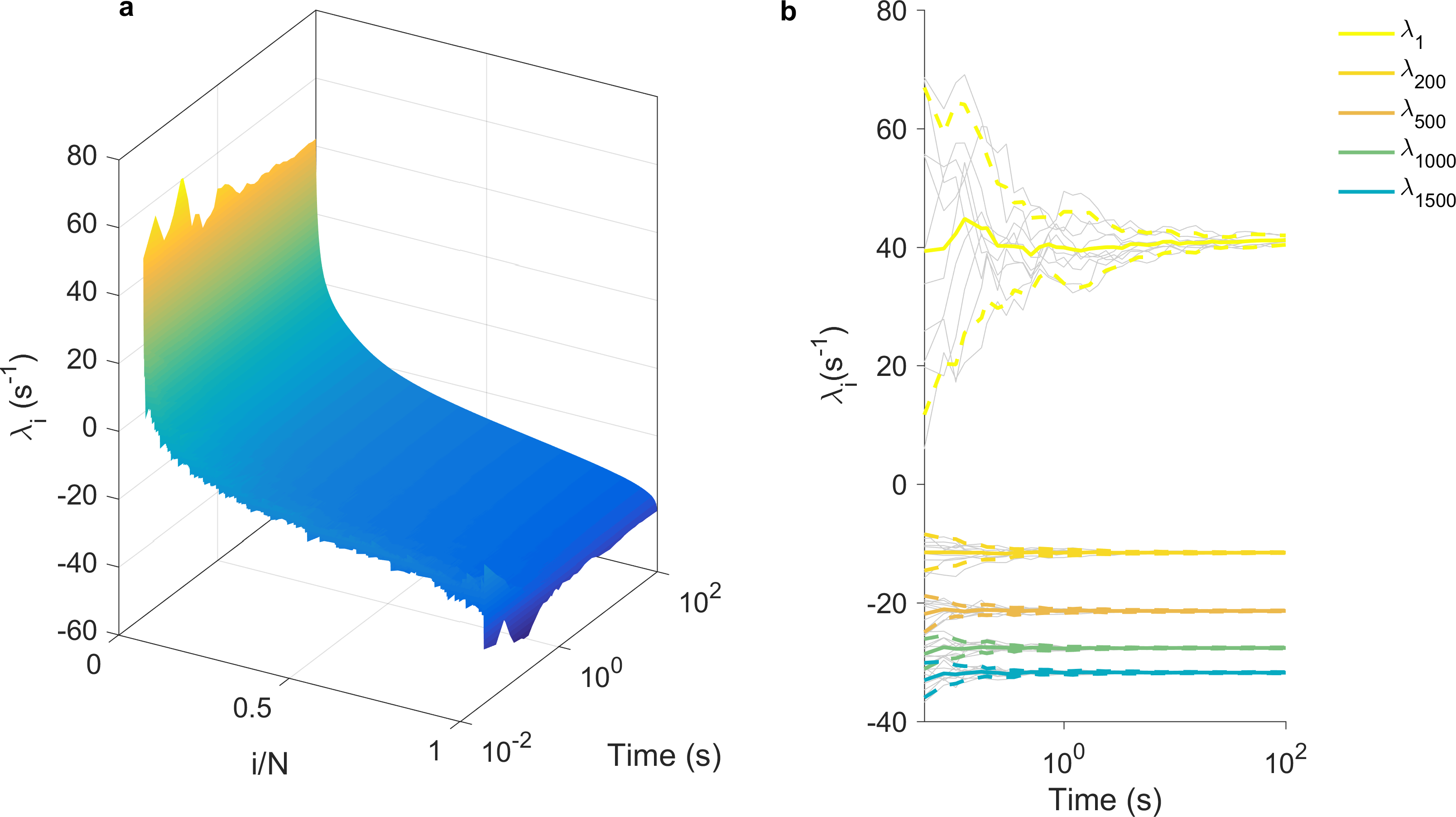}
\par\end{centering}
\caption[Convergence of Lyapunov spectra versus time in inhibitory networks.]{\textbf{}\label{fig:theta-inh.convergence_r=00003D3}\textbf{Convergence
of Lyapunov spectra over time in inhibitory networks with small AP
rapidness (r=3). }(logarithmic time scale) \textbf{a)} Convergence
of Lyapunov spectrum for one initial condition, \textbf{b)} gray lines:
some Lyapunov exponents for ten different network realizations, solid
color lines: averages, dotted color lines: averages $\pm$ double
standard errors (parameters: $N=2000$, $\bar{\nu}=1\,\mathrm{Hz}$,
$K=100$, $J_{0}=1$, $\tau_{\mathrm{m}}=10\,\mathrm{ms}$).}
\end{figure}

\begin{figure}[tp]
\begin{centering}
\includegraphics[clip,width=1\textwidth]{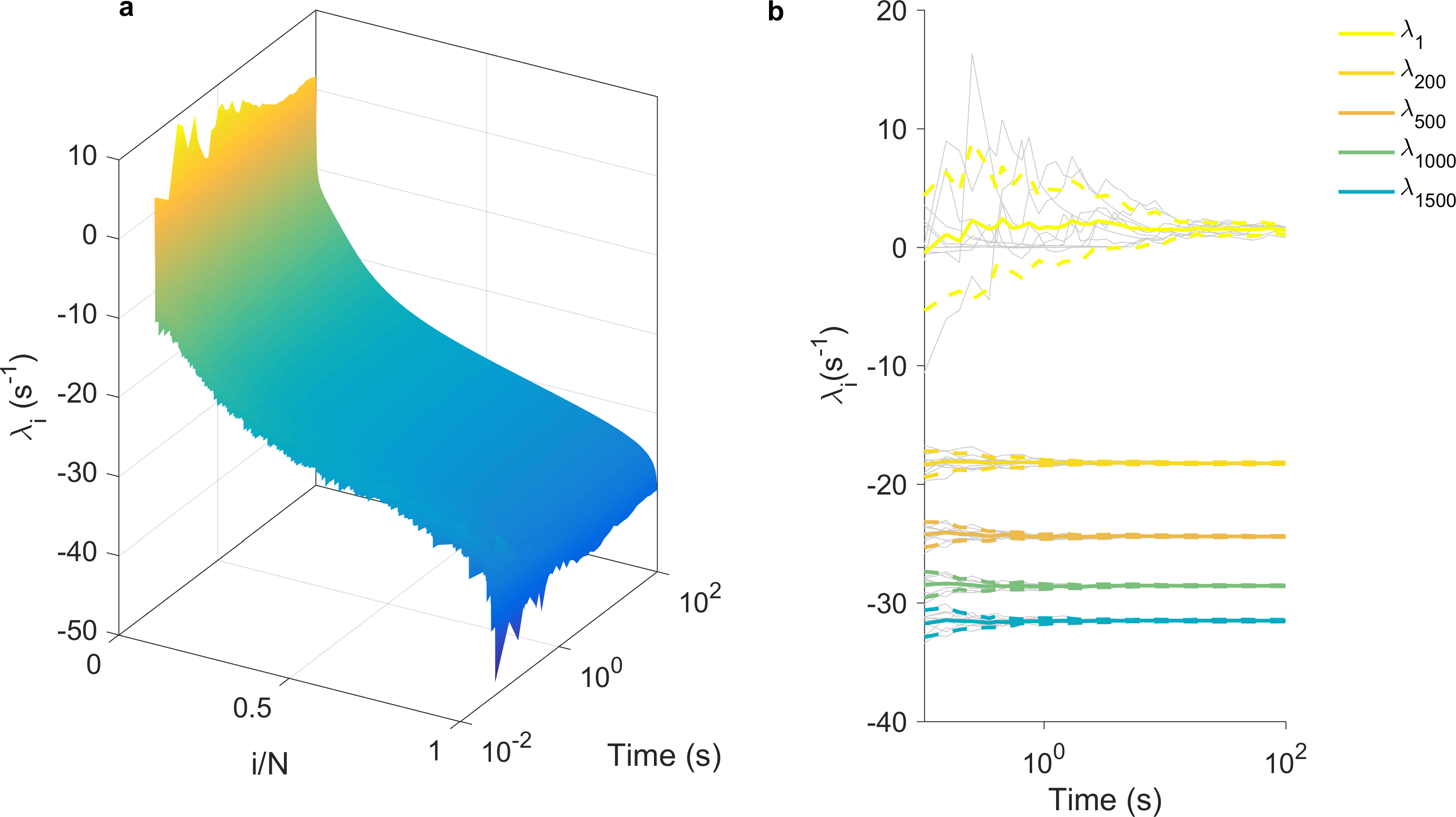}
\par\end{centering}
\caption[Convergence of Lyapunov spectra versus time in inhibitory networks.]{\textbf{}\label{fig:theta-inh.convergence_r=00003D100}\textbf{Convergence
of Lyapunov spectra versus time in inhibitory rapid theta networks
for large AP rapidness (r=100). }(logarithmic time scale) \textbf{a)}
Convergence of Lyapunov spectrum for one initial condition, \textbf{b)}
gray lines: some Lyapunov exponents for ten different network realizations,
solid color lines: averages, dotted color lines: averages $\pm$ double
standard errors (parameters: $N=2000$, $\bar{\nu}=1\,\mathrm{Hz}$,
$K=100$, $J_{0}=1$, $\tau_{\mathrm{m}}=10\,\mathrm{ms}$).}
\end{figure}

Note that the non-converged Lyapunov exponents generated during the
transient are meaningless, as they neither reflect the local nor finite-time
Lyapunov exponents. The converged Lyapunov exponents capture the network
dynamics on the (chaotic )attractor. \textbf{Fig.~\ref{fig:theta-inh.convergence_r=00003D3}a}
displays the convergence towards the full Lyapunov spectrum on a logarithmic
time scale. To ensure robustness, this calculation was repeated for
10 different initial phases \textbf{Fig.~\ref{fig:theta-inh.convergence_r=00003D3}b}
shows the results of ten such runs for six representative Lyapunov
exponents (gray lines), together with their averages $\lambda_{i}=\frac{1}{10}\sum_{n=1}^{10}\lambda_{i,n}$
(solid color lines) and confidence intervals (dotted color lines)
of the double standard error $2\triangle\lambda_{i}=2\sqrt{\frac{1}{10}\sum_{n=1}^{10}(\lambda_{i,n}-\lambda_{i})^{2}}$.
The Lyapunov spectrum was independent of the initial phases, and
in the limit of large networks, independent of network realizations.
Generally, all calculations of the Lyapunov spectra were repeated
ten times with different initial phases and network realizations.
Numerical errors were usually smaller than the symbol sizes in the
presented figures in the main manuscript.

\begin{figure}[tp]
\begin{centering}
\includegraphics[clip,width=1\textwidth]{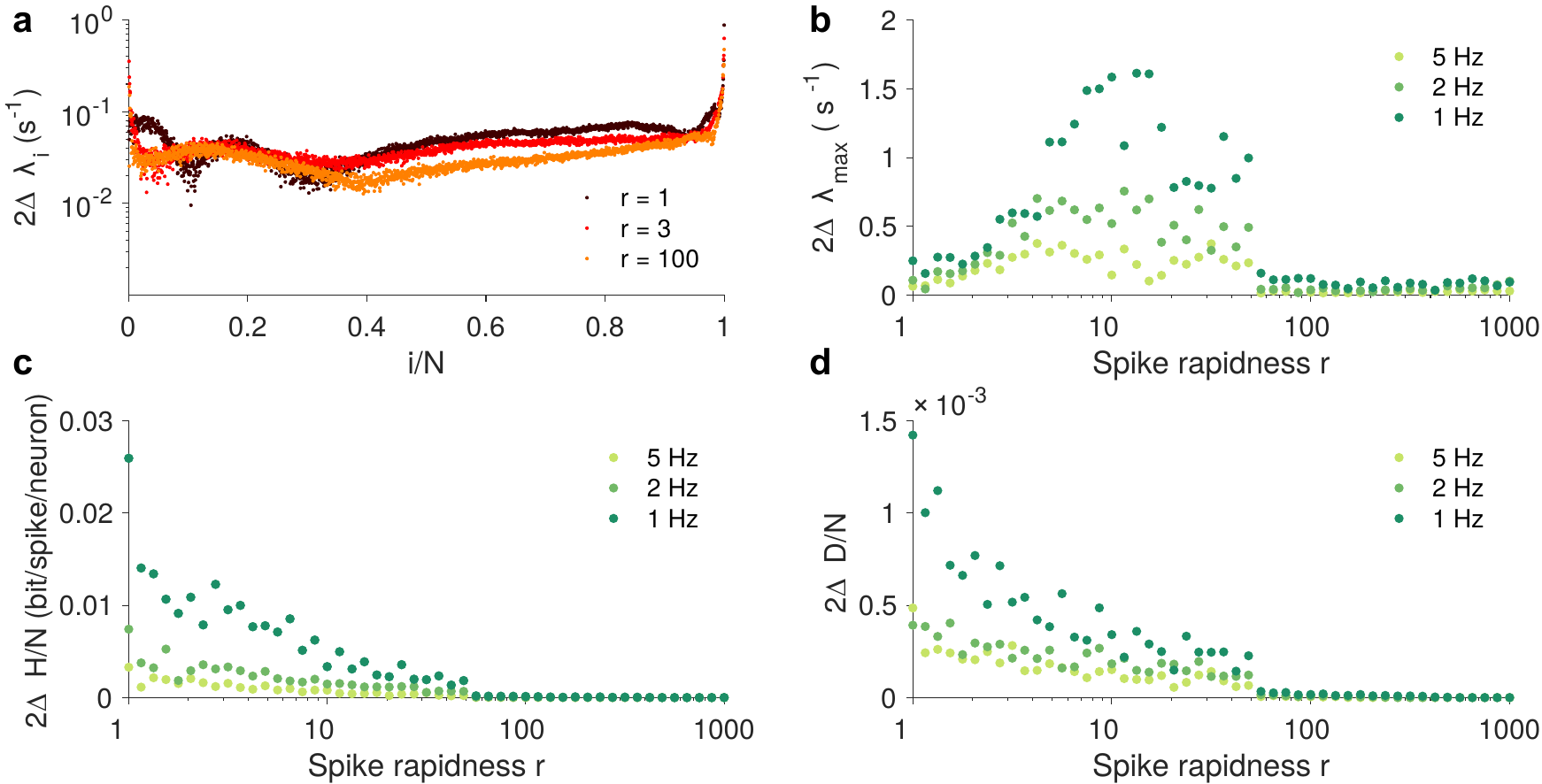} 
\par\end{centering}
\caption[Convergence of largest Lyapunov exponent in inhibitory networks versus
average number of spikes per neuron.]{\textbf{}\label{fig:theta-convergenceLEmax_inhibitory}\textbf{Small
double standard error of Lyapunov exponents, dynamical entropy rate
and attractor dimension in random inhibitory network a)} double standard
error of Lyapunov exponents across different network realizations
for different values of rapidness. \textbf{b)} double standard error
across network realizations of largest Lyapunov exponent for different
values of rapidness. \textbf{c)} same as \textbf{b} for dynamical
entropy rate. \textbf{d)} same as \textbf{b} for relative attractor
dimension. Note that all standard errors are orders of magnitude smaller
than the mean values for the respective quantities. ($N=2000$, $K=100$,
$J_{0}=1$, $\tau_{\mathrm{m}}=10\,\mathrm{ms}$).}
\end{figure}

\begin{figure}[tp]
\begin{centering}
\includegraphics[clip,width=1\textwidth]{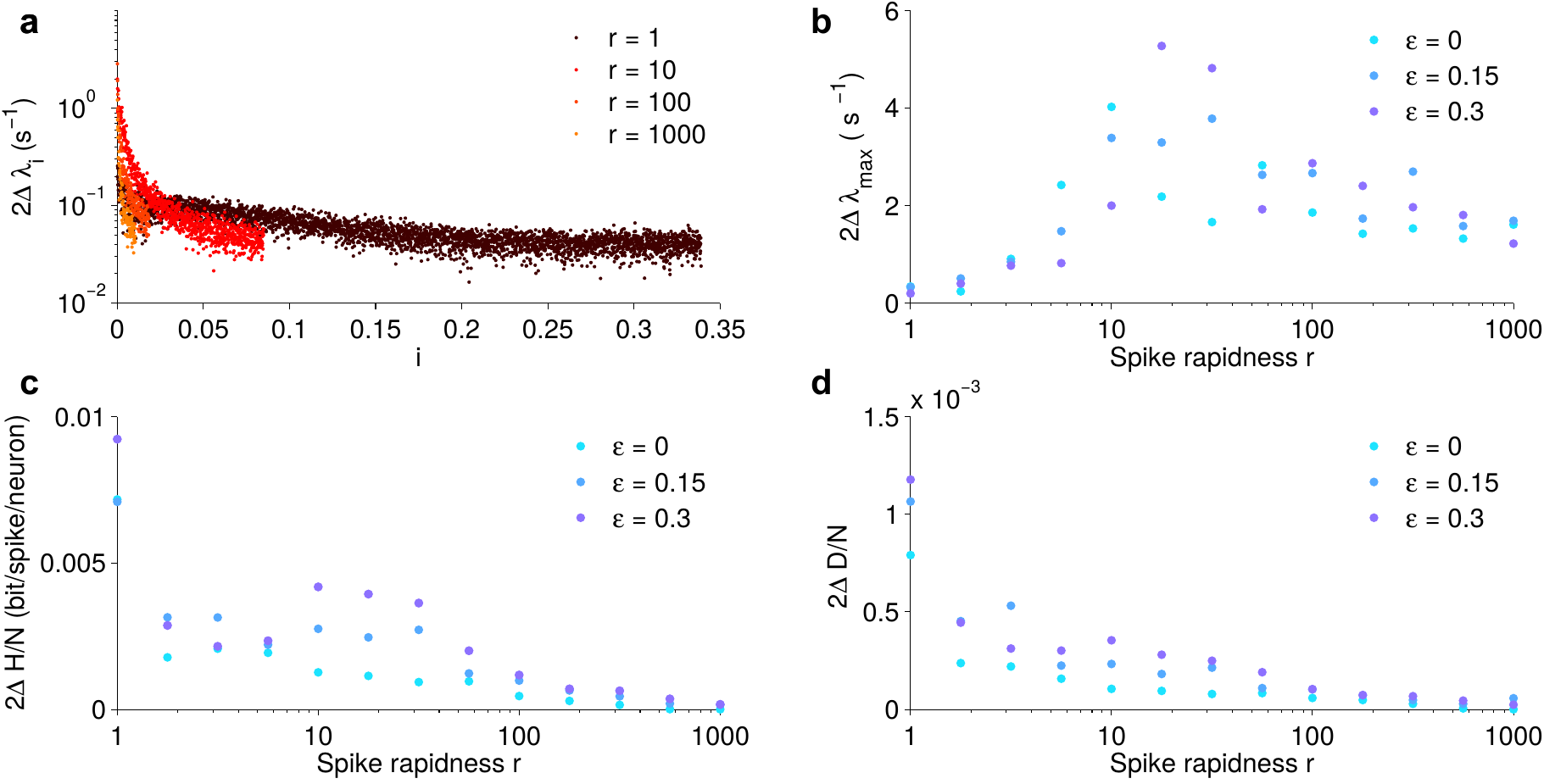} 
\par\end{centering}
\caption[Convergence of largest Lyapunov exponent in excitatory-inhibitory
networks versus average number of spikes per neuron.]{\textbf{}\label{fig:theta-convergenceLEmax_mixed}\textbf{Small double
standard error of Lyapunov exponents, dynamical entropy rate and attractor
dimension in random mixed network a)} double standard error of Lyapunov
exponents across different initial conditions for different values
of rapidness ($\epsilon=0.3$ ). \textbf{b)} double standard error
across network realizations of largest Lyapunov exponent for different
values of rapidness. \textbf{c)} same as \textbf{b} for dynamical
entropy rate. \textbf{d)} same as \textbf{b} for relative attractor
dimension. Note that all standard errors are orders of magnitude smaller
than the mean values for the respective quantities. ($N_{E}=8\,000$,
$N_{I}=2\,000$, $K=100$, $J_{0}=1$, $\tau_{\mathrm{m}}=10\,\mathrm{ms}$).\textbf{}}
\end{figure}
\textbf{Fig.~\ref{fig:theta-convergenceLEmax_inhibitory}} shows
the double standard error of Lyapunov exponents, dynamical entropy
rate and attractor dimension in random inhibitory networks across
10 trials. \textbf{Fig.~\ref{fig:theta-convergenceLEmax_mixed}}
shows the same for mixed networks. We identified two main contributions
to the variability of numerically calculated Lyapunov spectra. Firstly,
variability arises the fact that Lyapunov spectra are asymptotic properties
estimated from finite-time calculations. Secondly variability arising
from the quenched disorder in different random network topologies.
The first contribution is expected to vanish in the limit of infinitely
long simulations for ergodic systems. The second contribution is expected
to vanish in the limit of large network size due to self-averaging.
 Therefore, the Lyapunov spectrum of one realization of a large
network is representative for the whole ensemble. Hence, averaging
over many  network realizations is unnecessary in large networks. 

To validate our results, we applied three independent checks of our
semi-analytic numerically exact calculation of Lyapunov spectra. First,
the largest Lyapunov exponent can be calculated numerically by measuring
the exponential rate of divergence or convergence of nearby trajectories,
as described in \citep{key-Benettin1980,key-Wolf}. Secondly, in autonomous
systems that are not at a fixed point, there is always a neutral Lyapunov
vector in the direction of the flow, which has a zero corresponding
Lyapunov exponent, as the system can be shifted in time. Thirdly,
random matrix theory allows for the calculation of the mean Lyapunov
exponent, as shown in the next paragraph. All checks confirmed the
results obtained from our implementation of the semi-analytic calculation
of the full Lyapunov spectrum, validating the accuracy of our approach.

\section{Random matrix theory of the mean Lyapunov exponent\label{sec:Random-matrix-theory}}

Using the single spike Jacobian (Eq.~\ref{eq:theta-jacobianPRC}),
we derived a random matrix approximation for the mean Lyapunov exponent,
$\bar{\lambda}=\frac{1}{N}\sum_{i=1}^{N}\lambda_{i}$. The mean Lyapunov
exponent describes the rate of phase space volume compression, which
is captured by the determinant of the long-term Jacobian $\mathbf{T}=\mathbf{D}(t_{s})\cdots\mathbf{D}(0)$:
\begin{eqnarray}
\bar{\lambda} & = & \frac{1}{N}\lim_{s\to\infty}\frac{1}{t_{s}}\ln\big(\det\mathbf{T}\big)\nonumber \\
 & = & \frac{1}{N}\lim_{s\to\infty}\frac{1}{t_{s}}\sum_{p=1}^{s}\ln\big(\det\mathbf{D}(t_{p})\big).\label{eq:theta-meanLE}
\end{eqnarray}
 The random matrix approximation is obtained by assuming that the
single spike Jacobians to be random matrices of the form Eq.~\ref{eq:theta-jacobianPRC},
with independent and identically distributed random elements that
are obtained from the function $d(V)$, (Eq.~\eqref{eq:rapid-d-inh-V}).
The probability distribution of these random elements is determined
by the stationary membrane potential distribution $P(V)$ in the network.

For inhibitory networks, the determinants of the random matrices can
be approximated by $\det\mathbf{D}=\prod_{i^{*}}d_{i^{*}}\approx d(V)^{K}$,
since on average there are $K$ diagonal elements $d_{i^{*}}$, one
for each postsynaptic neuron. We assume homogeneous coupling strengths
$J_{ij}\equiv-J_{0}$ between connected neurons and identical external
currents $I_{i}^{\mathrm{ext}}\equiv I_{0}$ for all neurons. The
number of network spike events per second is given by $\lim_{s\to\infty}\frac{1}{t_{s}}\sum_{p=1}^{s}1=N\bar{\nu}$.
Thus, in the random matrix approximation, the mean Lyapunov exponent
for inhibitory networks becomes 
\begin{eqnarray}
\bar{\lambda} & \approx & \frac{1}{N}N\bar{\nu}\int\ln\big(d(V)^{K}\big)P\big(V\big)\mathrm{d}V\nonumber \\
 & = & K\bar{\nu}\int\ln d(V)\,P\big(V\big)\mathrm{d}V.\label{eq:theta-meanLERMT}
\end{eqnarray}
 We obtain $d(V)$ from Eq.~\eqref{eq:rapid-d-inh} using Eq.~\eqref{eq:rapid-trafo2}

\begin{eqnarray}
d_{-}(V) & = & \begin{cases}
\frac{\left(V-V_{\mathrm{G}}\right)^{2}+\frac{I_{0}\sqrt{K}}{a_{\mathrm{S}}}}{\left(V-V_{\mathrm{G}}+C\sqrt{I}\right)^{2}+\frac{I_{0}\sqrt{K}}{a_{\mathrm{S}}}} & V\le V_{\mathrm{G}}\\
\frac{r^{2}\left(V-V_{\mathrm{G}}\right)^{2}+\frac{I_{0}\sqrt{K}}{a_{\mathrm{S}}}}{\left(V-V_{\mathrm{G}}+C\sqrt{I}\right)^{2}+\frac{I_{0}\sqrt{K}}{a_{\mathrm{S}}}} & V_{\mathrm{G}}<V<V_{-}\\
\frac{r^{2}\left(V-V_{\mathrm{G}}\right)^{2}+\frac{I_{0}\sqrt{K}}{a_{\mathrm{S}}}}{r^{2}\left(V-V_{\mathrm{G}}+C\sqrt{I}\right)^{2}+\frac{I_{0}\sqrt{K}}{a_{\mathrm{S}}}} & V_{-}\le V.
\end{cases}\label{eq:rapid-d-inh-1-1}
\end{eqnarray}
In the large $K$ limit, $d(V)$ can be approximated by

\begin{eqnarray}
d_{-}(V) & \stackrel{K\to\infty}{\simeq} & \begin{cases}
1+\frac{2a_{\mathrm{S}}J_{0}\left(V-V_{\mathrm{G}}\right)}{I_{0}K}+\mathcal{O}\big(K^{3/2}\big) & V\le V_{\mathrm{G}}\\
1+\frac{a_{\mathrm{S}}(r^{2}-1)\left(V-V_{\mathrm{G}}\right)^{2}}{I_{0}\sqrt{K}}+\frac{\left(2a_{\mathrm{S}}J_{0}-(r^{2}-1)(V-V_{\mathrm{G}})^{3}\right)\left(V-V_{\mathrm{G}}\right)}{I_{0}^{2}K}+\mathcal{O}\big(K^{3/2}\big) & V_{\mathrm{G}}<V<V_{-}\\
1+\frac{2a_{\mathrm{S}}r^{2}J_{0}\left(V-V_{\mathrm{G}}\right)}{I_{0}K}+\mathcal{O}\big(K^{3/2}\big) & V_{-}\le V.
\end{cases}\label{eq:rapid-d-inh-1-1-1}
\end{eqnarray}

These approximations, along with the balance equation (\ref{eq:balance equation}),
lead to 
\begin{eqnarray}
\bar{\lambda} & \stackrel{K\to\infty}{\simeq} & \frac{2a_{\mathrm{S}}\langle V_{V\leq V_{\mathrm{G}}}\rangle}{\tau_{\mathrm{m}}}\alpha+\frac{2_{\mathrm{U}}\langle V_{V>V_{\mathrm{G}}}\rangle}{\tau_{\mathrm{m}}}(1-\alpha)+\mathcal{O}\left(\frac{1}{\sqrt{K}}\right),\label{eq:meanLE_RMT_largeK}
\end{eqnarray}
 where $\langle V_{V\leq V_{\mathrm{G}}}\rangle$ ($\langle V_{V>V_{\mathrm{G}}}\rangle$
) denotes the average membrane potential below (above) $V_{\mathrm{G}}$,
and $\alpha$ is the fraction of neurons below $V_{\mathrm{G}}$.

\begin{figure}[tp]
\begin{centering}
\includegraphics[clip,width=1\textwidth]{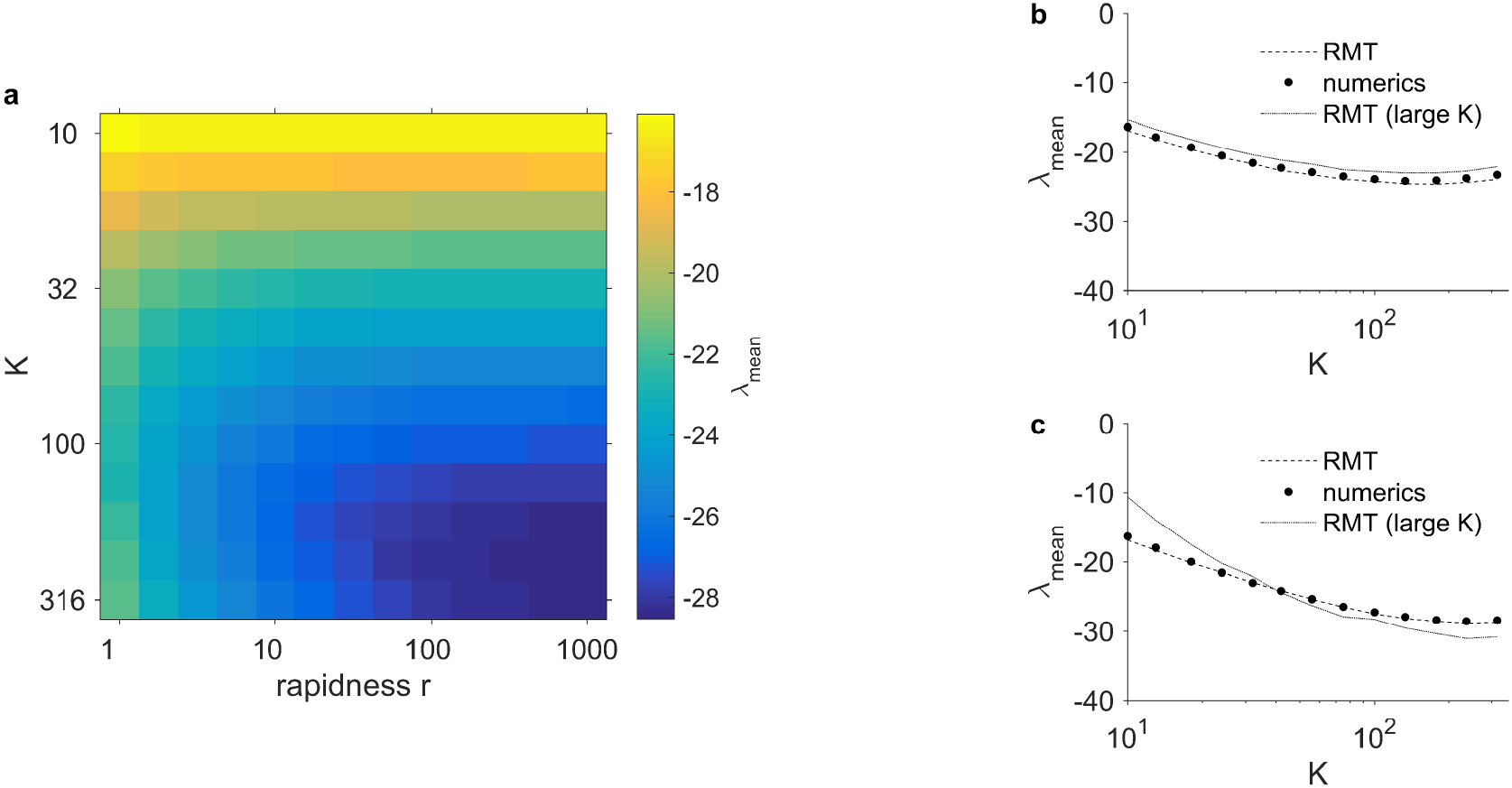} 
\par\end{centering}
\caption[Comparison of mean Lyapunov exponent in random matrix approximations
and from numerical simulations in balanced rapid theta neuron networks.]{\textbf{}\label{fig:theta-RMT}\textbf{Good match between mean Lyapunov
exponent in random matrix approximations and numerical simulations
in balanced rapid theta neuron networks. a)} Mean Lyapunov exponent
from numerical simulations for different $K$ and $r$ with $N=1000$.
\textbf{b)} $K$-dependence for $r=3$. Solid lines: numerical simulations,
dashed lines: random matrix approximations with full membrane potential
distributions, dotted lines: random matrix approximation in the large
$K$-limit \textbf{c)} $K$-dependence for $r=100$. Eq.~\eqref{eq:meanLE_RMT_largeK}
 (parameters: $\bar{\nu}=1\,\mathrm{Hz}$, $J_{0}=1$, $\tau_{\mathrm{m}}=10\,\mathrm{ms}$).}
\end{figure}

We have compared the derived random matrix approximations of the mean
Lyapunov exponent in inhibitory networks Eq.~\eqref{eq:theta-meanLERMT}
and the large-$K$ limits, Eq.~\eqref{eq:meanLE_RMT_largeK}, with
the results from simulations (\textbf{Fig.~\ref{fig:theta-RMT}}).
They are in very good agreement, indicating the validity of the random
matrix approximation. This likely results from the commutativity of
the determinants of the Jacobians, a property that generally does
not hold for the product of the Jacobians.

\section{Scaling of the largest Lyapunov exponent with network parameters}

\begin{figure}[tp]
\begin{centering}
\includegraphics[clip,width=1\textwidth]{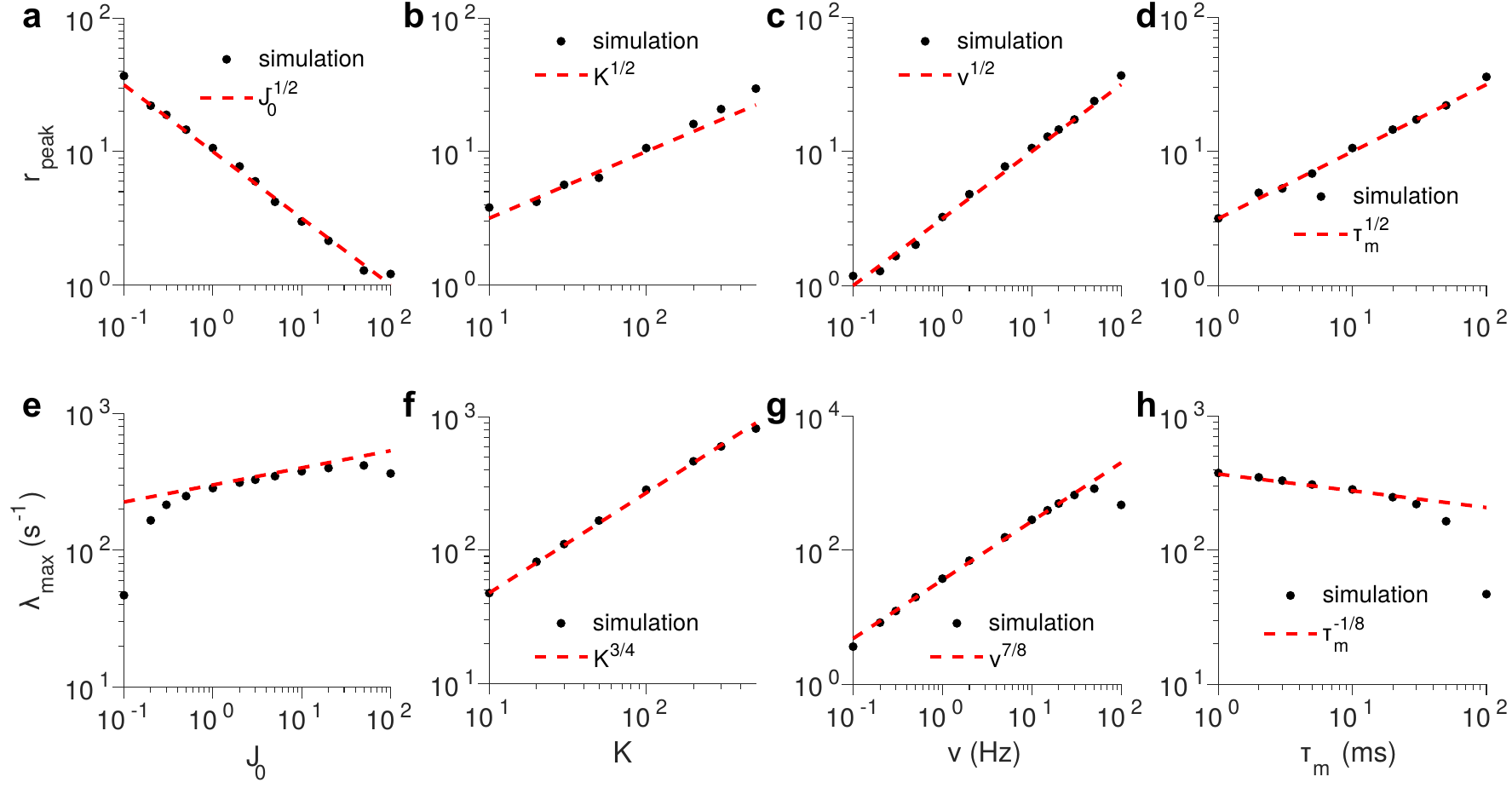} 
\par\end{centering}
\caption[Comparison of mean Lyapunov exponent in random matrix approximations
and from numerical simulations in balanced rapid theta neuron networks.]{\textbf{}\label{fig:r-peak-scaling}\textbf{Scaling of peak AP onset
rapidness }$r_{\mathrm{peak}}$ \textbf{with network parameters in
balanced inhibitory networks }$r_{\mathrm{peak}}\propto\sqrt{K\nu_{0}\tau_{\mathrm{m}}/J_{0}}$\textbf{.
}At the peak onset rapidness $r_{\mathrm{peak}}$ the largest Lyapunov
exponent has its maximum as a function of rapidness (\textbf{Fig.~2}
of main paper). \textbf{a)} peak onset rapidness $r_{\mathrm{peak}}$
vs. $J_{0}$, dashed red lines indicate power-law fit using the Levenberg-Marquardt
algorithm. \textbf{b)} peak onset rapidness $r_{\mathrm{peak}}$ vs.
$K$. \textbf{c)} peak onset rapidness $r_{\mathrm{peak}}$ vs. $J_{0}$.
\textbf{d)} peak onset rapidness $r_{\mathrm{peak}}$ vs. $J_{0}$.
\textbf{e)-h)} Second row: corresponding peak onset rapidness $\textbf{\ensuremath{\lambda_{\mathrm{\mathrm{max}}}}}$
(parameters: $\bar{\nu}=1\,\mathrm{Hz}$, $J_{0}=1$, $\tau_{\mathrm{m}}=10\,\mathrm{ms}$,
$N=1000$, $K=100$).}
\end{figure}

The largest Lyapunov exponent exhibits a single maximum as a function
of the action potential onset rapidness. The peak position, which
we call peak rapidness $r_{\mathrm{peak}}$, scales approximately
as\textbf{ }shown in \textbf{Fig.~\ref{fig:r-peak-scaling}}:

\begin{equation}
r_{\mathrm{peak}}\propto\sqrt{\frac{K\nu_{0}\tau_{\mathrm{m}}}{J_{0}}}\label{eq:r_peak_scaling}
\end{equation}

This behavior can be understood as a transition between two qualitatively
different scaling regimes for the largest Lyapunov exponent. For small
values of $K$ and $\bar{\nu}$ the largest Lyapunov exponent grows,
while for large values of $K$ and $\bar{\nu}$ the largest Lyapunov
exponent reaches a plateau. Qualitative differences between these
two collective network states become clear in the localization of
the first covariant Lyapunov vector, the local Lyapunov exponents
and in a spatiotemporal analysis of the network chaos in section \ref{sec:Participation-ratio-and}.

\begin{figure}[tp]
\begin{centering}
\includegraphics[clip,width=1\textwidth]{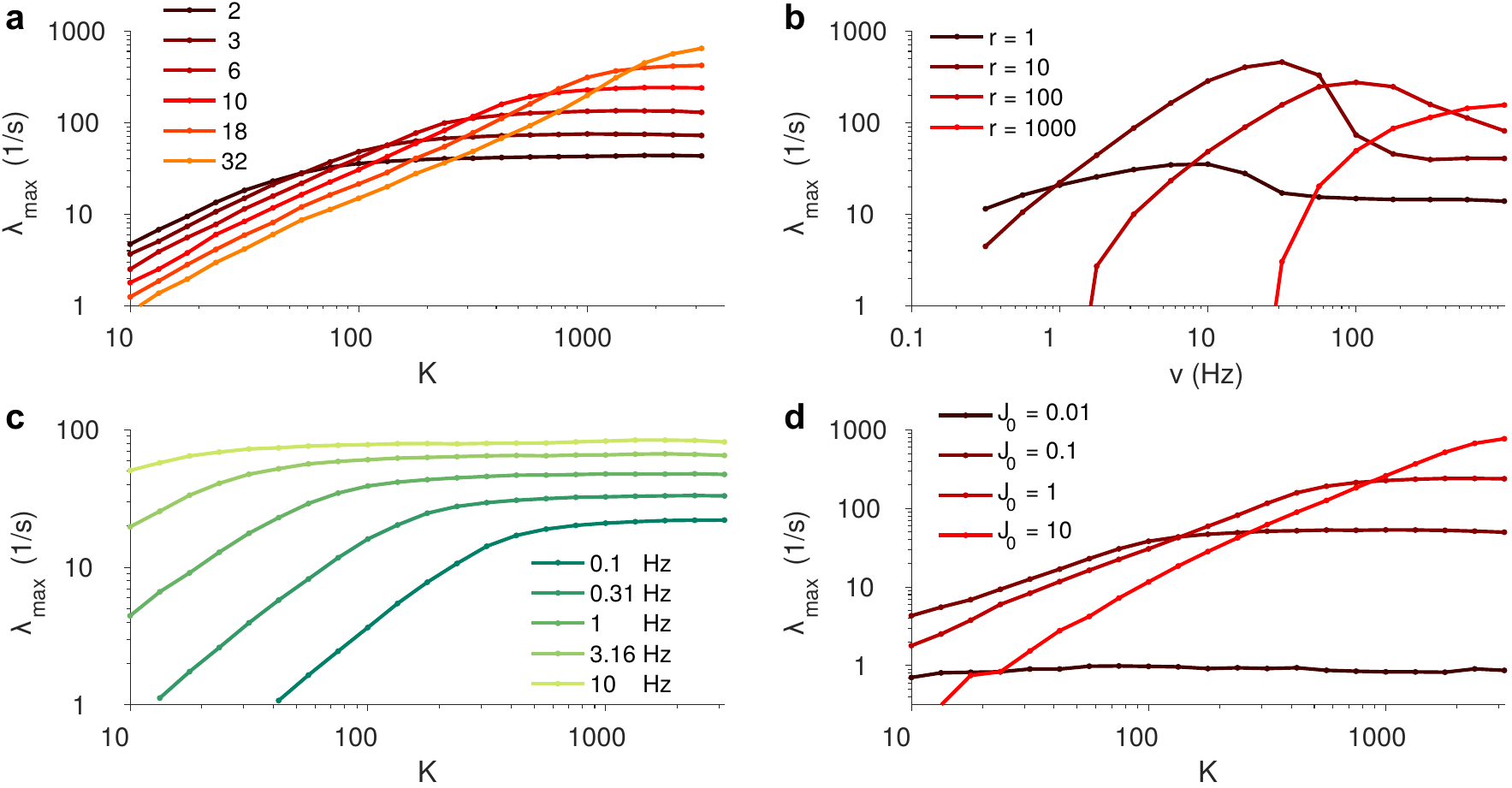} 
\par\end{centering}
\caption[Comparison of mean Lyapunov exponent in random matrix approximations
and from numerical simulations in balanced rapid theta neuron networks.]{\textbf{}\label{fig:lambda_max_K-r}\textbf{Different scaling regimes
explain peak in the largest Lyapunov exponent $\textbf{\ensuremath{\lambda_{\mathrm{\mathrm{max}}}}}$.
a)} K-dependence of largest Lyapunov exponent from for different values
of rapidness $r$ with $N=10000$. \textbf{b)} $\nu$-dependence of
largest Lyapunov exponent for different rapidness $r$. \textbf{c)}
$K$-dependence of largest Lyapunov exponent for mean firing rate
$\nu_{0}$ with $N=1000$. \textbf{d)} $K$-dependence of largest
Lyapunov exponent for different coupling strength $J_{0}$ (parameters:
$\bar{\nu}=10\,\mathrm{Hz}$, $J_{0}=1$, $\tau_{\mathrm{m}}=10\,\mathrm{ms}$,
$N=10000$, $K=100$).}
\end{figure}

\begin{figure}[tp]
\begin{centering}
\includegraphics[clip,width=1\textwidth]{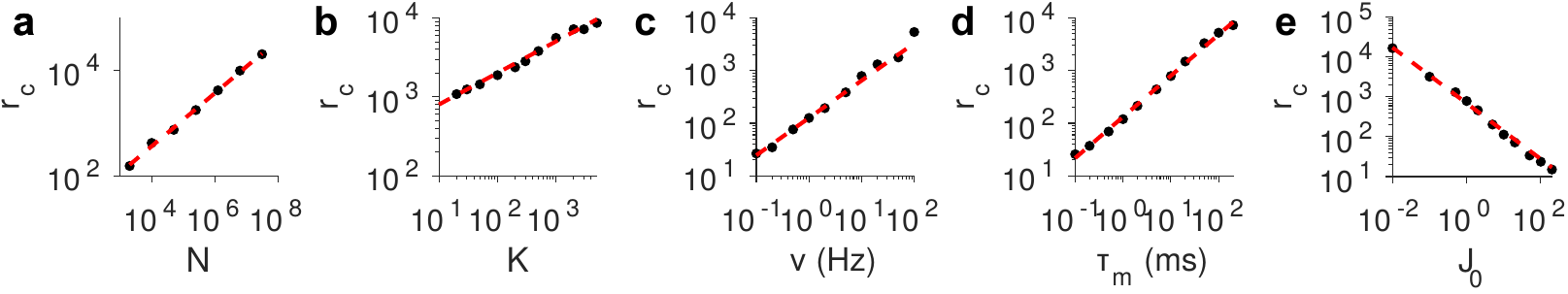} 
\par\end{centering}
\caption[Comparison of mean Lyapunov exponent in random matrix approximations
and from numerical simulations in balanced rapid theta neuron networks.]{\textbf{}\label{fig:r-crit-scaling}\textbf{Scaling of critical AP
onset rapidness }$r_{c}$ \textbf{between stable and chaotic dynamics
in balanced inhibitory networks }$r_{\mathrm{crit}}\propto N^{0.5}K^{0.4}\bar{\nu}^{0.8}\tau_{\mathrm{m}}J_{0}^{-0.7}$\textbf{.
}The critical spike rapidness $r_{\mathrm{crit}}$ separates chaotic
dynamics (above) from stable dynamics (below). \textbf{a)} $r_{\mathrm{crit}}$
vs. $N$, \textbf{b)} $r_{\mathrm{crit}}$ vs. $K$, \textbf{c)} $r_{\mathrm{crit}}$
vs. $\bar{\nu}$, \textbf{d)} $r_{\mathrm{crit}}$ vs. $\tau_{\mathrm{m}}$,
\textbf{e)}$r_{\mathrm{crit}}$ vs. $J_{0}$ Power-law fits done using
the Levenberg-Marquardt algorithm (parameters: $\bar{\nu}=10\,\mathrm{Hz}$,
$J_{0}=1$, $\tau_{\mathrm{m}}=10\,\mathrm{ms},$ $N=2000$, $K=100$).}
\end{figure}

The critical spike onset rapidness $r_{\mathrm{crit}}$ separates
the chaotic dynamics from the stable dynamics. This transition with
network parameters has the following scaling behavior (\textbf{Fig.~\ref{fig:r-crit-scaling}}):

\begin{equation}
r_{\textrm{crit}}=N^{0.5}K^{0.4}\bar{\nu}^{0.8}\tau_{m}^{0.8}J_{0}^{-0.7}\label{eq:r_crit_scaling}
\end{equation}
This scaling indicates that in the thermodynamic limit of large $K$
and large $N$, the critical rapidness $r_{\textrm{crit}}$ diverges
and the network is always chaotic. This is in agreement with the scaling
of the flux tube radius of networks of leaky integrate and fire neurons,
which goes to zero in the limit of large networks. Note however that
even in the thermodynamic limit it is possible to bring the largest
Lyapunov exponent and thus also the Kolmogorov Sinai entropy rate
arbitrarily close to zero by increasing the AP onset rapidness $r$.
 
\paragraph*{Extensivity of Lyapunov spectrum and asymptotic form:}

\begin{figure}[tp]
\begin{centering}
\includegraphics[clip]{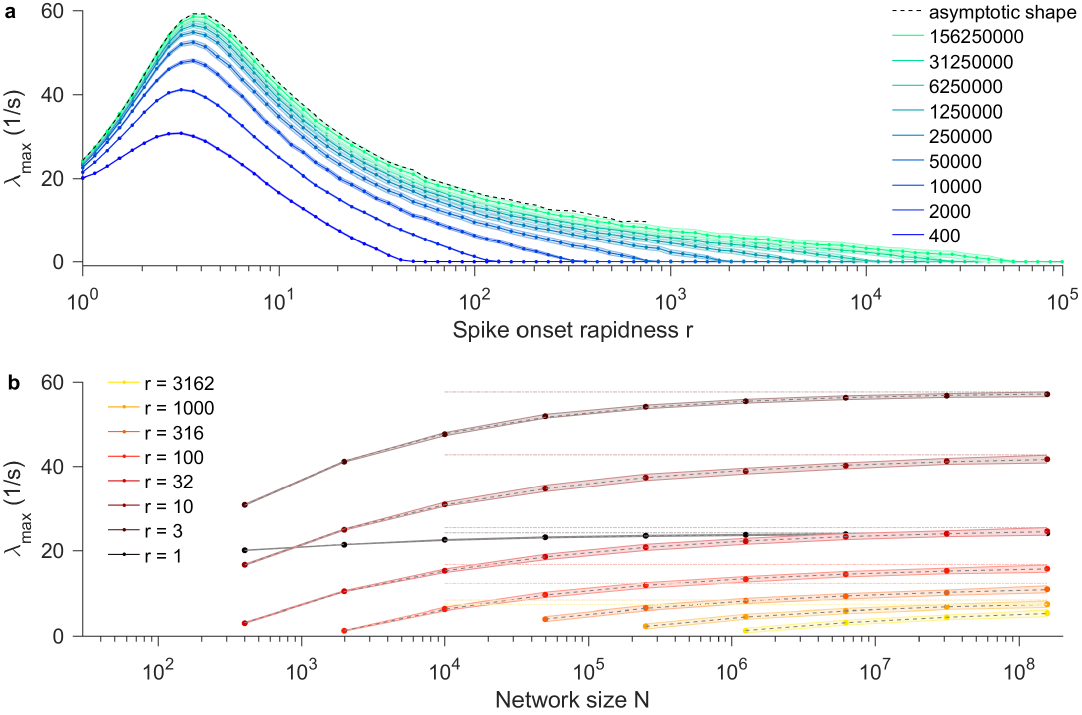} 
\par\end{centering}
\caption[Comparison of mean Lyapunov exponent in random matrix approximations
and from numerical simulations in balanced rapid theta neuron networks.]{\textbf{}\label{fig:maxN}\textbf{Largest Lyapunov exponent converges
to asymptotic shape for very large network size $N$: a)} The largest
Lyapunov exponent exhibits a slow convergence with network size $N$
($\bar{\nu}=1\,\mathrm{Hz}$). \textbf{b)} The convergence of $\lambda_{\mathrm{max}}(N)$
can be accurately fitted by $\lambda_{\mathrm{max}}(N)=\lambda_{\mathrm{max}}(\infty)-c\cdot N^{-\frac{1}{\gamma}}$,
where $\lambda_{\mathrm{max}}(\infty)=\lim_{N\to\infty}$ $\lambda_{\mathrm{max}}(N)$
(median (dots) across 10 topologies, shaded error shadings indicate
median bootstrapped 95\% confidence intervals). The estimated $\lambda_{\mathrm{max}}(\infty)$
is indicated by a black dashed line. ($r=10$,~$\bar{\nu}=10\,\mathrm{Hz}$)
(parameters: $J_{0}=1$, $\tau_{\mathrm{m}}=10\,\mathrm{ms},$ $K=100$).}
\end{figure}

\textbf{Figure 2} and \textbf{3} of the main paper show that for sufficiently
large networks the dynamical entropy rate and Kaplan-Yorke attractor
dimension scales linear with network size $N$ indicating extensive
deterministic chaos. While the dynamical entropy rate and Kaplan-Yorke
attractor dimension converge to a linear scaling with $N$ already
for moderate network size, the largest Lyapunov exponent exhibited
a slower convergence to its large $N$ limit (\textbf{Fig.}~\textbf{\ref{fig:maxN}}).
While the peak rapidness $r_{\mathrm{peak}}$ is independent of $N$,
$\lambda_{\mathrm{max}}(N)$ converges exponentially towards its large
$N$ limit. This exponential convergence allows an estimate of the
asymptotic value $\lim_{N\to\infty}$ $\lambda_{\mathrm{max}}(N)$,
indicated in \textbf{Fig.}~\textbf{\ref{fig:maxN}b} by a dashed
line. Note that the critical rapidness $r_{\mathrm{crit}}$ diverges
with $N$ as shown in \textbf{Fig.}~\textbf{\ref{fig:r-crit-scaling}}.

\section{Participation ratio and localization of chaos\label{sec:Participation-ratio-and}}

\textbf{}
\begin{figure}
\textbf{\includegraphics[clip,width=0.5\columnwidth]{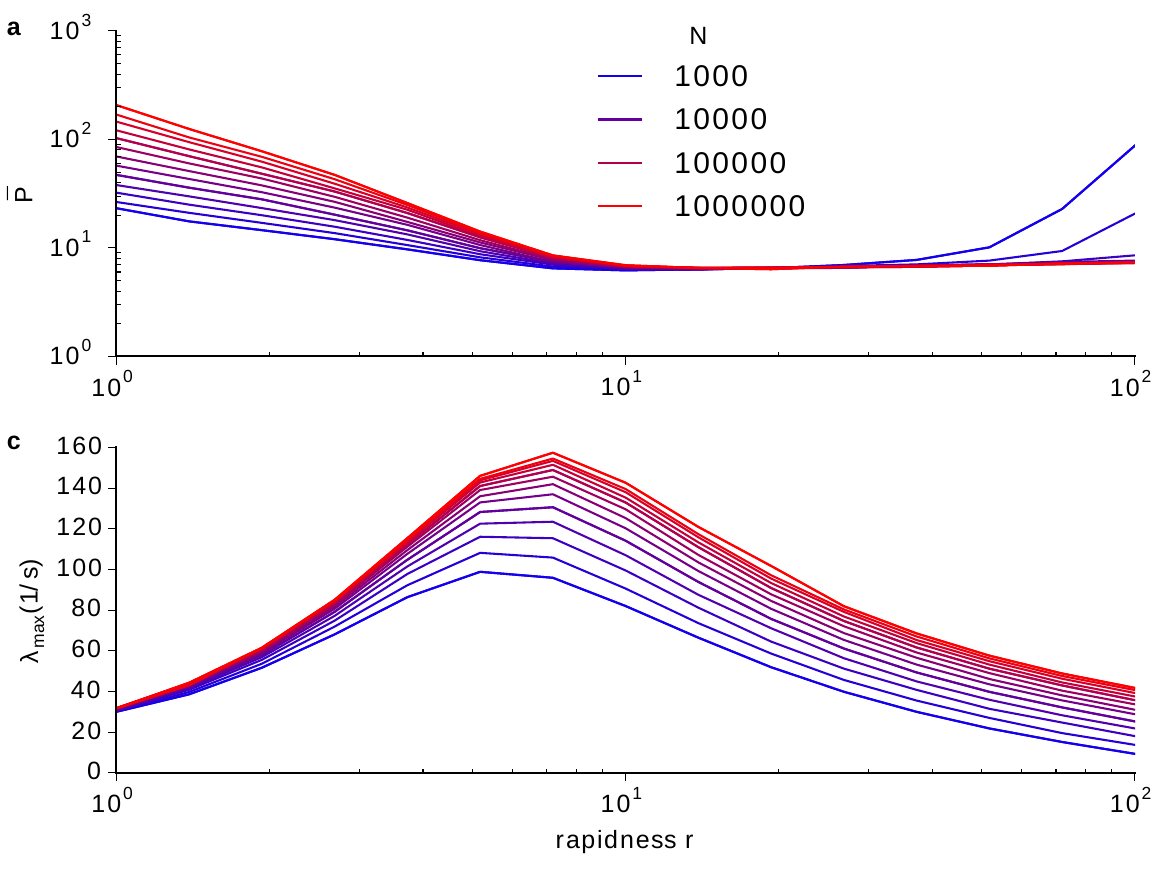}}\includegraphics[clip,width=0.5\columnwidth]{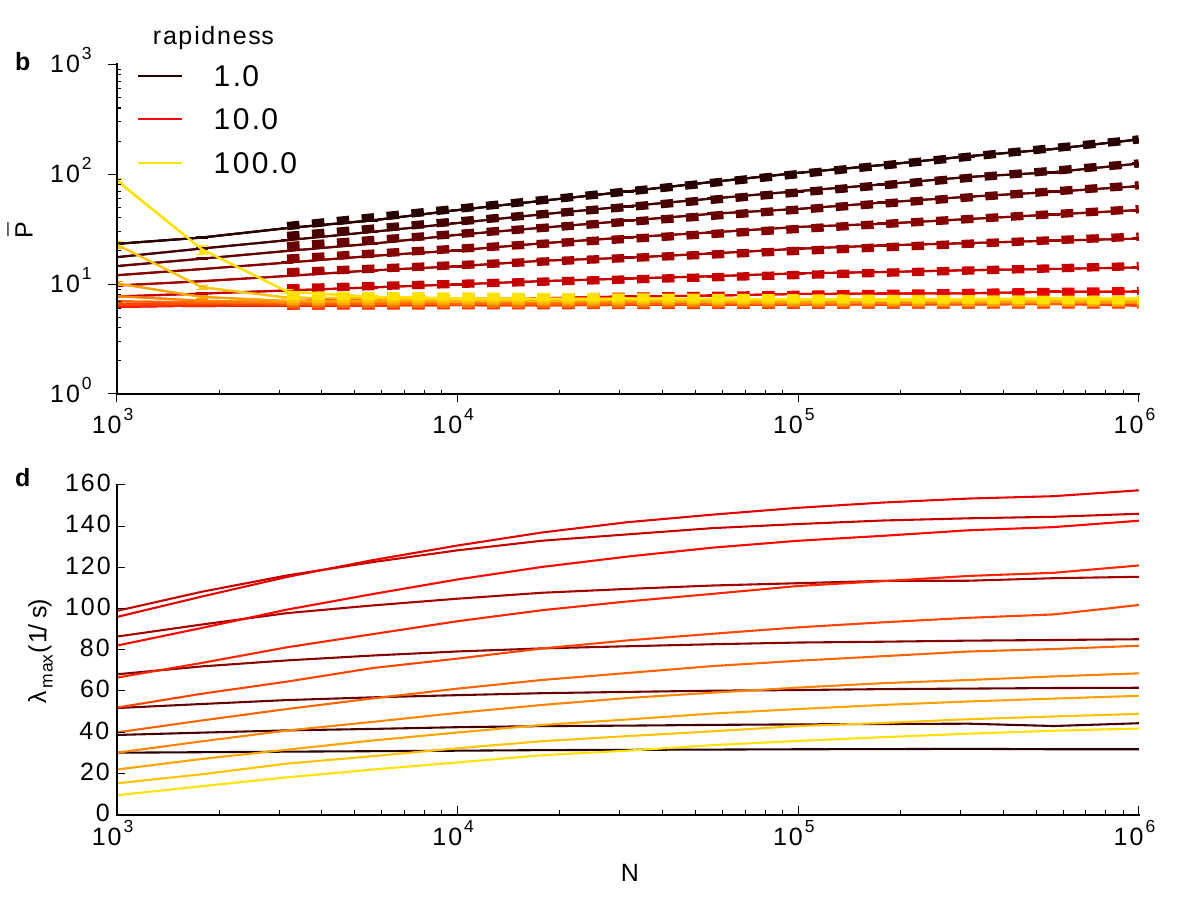}

\caption[Comparison of mean Lyapunov exponent in random matrix approximations
and from numerical simulations in balanced rapid theta neuron networks.]{\textbf{}\label{fig:ParticipationRatio}\textbf{Average participation
ratio vs rapidness $r$ and network size $N$ reveals localization
of first covariant Lyapunov vector above peak rapidness: a)} The mean
participation ratio $\bar{P}$ versus rapidness $r$, colors encode
network size~$N$. \textbf{b)} $\bar{P}$ versus $N$, colors encode
rapidness $r$. $\bar{P}$ shows power-law scaling $\bar{P}\sim N^{\alpha}$,
where the exponent $\alpha$ decreases as function of rapidness $r$,
dashed lines indicate power-law fits using the Levenberg-Marquardt
algorithm (see \textbf{Fig.}~\textbf{\ref{fig:LV-2-1}} for comprehensive
scaling of $\alpha$). \textbf{c)} The largest Lyapunov exponent vs
rapidness $r$ exhibits a peak approximately where the participation
ratio becomes independent of network size $N$ in \textbf{a}. \textbf{d)}
The largest Lyapunov exponent vs. $N$ converges exponentially to
asymptotic limit (parameters: $\bar{\nu}=3\,\mathrm{Hz}$, $J_{0}=1$,
$\tau_{\mathrm{m}}=10\,\mathrm{ms},$ $K=100$).}
\end{figure}
To quantify how many neurons contribute to the chaotic dynamics at
each and every moment in time, we investigated properties of the covariant
Lyapunov vectors $\vec{\delta\phi^{c}}(t)$. The first Lyapunov vector,
which corresponds to the first Gram-Schmidt vector ($\sum_{i=1}^{N}\delta\phi_{i}(t)^{2}=1$)
gives at any point in time the direction in which almost all initial
infinitesimal perturbations grow with average rate $\lambda_{\textrm{max}}$.
The number of neurons contributing to the maximally growing direction
at time $t$ can be measure by the participation ratio $P(t)=\left(\sum_{i=1}^{N}\delta\phi_{i}(t)^{4}\right)^{-1}$
\citep{key-Wegner1980,key-Kaneko,key-Cross-1993}. If all neurons
contribute similarly to the Lyapunov vector $\left|\delta\phi_{i}(t)\right|=1/\sqrt{N}$
the participation ratio is $P(t)=1/(N/N^{2})=N$. If only one neurons
contributes to the Lyapunov vector the participation ratio is $P(t)=1.$
We found that the participation ratio strongly depends on the spike
onset rapidness. Increasing rapidness generally reduced the participation
ratio (\textbf{Fig.~\ref{fig:ParticipationRatio}}) for large networks.
This means that for increasing rapidness fewer neurons contribute
to the most unstable direction. On the level of the single neuron
dynamics, this can be explained by a decreasing fraction of (postsynaptic)
neurons that are in the unstable regime with voltages above the glue
point $V_{\mathrm{G}}$ for increasing values of $r$. Thus, there
exist on average fewer entries in the Jacobian with large entries
$d(V_{i})$. Therefore, the first Lyapunov vector is on average expected
to have few entries with large values. 

To further characterize the nature of the chaotic collective network
state, it is crucial to investigate the scaling of the mean participation
ratio $\bar{P}=\left\langle P(t)\right\rangle $ with network size.
Whether the Lyapunov vector is called localized or delocalized depends
on how $\bar{P}$ scales as a function of network size $N$. A delocalized
state is indicated by a linear scaling $\bar{P}\sim N$, while in
case of a localized state, the participation ratio would be independent
of $N$. 

\begin{figure}
\includegraphics[clip,width=0.4\columnwidth]{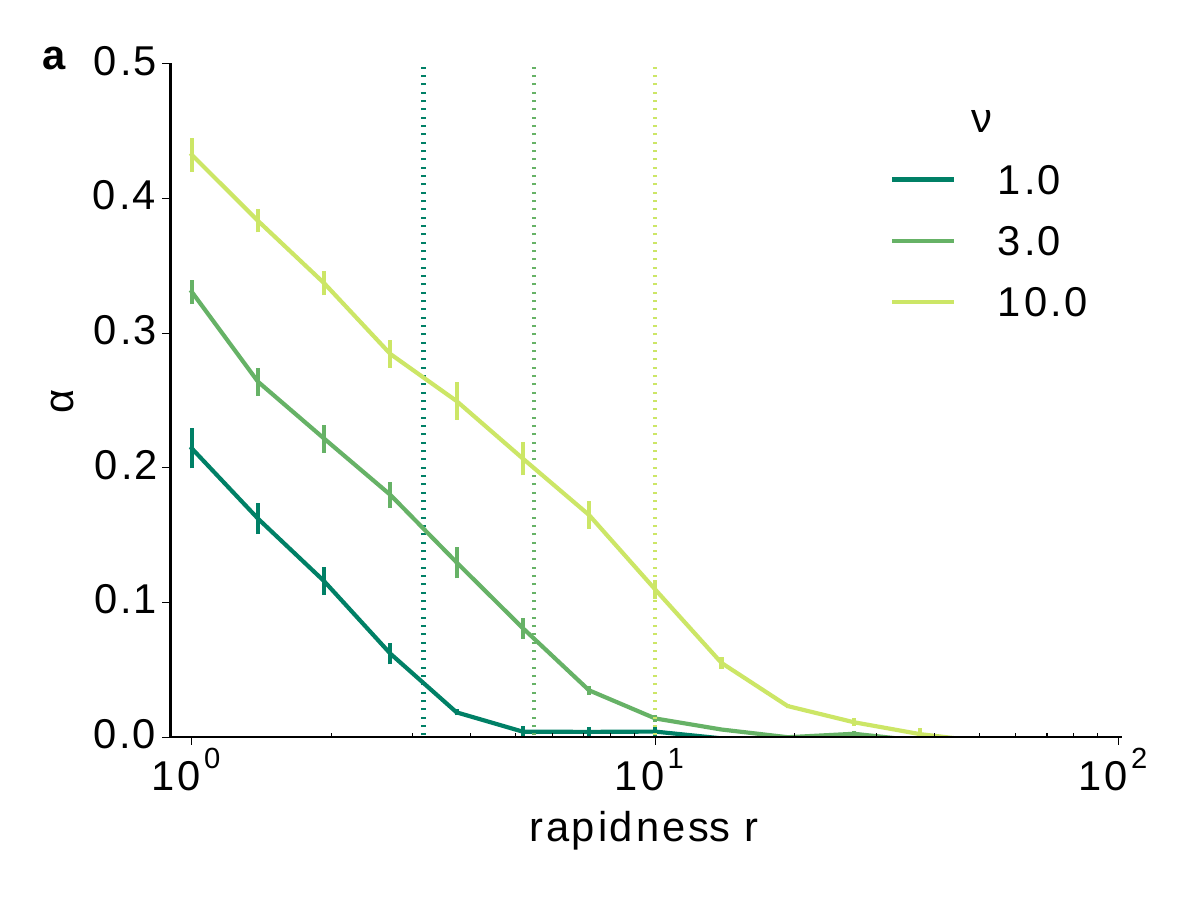}\includegraphics[clip,width=0.4\columnwidth]{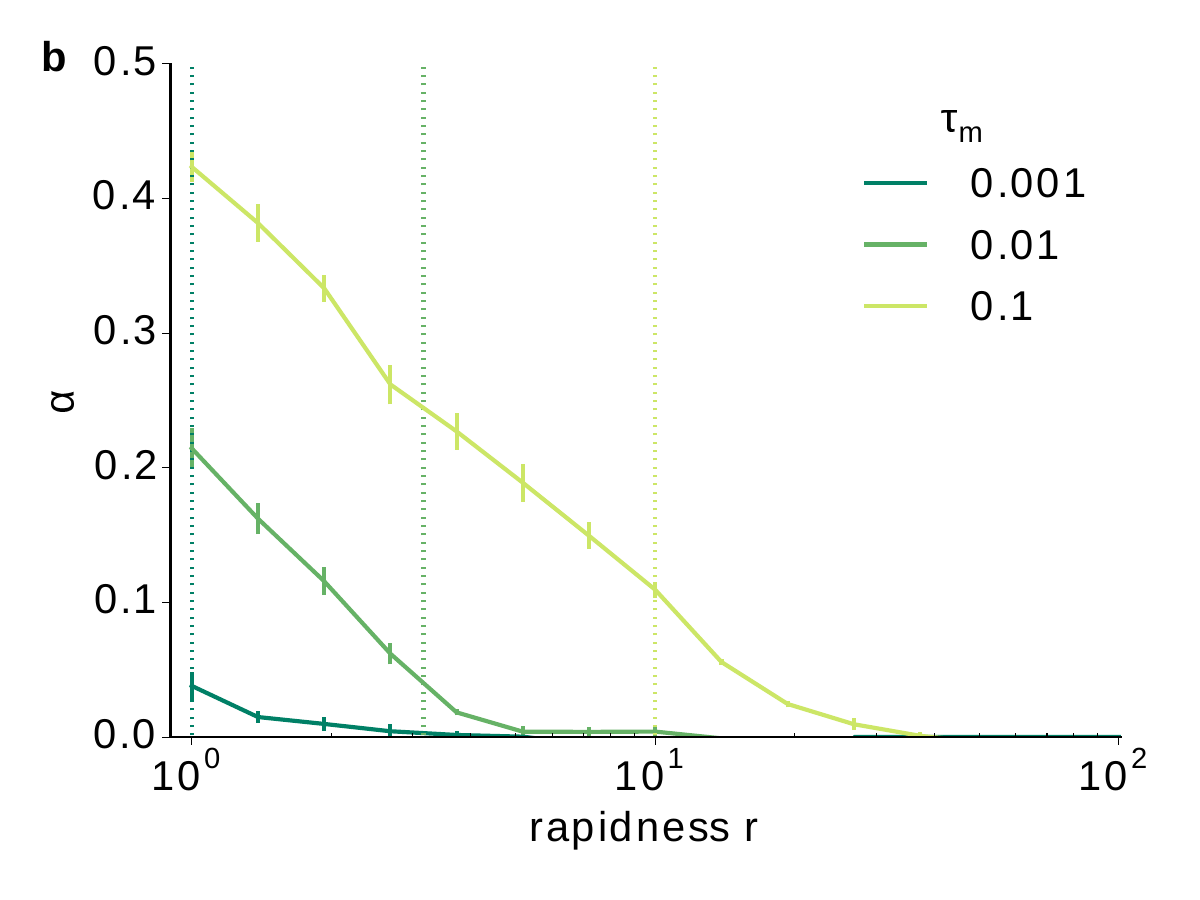}

\includegraphics[clip,width=0.4\columnwidth]{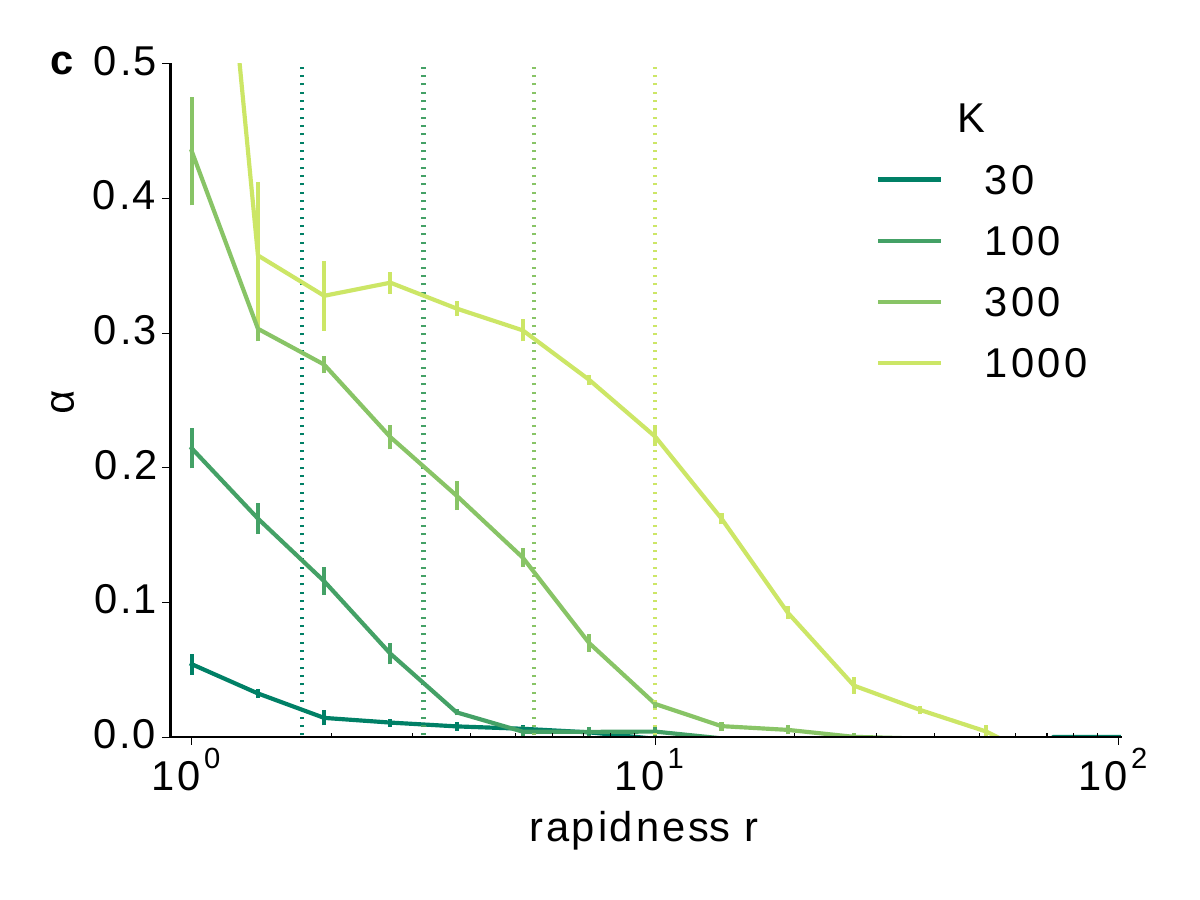}\includegraphics[clip,width=0.4\columnwidth]{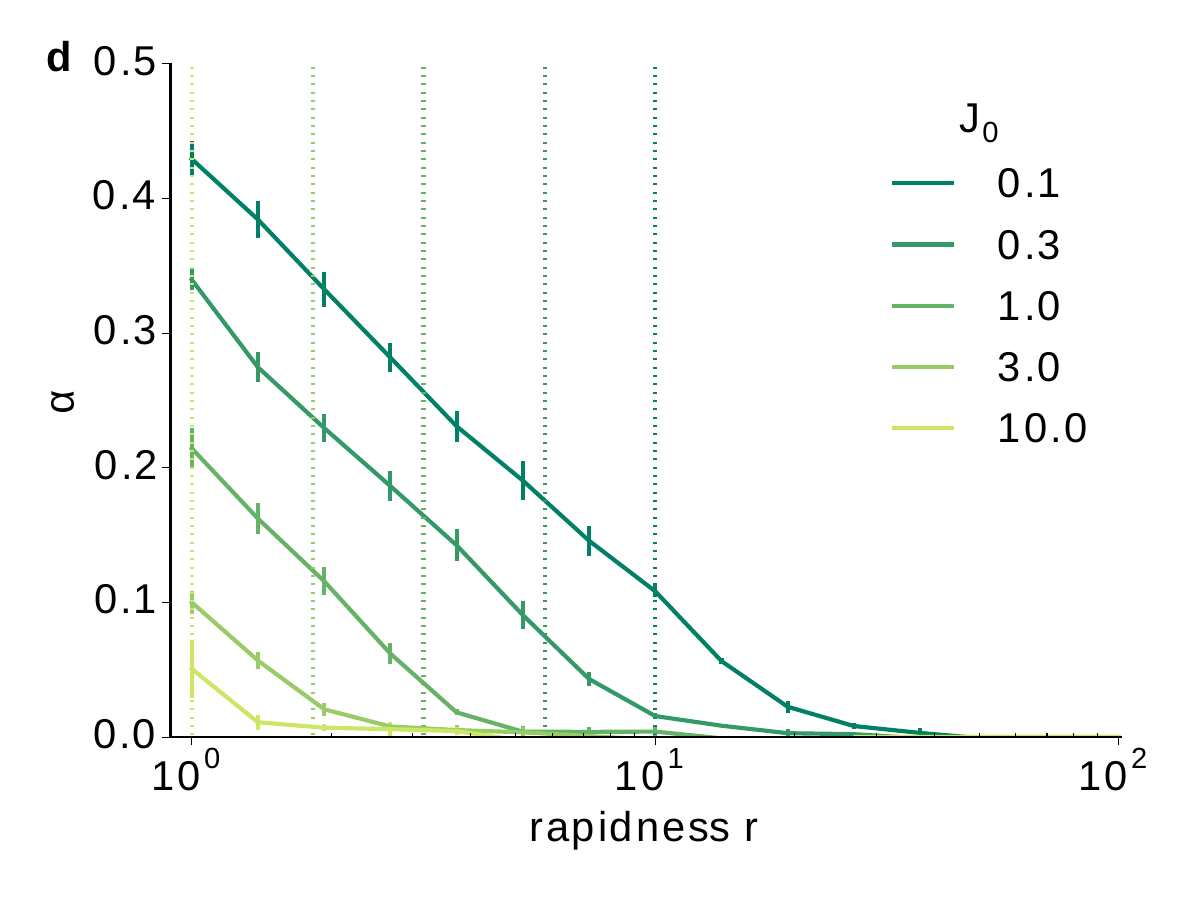}

\caption{\label{fig:LV-2-1}\textbf{Scaling of participation ratio with network
parameters shows that localization of first covariant Lyapunov vector
has same $r$-scaling as peak rapidness: a)} power-law scaling exponent
$\alpha$ from $\bar{P}\sim N^{\alpha}$ fits mean firing rate $\bar{\nu}$
decreases approximately logarithmically as function of $r$. Dashed
vertical lines indicate corresponding values of peak rapidness $r_{\mathrm{peak}}$.
\textbf{b)} Same for different membrane time constants $\tau_{\mathrm{m}}$.
\textbf{c)} Same for different number of synapses per neuron $K$.
\textbf{d)} Same for different coupling strengths $J_{0}$ (parameters:
$N=10^{3}-10^{6}$, $K=100$, $\bar{\nu}=1\textrm{ Hz}$, $J_{0}=1$,
$\tau_{\mathrm{m}}=10\mathrm{ms}$, $\varepsilon=0.3$).}
\end{figure}

In an earlier study of theta neurons, corresponding to $r=1$, it
was found that the Lyapunov vector was dominated by subsets of neurons
that changed over time \citep{key-Monteforte2010}. The participation
ratio exhibited a sublinear scaling $\bar{P}\sim N^{\alpha}$, with
$0<\alpha<1$. Here, we found a strong dependence of the participation
ratio and localization on the spike onset rapidness $r$. For increasing
spike onset rapidness $r$, the exponential scaling parameter $\alpha$
deceased approximately logarithmically as function of $r$ and turned
zero at a certain value of rapidness that depends on firing rate $\bar{\nu}$,
coupling strength $J_{0}$, number of synapses per neuron $K$, and
membrane time constant $\tau_{m}$ (\textbf{Fig.~\ref{fig:LV-2-1}}).
For larger values of rapidness, there was on average a fixed number
of neurons contributing to the first Lyapunov vector, which was independent
of network size $N$ for sufficiently large networks. An extensive
analysis of the scaling of the onset rapidness with network parameters
where the localization occurred coincided with the scaling of the
rapidness where the Lyapunov exponents peaks (\textbf{Fig.~\ref{fig:LV-2-1}})
\begin{equation}
r_{\mathrm{localization}}\propto\sqrt{\frac{K\nu_{0}\tau_{\mathrm{m}}}{J_{0}}}.\label{eq:r_localization_scaling}
\end{equation}

The scaling behavior indicates that the localization of the first
Lyapunov vector occurs when $K$, $\nu_{0}$ or $\tau_{\mathrm{m}}$
is sufficiently small or when $J_{0}$ is sufficiently large. Thus,
localization occurs, when there are few postsynaptic potentials per
spike ($K$ small), which occur infrequent ($\nu_{0}$ or $\tau_{\mathrm{m}}$
small), and/or are strong ( $J_{0}$ large). 

Note that despite the same scaling, localization occurs always at
slightly larger values of rapidness than the peak in the largest Lyapunov
exponent. So the first Lyapunov vector seems to localize not until
the largest Lyapunov exponent is in the regime where it is independent
of the number of synapses per neuron $K$.

\begin{figure}
\textbf{\includegraphics[clip,width=0.5\columnwidth]{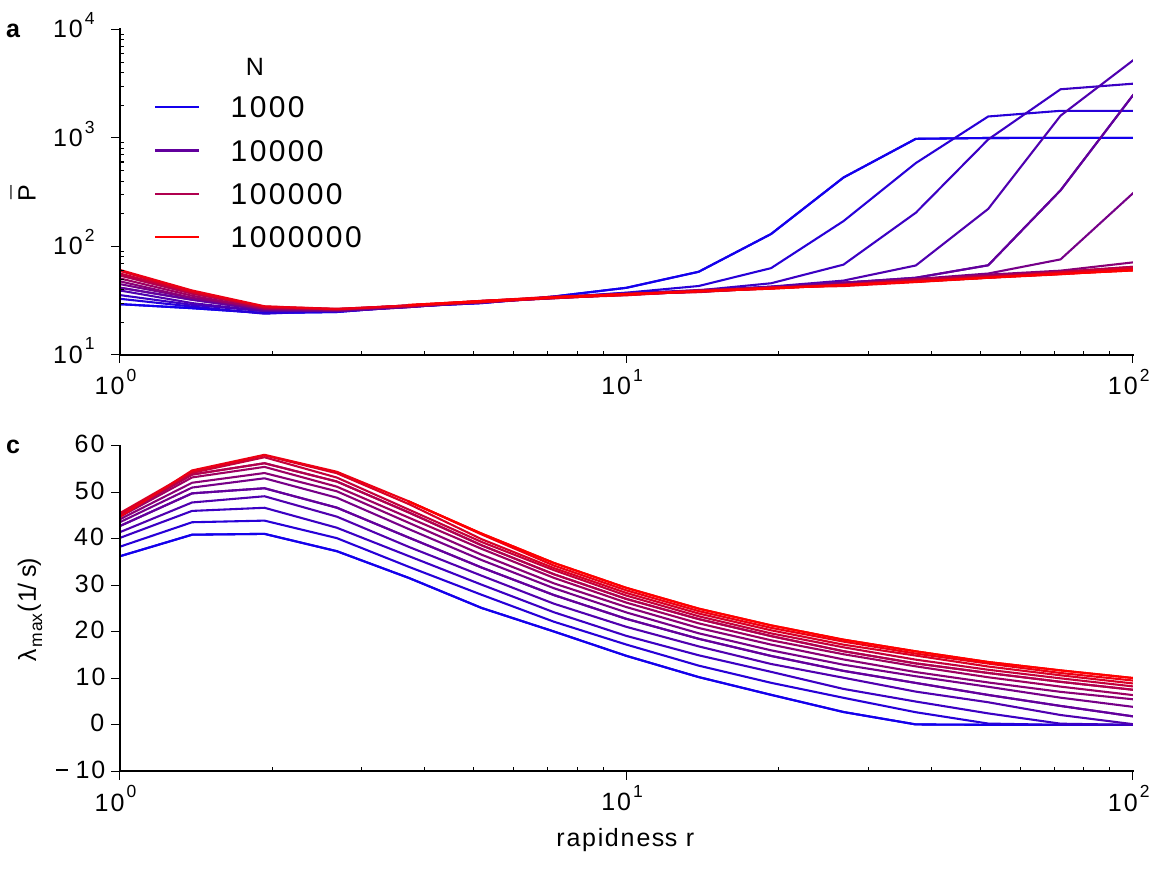}}\includegraphics[clip,width=0.5\columnwidth]{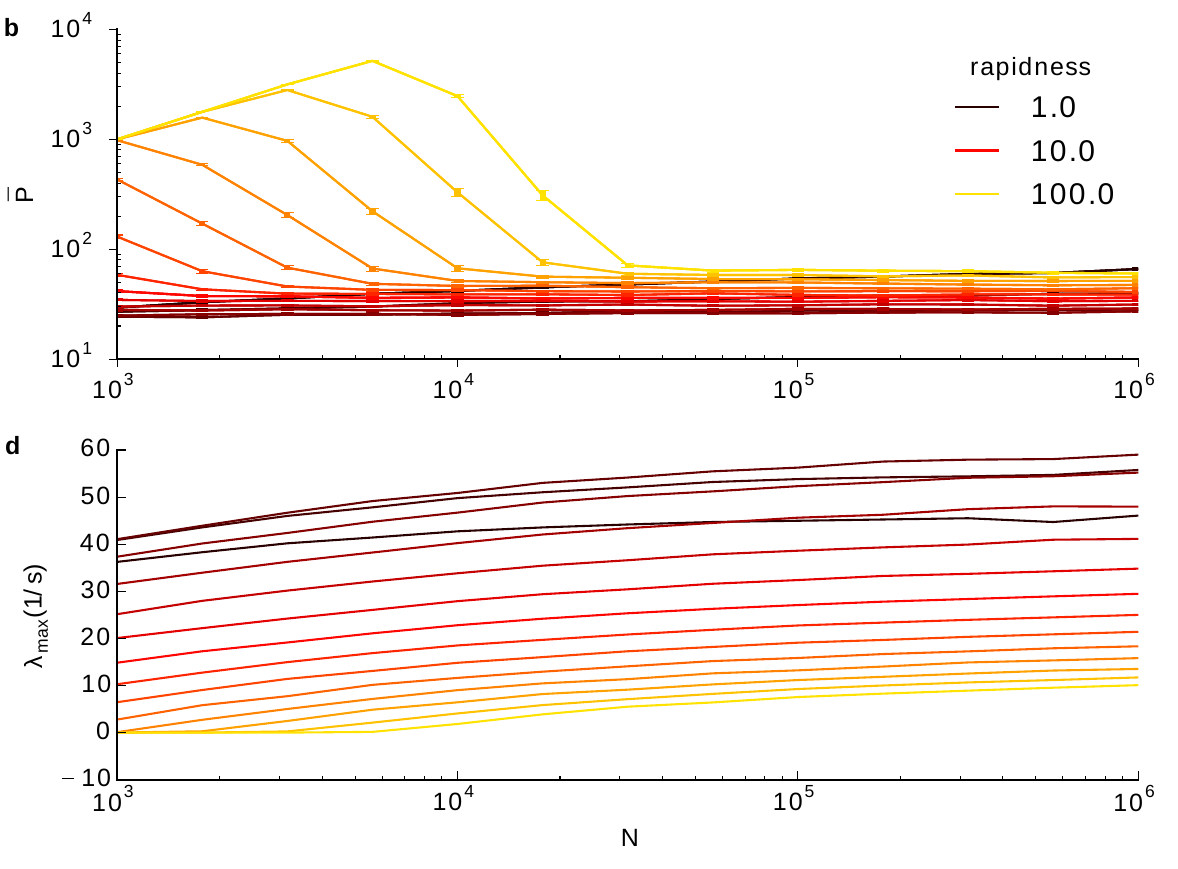}

\caption[Comparison of mean Lyapunov exponent in random matrix approximations
and from numerical simulations in balanced rapid theta neuron networks.]{\textbf{}\label{fig:PartiRatio}\textbf{Average participation vs
rapidness $r$ and network size $N$ reveals delocalization of first
covariant Lyapunov vector at critical rapidness: }Same as \textbf{Fig.}~\textbf{\ref{fig:ParticipationRatio}}
for $\bar{\nu}=1\,\mathrm{Hz}$, $J_{0}=3$. Note that for large spike
onset rapidness $r$ the mean participation $\bar{P}$ increases in
small networks and saturates at $\bar{P}=N$, when the largest Lyapunov
exponent becomes zero (parameters: $\bar{\nu}=1\,\mathrm{Hz}$, $J_{0}=3$,
$\tau_{\mathrm{m}}=10\,\mathrm{ms},$ $K=100$).}
\end{figure}

For large values of spike onset rapidness $r$ at a fixed network
size $N$, we found that the participation ratio increases with rapidness
until it saturates at $\bar{P}=N$ (\textbf{Fig.~\ref{fig:PartiRatio}a}).
This saturation occurred exactly at the critical spike onset rapidness
$r_{\mathrm{crit}}$, when the largest Lyapunov exponent becomes zero
(\textbf{Fig.~\ref{fig:PartiRatio}c}).

\section{Cellular dynamics propel network chaos at instability events}

\begin{figure}
\textbf{\includegraphics[clip,width=0.5\columnwidth]{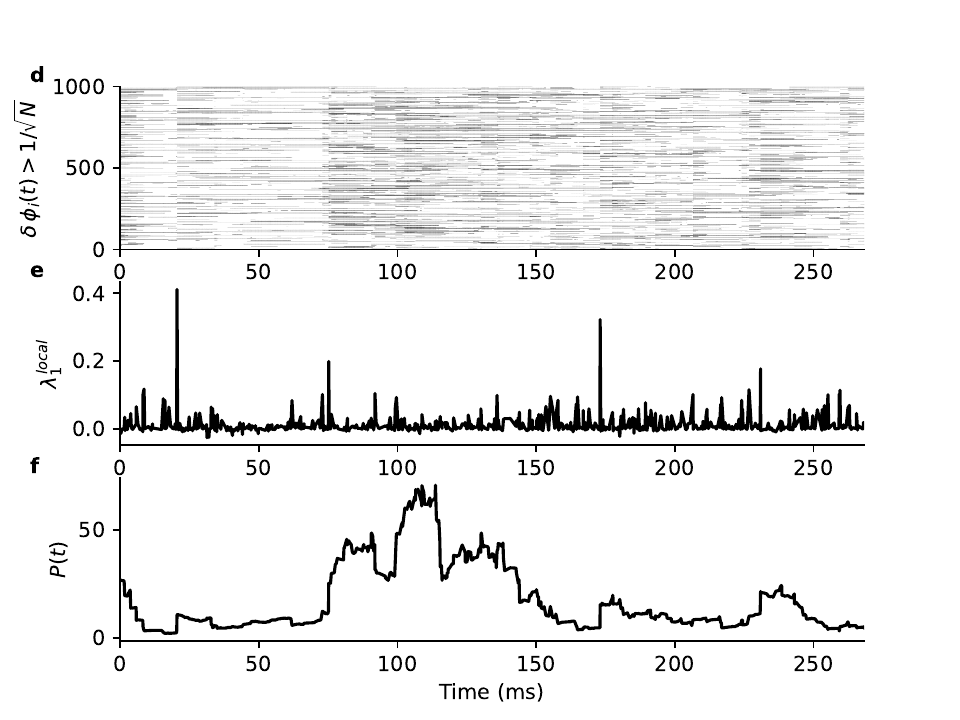}\includegraphics[clip,width=0.5\columnwidth]{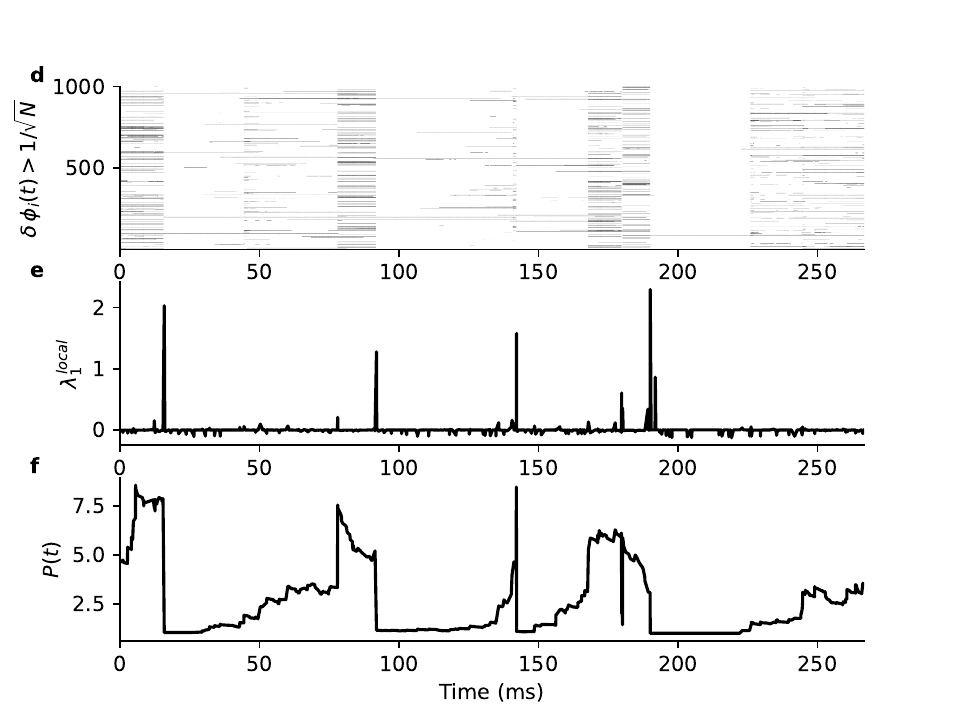}}

\caption[Cellular dynamics propel network chaos at instability events]{\textbf{}\label{fig:Timeseries_lvec_lle_P}\textbf{Cellular dynamics
propel network chaos at instability events: a)} First covariant Lyapunov
vector $\delta\vec{\phi}(t)$ (black whenever $|\delta\vec{\phi}(t)|>1/\sqrt{N}$).
\textbf{b)} First local Lyapunov exponent $\lambda_{1}^{\mathrm{local}}(t),$\textbf{
c)} participation ratio $P(t)$ of first covariant Lyapunov vector
for $r\approx1.33$. \textbf{d), e), f) }same as \textbf{a, b, c}
for $r\approx31.6$. Note that the largest Lyapunov exponent is approximately
the same for the left and right column, but the network is in the
two different regimes described in section \ref{sec:Participation-ratio-and}.
Parameters as in \textbf{Fig.}~\textbf{\ref{fig:ParticipationRatio}
with} $N=1000$).}
\end{figure}

\begin{figure}
\textbf{\includegraphics[clip,width=0.5\columnwidth]{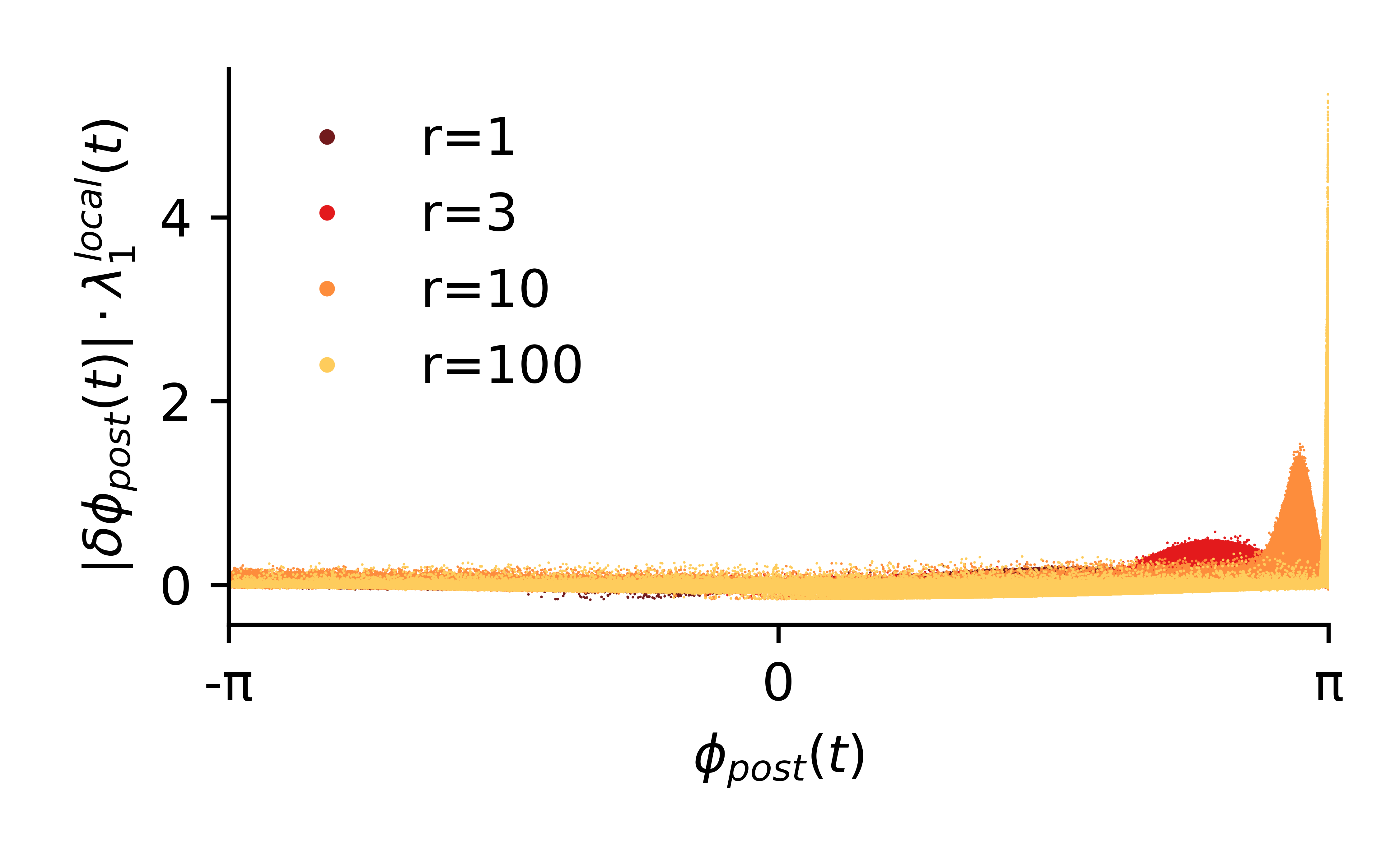}}

\caption[Network chaos mostly sustained by postsynaptic neurons close to spiking]{\textbf{}\label{fig:TheFigure-phi_post_lle}\textbf{Network chaos
is mostly sustained by postsynaptic neurons close to spiking: }Local
Lyapunov exponent projected on entries of first Lyapunov vector of
postsynaptic neurons vs. phase of postsynaptic neurons for different
rapidness $r$. Large local Lyapunov exponents are associated with
postsynaptic neurons with $\phi_{\mathrm{post}}>\phi_{\mathrm{G}}$.
Larger spike onset rapidness $r$ leads to shorter but more unstable
episodes of postsynaptic neurons in susceptible regime. The envelop
of this figure is captured by the derivative of the phase-response
curve. Parameters as in \textbf{Fig.}~\textbf{\ref{fig:ParticipationRatio}
with} $N=1000$ and $T_{\mathrm{total}}=600$~s). }
\end{figure}

To scrutinize how the collective network chaos depends on the single
neuron dynamics, we performed a time-resolved analysis of the first
covariant Lyapunov vector and the state of postsynaptic neurons.
The first covariant Lyapunov vector is ruled by a small and temporally
varying fraction of neurons (\textbf{Fig.~\ref{fig:Timeseries_lvec_lle_P}a}).
 We found different behavior of the local Lyapunov exponent in the
localized regime: There were infrequent \emph{instability events}
with large local Lyapunov exponents, which coindicided with a sudden
decrease of the participation ratio $P(t)$. When these\emph{ instability
events} occured, a postsynaptic neuron was in the instable regime
of its internal dynamics \textbf{Fig.~\ref{fig:TheFigure-phi_post_lle}}.
Larger spike onset rapidness $r$ leads to shorter but more unstable
episodes of postsynaptic neurons in the susceptible regime. The envelop
of the local Lyapunov vector weighted by the entries of the first
Lyapunov vector as function of the postsynaptic phase is captured
by the derivative of the phase-response curve. This establishes a
direct link between the collective network chaos and the single neuron
dynamics.

\section{Pairwise correlations in rapid theta networks}

\paragraph*{Figure 3 e-g of main paper. }

We measured spiking correlations using zero-lag pairwise Pearson spike
count correlations. First we obtained spike counts $n(t)$ by binning
the spike train of each neuron $i$ into bins of window size $T_{\mathrm{win}}$.
The pairwise Pearson spike count correlation between spike trains
of neuron $i$ and $j$ was then calculated using the standard expression~\citep{key-Renart}:

\begin{equation}
r_{ij}=\frac{\mathrm{cov}(n_{i},n_{j})}{\sigma_{i}\sigma_{j}},\label{eq:rapid-pearson}
\end{equation}
where $\mathrm{cov}(n_{i},n_{j})$ is the covariance between the spike
counts of cells $i$ and j and $\sigma_{i}$,$\sigma_{j}$ are the
respective standard deviations. By definition, $r_{ij}\in[-1,1]$.
For figure 3 of the main paper, we chose a window size of $T_{\mathrm{win}}=20$~ms. 

To estimate the effect of our limited sampling time, we generated
jittered spike trains, where the total number of spikes per neuron
was fixed but the spike times drawn uniformly in the interval $[0,\,T_{\mathrm{total}}]$,
where $T_{\mathrm{total}}$ is the total simulation time. We calculated
the correlation of the shuffled spike trains $r_{ij}^{\mathrm{shuffled}}$
using the same binning and definition for the correlation. 

To provide a fair comparison of pairwise correlations for networks
with different spike onset rapidness $r$, the mean firing rate $\bar{\nu}$
was fixed by adapting the external current $I_{\mathrm{ext}}$.

\paragraph{Additional results: }

\begin{figure*}
\includegraphics[clip,width=1\columnwidth]{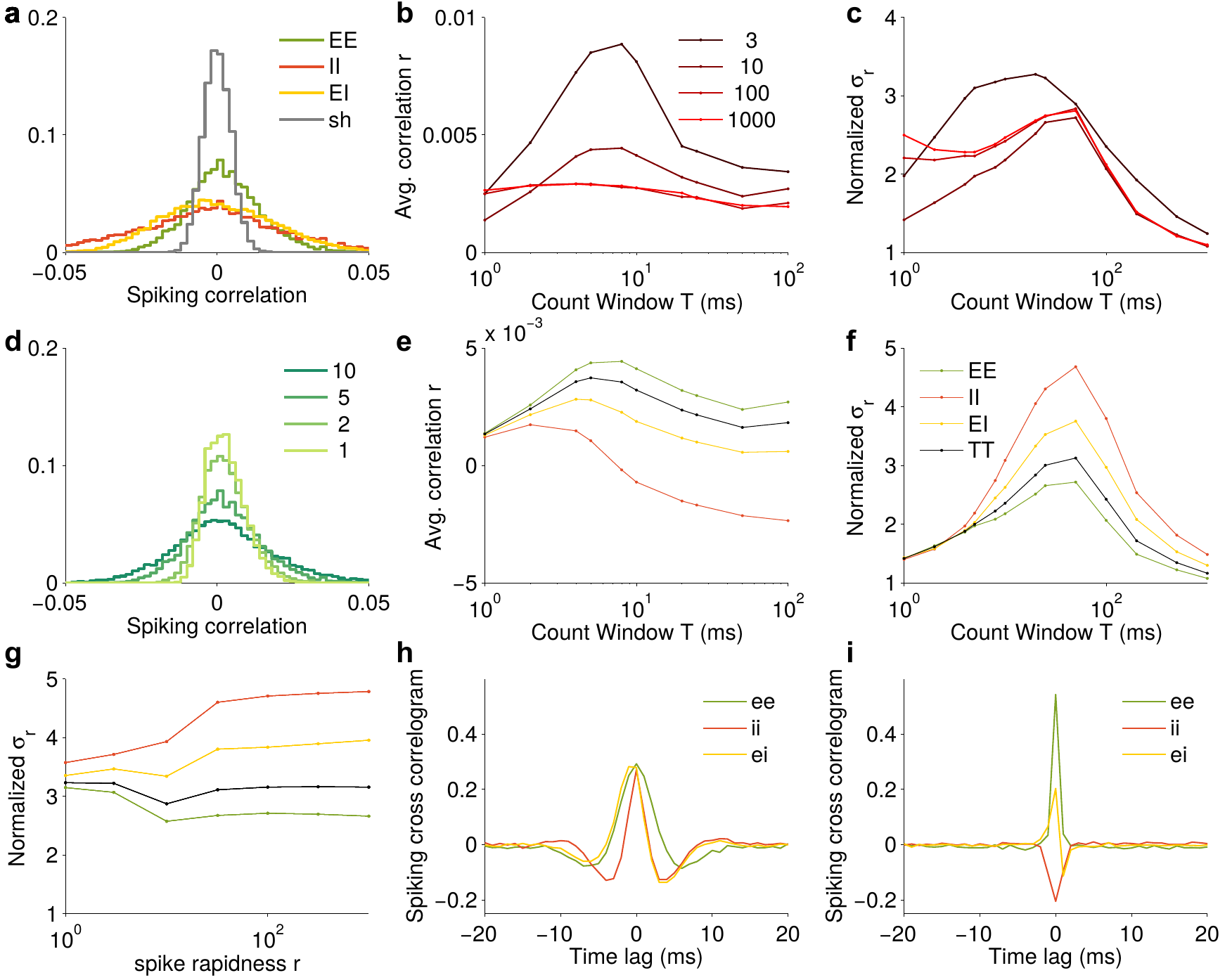} 

\caption{\label{fig:correlations}\textbf{Pairwise correlations in balanced
spiking networks with different values of spike onset rapidness:}
\textbf{a)} Histograms of the pairwise count spike correlation $r_{ij}$
for different cell pairs from excitatory and inhibitory populations
(count window $T_{\mathrm{win}}=20\,$ms. Jittered spike trains were
generated by Poisson process of same rate. \textbf{b)} Mean of pairwise
spike count correlation vs. count window sizes $T_{\mathrm{win}}$
for different values of rapidness. \textbf{c)} Standard deviation
of spike count correlations vs. $T_{\mathrm{win}}$ for different
values of rapidness. \textbf{d)} Histogram of pairwise correlations
for different mean firing rates. \textbf{e)} Standard deviation of
pairwise spiking correlation as a function of rapidness for $T_{\mathrm{win}}=20\,$ms.
\textbf{f)} Average spike cross correlogram of different pair types
for rapidness $r=10$ \textbf{g)} same as \textbf{f} for $r=1000$.
\textbf{h)} Spiking cross correlogram of different pair types for
rapidness $r=10$. \textbf{i}) same as \textbf{h} for $r=1000$ (parameters
as in \textbf{Fig.}~\textbf{3} of main paper).}
\end{figure*}

We characterized the effect of different count window sizes $T_{\mathrm{win}}$
on the mean and standard deviation of the pairwise spike count correlation
(\textbf{Fig.~\ref{fig:correlations}}).

At large count window size, the mean pairwise correlations tend to
be smaller for large rapidness. This can be explained by the faster
inhibitory feedback for large action potential onset rapidness. As
for large rapidness, rapid theta neurons are capable of tracking input
changes more quickly (\textbf{Figure 1d,~e} of main paper). Therefore,
the dynamic decorrelation of balanced networks, which was described
earlier \citep{key-Renart}, is more effective. For small bin size,
the mean pairwise correlations of rapid theta neurons with large rapidness
are larger. This is because rapid theta neurons which receive shared
excitatory input and are kicked across the unstable fixed point will
spike almost instantaneously, which results in moderately increased
correlations on very short time scales (\textbf{Fig.}~\textbf{\ref{fig:correlations}b}).

\paragraph{Cross-correlograms: }

The cross-correlogram between the pairs of the binned the spike trains
$n_{i}(t)$ and $n_{j}(t)$ was calculated as:

\begin{equation}
c_{ij}(\tau)=\frac{\mathrm{\mathrm{cov}}(n_{i}(t),n_{j}(t+\tau))}{\bar{\nu}_{i}\bar{\nu}_{j}}-1.\label{eq:rapid-corr}
\end{equation}
where $\mathrm{cov}(n_{i},n_{j})$ is the covariance between the spike
counts of cells $i$ and j and $\bar{\nu}_{i}$,~$\bar{\nu}_{j}$
are the respective mean firing rates. By definition $c_{ij}\in[-\infty,\infty],$
but $\underset{t\rightarrow\infty}{\lim}c_{ij}=0$ for shuffled spikes.

\section{Attractor dimension and \textquotedblleft entangled\textquotedblright{}
statistics}

In the main paper, we show both an upper and lower bound for the attractor
dimension which both strongly depend on the spike onset rapidness
(\textbf{Fig.~3}). For increasing spike rapidness, the network dynamics
has a transition from chaotic to stable dynamics. While chaotic dynamics
is accompanied in continuous dynamical systems by a fractal attracting
set with dimension $\mathrm{D>2}$, stable dynamics has an attractor
dimension $\mathrm{D=1}.$ Here, we compare this to Gaussian estimates
of the dimensionality, based on pairwise correlations, to evaluate
higher order correlations. Assuming Gaussian statistics, all dependencies
between neurons would be captured by the pairwise correlations. Based
on the correlation matrix, an estimate of the dimensionality can be
obtained. We use two common estimators of the dimensionality: the
number of principle components needed explain 95\% of the correlation
matrix\textquoteright s variance and a dimensionality based on the
participation ratio of the correlation matrix that measures the effective
number of degrees of freedom over which power is distributed.

\begin{figure*}
\includegraphics[clip,width=1\columnwidth]{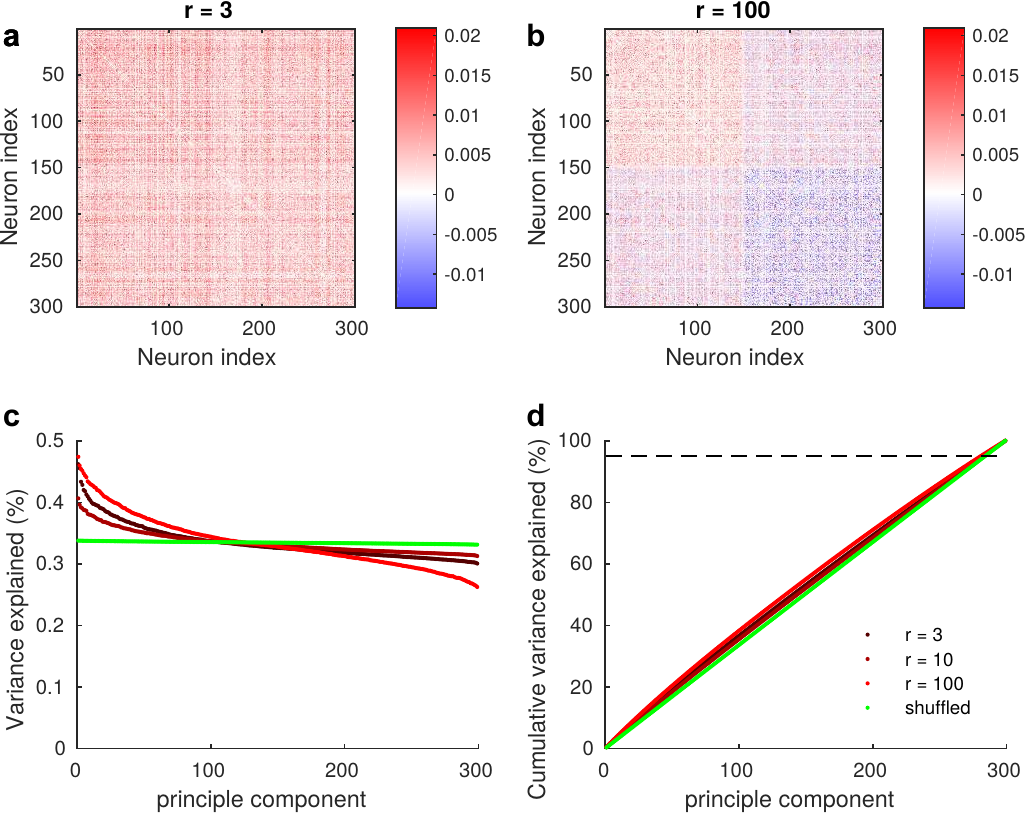} 

\caption{\label{fig:correlation-dimensionality}\textbf{Weakly structured correlation
matrix in balanced spiking networks with different values of spike
onset rapidness:} \textbf{a)} Matrix of pairwise spike count correlations
$r_{ij}$ for rapidness $r=3$. First 150 neurons are excitatory,
others inhibitory. \textbf{b)} same for $r=100$. \textbf{c)} Variance
explained per principal component for different rapidness. Jittered
spike trains were generated by a Poisson process of the same rate.
\textbf{d)} Cumulative variance explained for different values of
rapidness, (parameters: $T_{\mathrm{win}}=20$~ms, $T_{\mathrm{total}}=1000$~s,
other parameters as in \textbf{Fig.~\ref{fig:correlations}}).}
\end{figure*}

\textbf{Figure \ref{fig:correlation-dimensionality}a+b} shows the
correlation matrix for two different values of spike onset rapidness.
Principal component analysis (PCA) yields the percentage of the total
variance explained by each principal component (\textbf{Fig.~\ref{fig:correlation-dimensionality}c}).
The number of principal components necessary to account for 95\% of
the total variance gives an estimate of the number of degrees of freedom
of the underlying dynamics. If few principal components would explain
most of the variance, most of the dynamics is constrained to a hyperellipsoid
with few large axes. If many principal components are necessary to
explain most of the variance, no such collective structures are detected.
This excludes the possibility that the dynamics is explained solely
by pairwise correlations. A different estimate of the dimensionality
based on the correlation matrix is the inverse participation ratio
of the eigenvalue spectrum of the correlation matrix. The inverse
participation ratio is defined as the normalized inverse squared sum
of eigenvalues of the correlation matrix:

\begin{equation}
D_{\mathrm{PR}}=\frac{(\sum\lambda_{i})^{2}}{\sum(\lambda_{i}^{2})}\label{eq:participation-ratio}
\end{equation}
where $\mathrm{\lambda_{i}}$ is the $i^{\textrm{th}}$ eigenvalue
of the correlation matrix. Thus, $D_{\mathrm{PR}}$ is 1 if one eigenvalue
is dominating while the others are zero. If all eigenvalues contribute
equally $\mathrm{\lambda_{i}}=\frac{1}{N}$, the dimension is $N$
\citep{key-Monteforte2010,key-Surya,key-Engelken2023,key-Clark2023}. The dimensionality estimate
based on the participation ratio also shows that the pairwise correlations
have very little localized structure independent of spike onset rapidness
(\textbf{Fig.~\ref{fig:correlations-dimensionality-varyTtotal-varyR}b+d}).
Furthermore, we show that this result is largely insensitive to the
spike count window of the correlation matrix (\textbf{Fig.~\ref{fig:correlations-dimensionality-varyTtotal-varyR}a+c}).
To conclude, we find a low attractor dimensionality based on the Kaplan-Yorke
dimension and our lower bound estimate coming from the number of positive
Lyapunov exponents, despite low and weakly structured pairwise correlations.
To obtain a precise estimate of the pairwise spike count correlations,
we averaged the correlation matrix over 100 runs with different initial
conditions but identical network topology each with $T_{\mathrm{total}}=1000$~s.

\begin{figure*}
\includegraphics[clip,width=1\columnwidth]{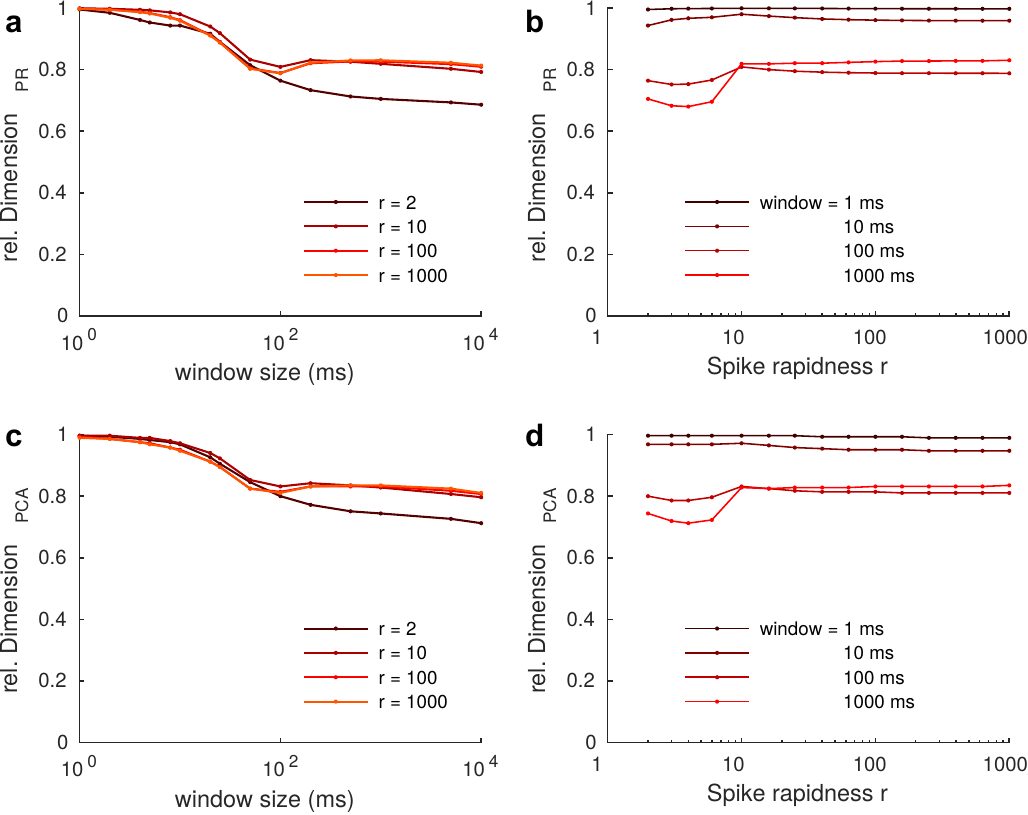} 

\caption{\label{fig:correlations-dimensionality-varyTtotal-varyR}\textbf{Correlation
based dimensionality estimates of dynamics based on pairwise correlations
indicates weakly structured correlations:} \textbf{a)} relative dimension
based on participation ratio (Eq.~\ref{eq:participation-ratio})
vs. window size $T_{\mathrm{win}}$. \textbf{b)} same as \textbf{a}
as a function of spike rapidness $r$. \textbf{c)+d)} same as \textbf{a+b}
for relative dimension based on PCA. (parameters as in \textbf{Fig.~3})}
\end{figure*}

\begin{figure*}
	\includegraphics[clip,width=0.49\columnwidth]{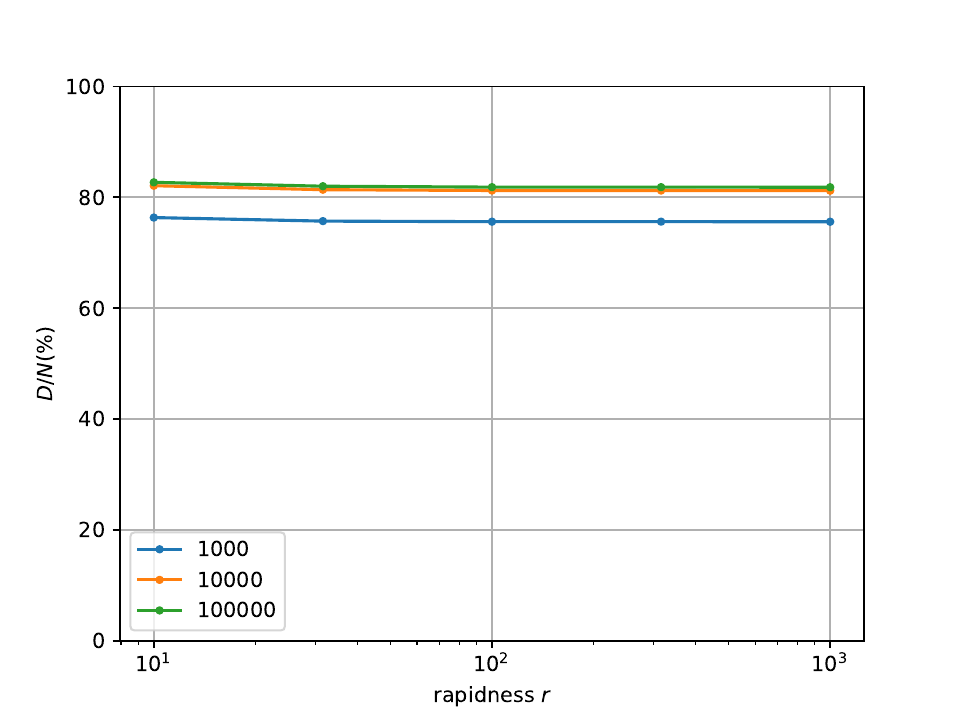} 
	\includegraphics[clip,width=0.49\columnwidth]{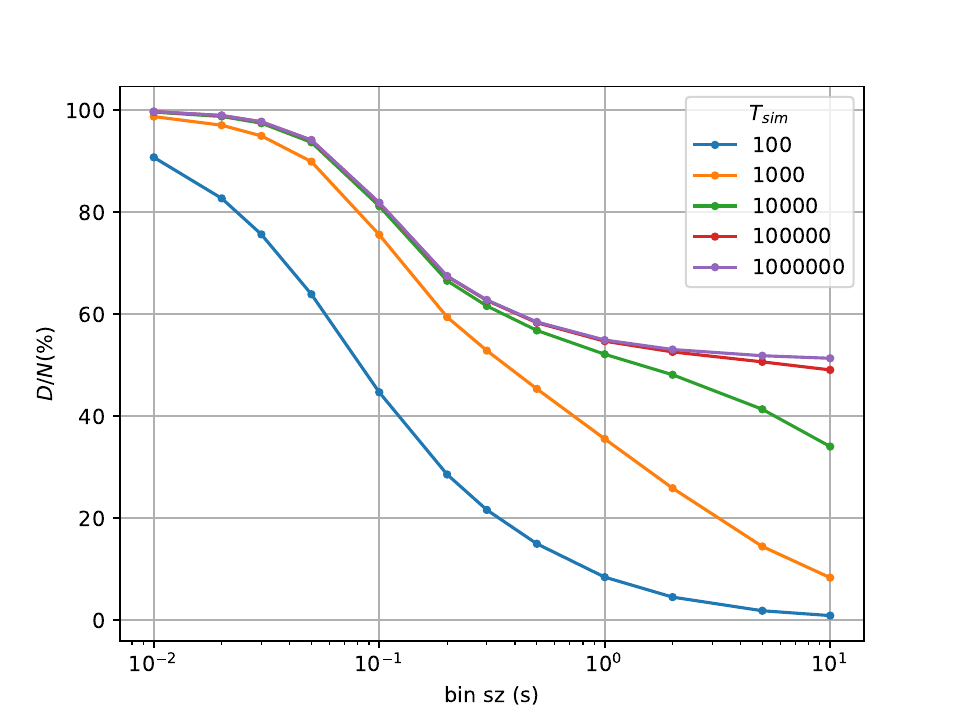}
	\includegraphics[clip,width=0.49\columnwidth]{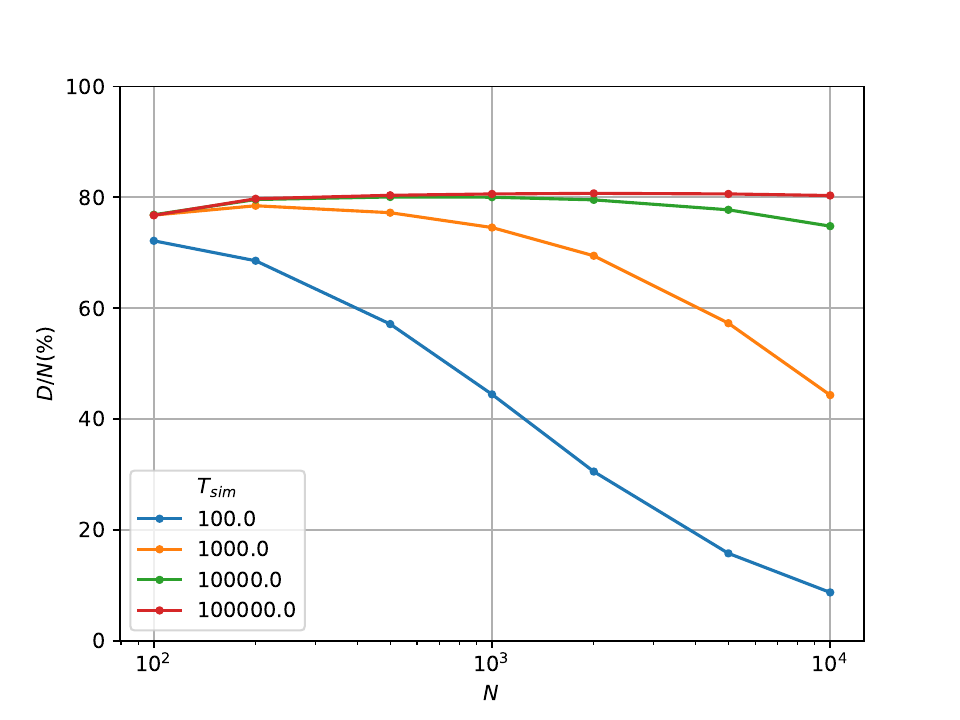}	
	\includegraphics[clip,width=0.49\columnwidth]{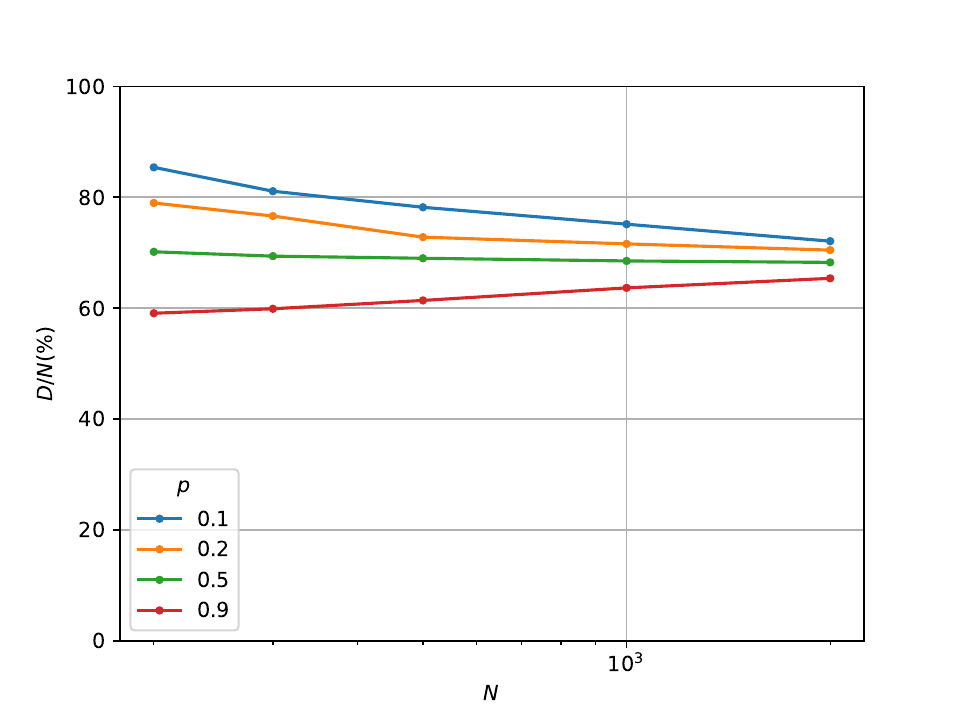}		
	
	\caption{\label{fig:correlations-dimensionality-varyTtotal-varyR}\textbf{Correlation
			based dimensionality estimates of dynamics based on pairwise correlations
			indicates weakly structured correlations (Fixed indegree):} \textbf{Top right)} relative dimension
		based on participation ratio (Eq.~\ref{eq:participation-ratio})
		rapidness for fixed synaptic indegree $K$. PCA-dimension (participation ratio) is largely independent of rapidness for inhibitory balanced rapid theta networks. Colors are different simulation time $T_{sim}$.\textbf{top right)} same as \textbf{a}
		as a function of window bin size (in s) for fixed indegree. Colors are different simulation time $T_{sim}$. \textbf{bottom left)} PCA-dimension (participation ratio)
		for varying network size $N$ with fixed synaptic indegree $K$. Colors are different simulation time $T_{sim}$.  \textbf{bottom right)} PCA-dimension (participation ratio)
		for varying network size $N$ with fixed connection probability $p=K/N$. Colors are different simulation time $p$, simulation time is $T=10^5s$ and $r=1000$. (Other parameters if not mentioned: $K=30$, $\nu=1Hz$, $J_0=1$, $r=1000$, bin $size=100ms$)}
\end{figure*}

\section{Network synchrony}
\begin{figure*}
	\includegraphics[clip,width=1\columnwidth]{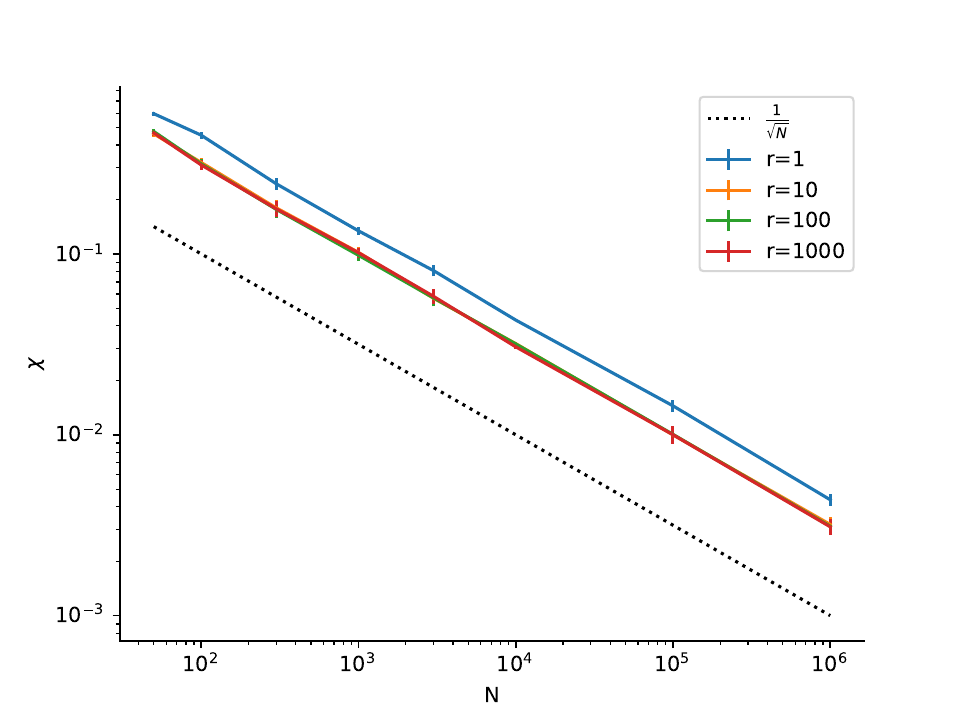} 
	
	\caption{{\bf {Asynchronous network dynamics in balanced rapid theta networks}}\newline
		Network synchrony measure $\chi$ as a function of number of neurons $N$ for different values of rapidness $r$.  We find asynchronous network dynamics indicated by $\chi\propto \frac{1}{\sqrt{N}}$ for various values of $r$.
		\vspace{-0.5cm}
	}
	\label{fig:Chi}
\end{figure*}

We quantify network synchrony network by using the network synchrony measure \cite{key-Hansel2003}: $\chi^2=\frac{\sigma_{\phi(t)}^2}{\sum_{i=1}^N \sigma^2_{\phi_i(t)}/N}$, where $\phi(t)=\sum_{i=1}^N \phi_i(t)/N$ is the mean activity in a phase representation. 
We find that networks are in an asynchronous regime with $\chi\propto \frac{1}{\sqrt{N}}$ for various values of $r$ across 4 order of magnitude (see \textbf{Fig.}~\textbf{\ref{fig:Chi}}).
We note that balanced networks of theta neurons ($r=1$) have well-known pathological network oscillations \cite{key-Monteforte2011} for large $K$ and strong coupling $J_0$/low firing rate that were recently described in a mean-field approach  \cite{key-Goldobin2024}. We observed in rapid theta (and also EIF) networks, that increasing rapidness $r$ causes oscillations to vanish -or requires a larger $K$ for oscillations to emerge \cite{key-Monteforte2011}. A more detailed account will be given elsewhere.

\section{Flux-tube structure of phase space, stable chaos and single-spike
perturbations}

\paragraph{Figure 3 d,h of main paper: }

For sufficiently large spike onset rapidness, we find that infinitesimal
perturbations decay exponentially but sufficiently strong perturbation
lead to an exponential decorrelation of neighboring trajectories.
This exotic phase space structure was first described in balanced
purely inhibitory networks of pulse-coupled leaky integrate and fire
neurons earlier and termed flux tubes \citep{key-Monteforte2012}.
Following Ref.~\citep{key-Monteforte2012}, we find the critical
perturbation strength $\varepsilon_{\mathrm{ft}}$ that is sufficient
for an exponential decorrelation of trajectories by fitting the probability
that a perturbation of strength $\varepsilon$ causes an exponential
state separation to the function

\begin{equation}
P_{\mathrm{s}}(\varepsilon)=1-\exp(-\varepsilon/\varepsilon_{\mathrm{ft}})\label{eq:def_epsilonFT}
\end{equation}
$\varepsilon_{\mathrm{ft}}$ is the average radius of a basin of attraction,
called flux tube radius. We found that this flux tube radius strongly
depends on the spike onset rapidness $r$. For values of $r$ only
slightly larger than the critical rapidness $r_{\mathrm{crit}}$,
the flux tube radius is small, but large $r$ yield larger basins
of attraction. This is depicted in \textbf{Fig.~\ref{fig:correlations}}.

\begin{figure*}
\includegraphics[clip,width=1\columnwidth]{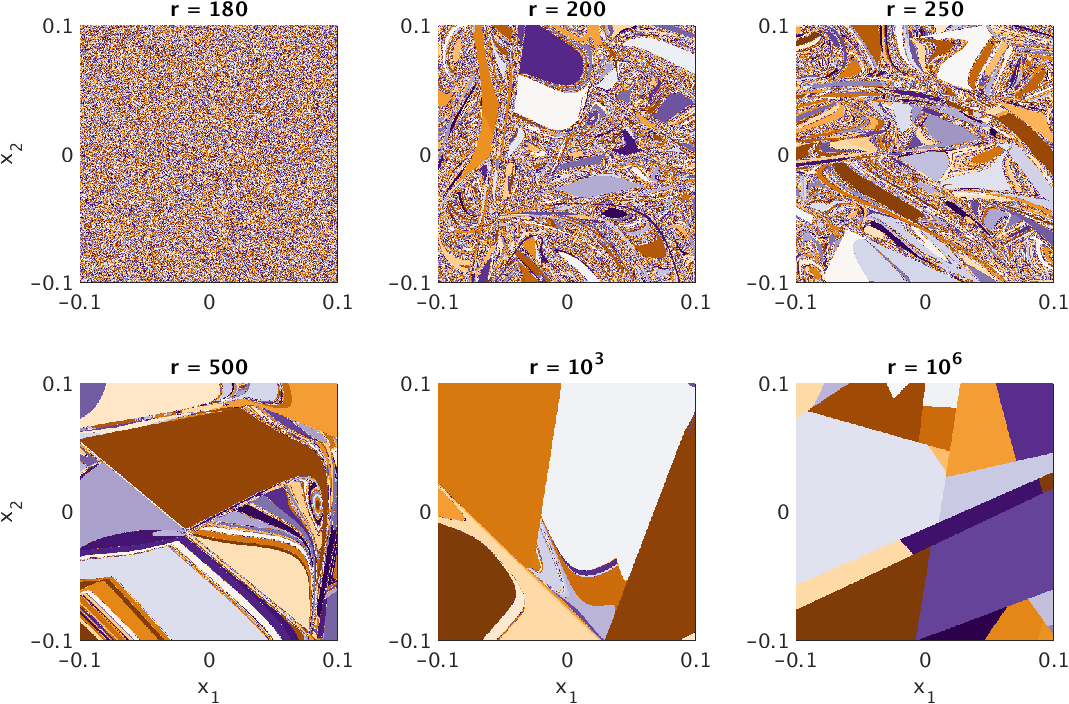} 

\caption{\label{fig:fluxtube-cross-sections}\textbf{Random cross section through
N-dimensional phase space for different values of spike onset rapidness
}$r$\textbf{:} Phase space cross sections spanned by two random N-dimensional
vectors $x_{1}$and $x_{2}$ orthogonal to the trajectory $\vec{1}$
and to each other. Initial conditions that converge to the same trajectory
are drawn in the same color (parameters: $\bar{\nu}=10\,\mathrm{Hz}$,
$J_{0}=1$, $\tau_{\mathrm{m}}=10\,\mathrm{ms},$ $N=200$, $K=100$).}
\end{figure*}

We perturbed initial conditions along two random N-dimensional vectors
orthogonal to the flow of the dynamics $\vec{1}$. These two vectors
span a two-dimensional cross section of the N-dimensional phase space.
Each pixel is a different initial condition for a simulation. Neighboring
initial conditions that converge to the same trajectory are assigned
same colors. For increasing values of $r$, the flux tubes radius
increases and the boundaries become straighter. In the limit of very
large $r$, the flux tubes are similar to those in the leaky integrate
and fire model (compare to \textbf{Fig.~7} in \citep{key-Monteforte2012}).
 Overall, flux tube radii get smaller with increasing network size
$N$, number of synapses per neuron $K$, mean firing rate $\bar{\nu}$
and decreasing spike rapidness $r$. At the critical rapidness $r_{\mathrm{crit}}$
the flux tubes vanish.

\section{Poincaré maps of chaotic networks}

A Poincaré map is the intersection of the trajectory of a $N$ degree
of freedom dynamical system with a $N-1$ dimensional subspace called
the Poincaré surface or section.

\begin{figure*}
\includegraphics[clip,width=1\columnwidth]{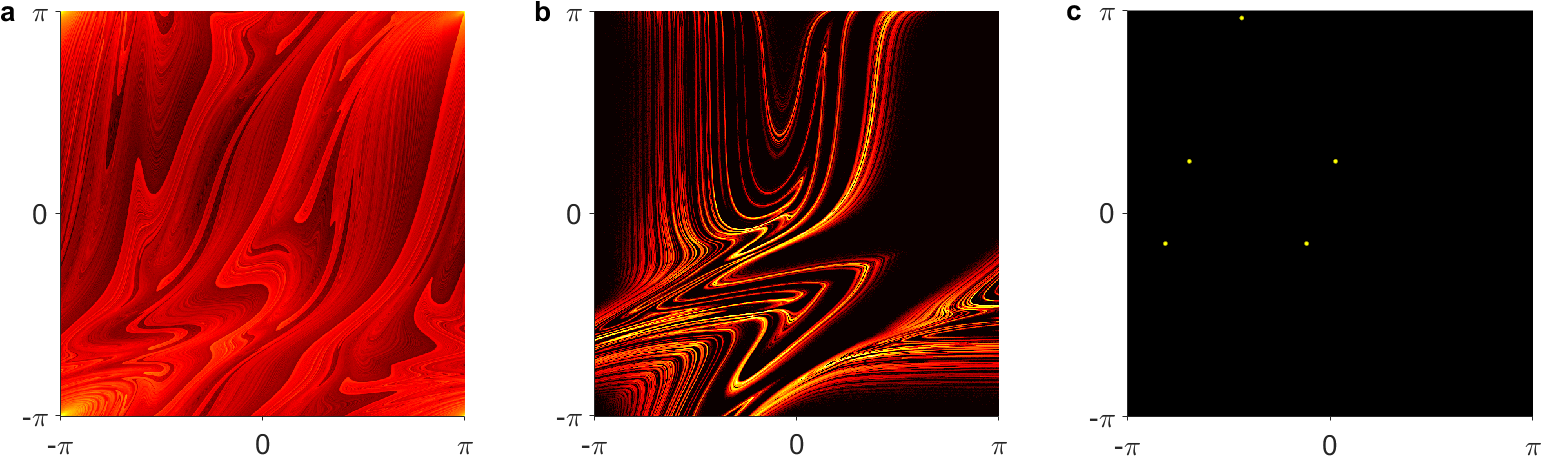} 

\caption{\label{fig:fluxtube-poincare-sections-N3}\textbf{Poincaré sections
through phase space reveal reorganization of chaotic strange attractor
by AP onset rapidness $r$ in small networks }$N=3$\textbf{: a)}
Poincare section of the phases of neuron 2 and 3 whenever neuron 1
spikes for low rapidness ($r=1$). The relative density of points
is represented using a heat map, where hot colors indicate high densities.
\textbf{b)} Same as \textbf{a} for $r=4$. \textbf{c)} Same as \textbf{a}
for $r=25$ (parameters: $\bar{\nu}=14.5\,\mathrm{Hz}$, $J_{0}=1$,
$\tau_{\mathrm{m}}=10\,\mathrm{ms},$ $N=3$, $K=1$, $r=1,\,4\,,25$).}
\end{figure*}

Increasing the spike rapidness leads to a thinning of stable phase
space regions (\textbf{Fig.~\ref{fig:fluxtube-poincare-sections-N3}}).
Beyond the critical spike rapidness, the network settles after a transient
period into a periodic orbit. Therefore, there is only a finite number
of unique points in the Poincare section. For increasing network size,
the Poincaré sections did not capture the exotic structure of the
phase space. This stresses again that the attractor is a high-dimensional
object, which is hard to visualize in two dimensions.

\begin{figure*}
\includegraphics[clip,width=1\columnwidth]{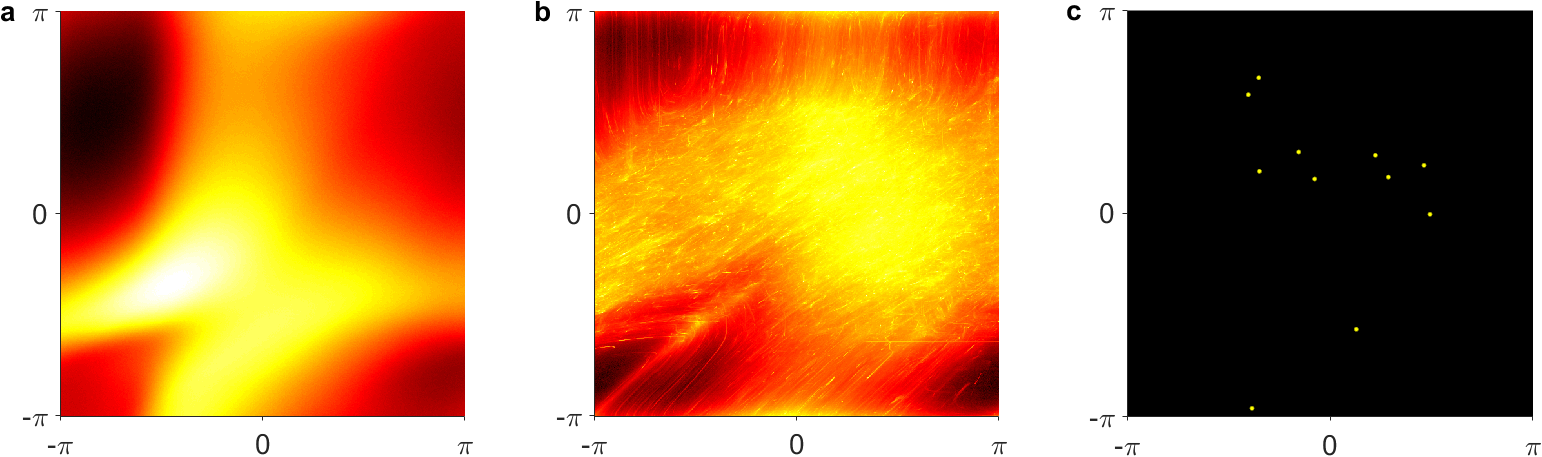} 

\caption{\label{fig:fluxtube-poincare-sections-N20}\textbf{Two-dimensional
sections through phase space for different values of spike onset rapidness
}$r$\textbf{ for }$N=20$\textbf{:} Surface of the phases of neuron
2 and 3 whenever neuron 1 spikes for low rapidness ($r=1$). The relative
density of points is represented using a heat map, where hot colors
indicate high densities. \textbf{b)} Same as \textbf{a} for $r=10$.
\textbf{c)} Same as \textbf{a} for $r=25$ (parameters: $\bar{\nu}=14.5\,\mathrm{Hz}$,
$J_{0}=1$, $\tau_{\mathrm{m}}=10\,\mathrm{ms},$ $N=3$, $K=1$,
$r=1,\,10\,,25$).}
\end{figure*}
\begin{figure*}
\includegraphics[clip,width=1\columnwidth]{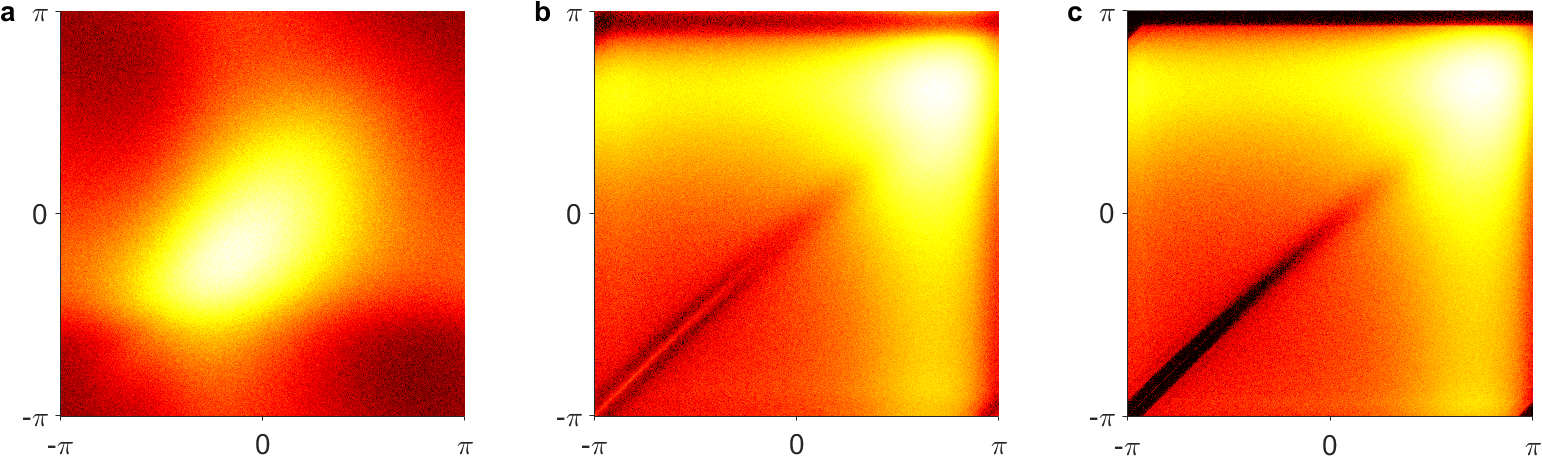} 

\caption{\label{fig:fluxtube-poincare-sections-N200}\textbf{Two-dimensional
sections through phase space for different values of spike onset rapidness
}$r$\textbf{ for }$N=200$\textbf{:} Surface of the phases of neuron
2 and 3 whenever neuron 1 spikes for low rapidness ($r=1$). The relative
density of points is represented using a heat map, where hot colors
indicate high densities. \textbf{b)} Same as \textbf{a} for $r=25$.
\textbf{c)} Same as \textbf{a} for $r=250$ (parameters: $\bar{\nu}=14.5\,\mathrm{Hz}$,
$J_{0}=1$, $\tau_{\mathrm{m}}=10\,\mathrm{ms},$ $N=3$, $K=1$,
$r=1,\,25\,,250$).}
\end{figure*}

\section{Local Lyapunov exponents reveal stable and unstable phase space regions}

\begin{figure*}
\includegraphics[clip,width=1\columnwidth]{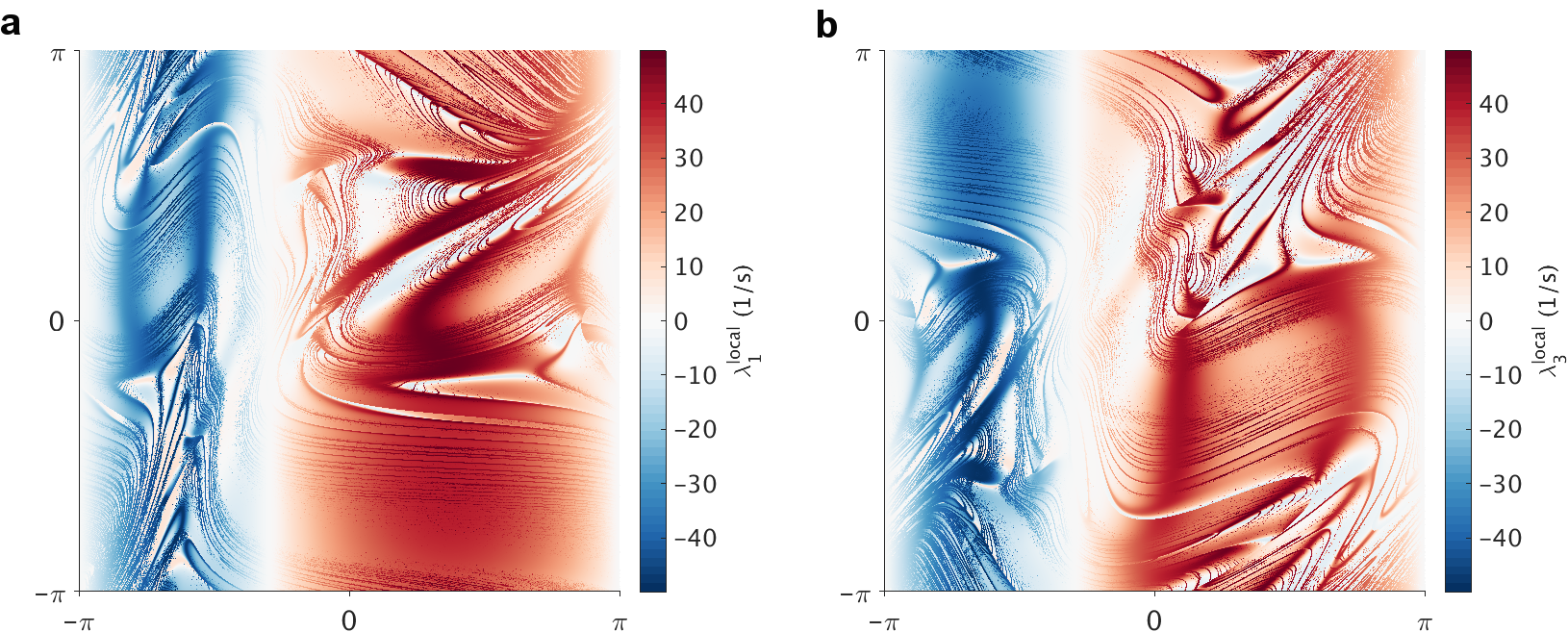} 

\caption{\label{fig:fluxtube-poincare-sections-N3-LLE}\textbf{Local Lyapunov
exponent reveal stable and unstable phase space regions of chaotic
attractor in small networks }$N=3,\,r=1$\textbf{:} \textbf{a)} Poincaré
section of the phases of neuron 2 and 3 whenever neuron 1 spikes.
The first local Lyapunov exponents (LLE) at each point is color-coded,
red colors indicate local instability, blue indicates local stability.
\textbf{b)} Same as \textbf{a} for third LLE. The second LLE is trivially
zero (neutral direction, not shown). The associated density of states
is depicted in \textbf{Fig.~\ref{fig:fluxtube-poincare-sections-N3}a}
(parameters: $\bar{\nu}=14.5\,\mathrm{Hz}$, $J_{0}=1$, $\tau_{\mathrm{m}}=10\,\mathrm{ms},$
$K=1$, $r=1$).}
\end{figure*}

\begin{figure*}
\includegraphics[clip,width=1\columnwidth]{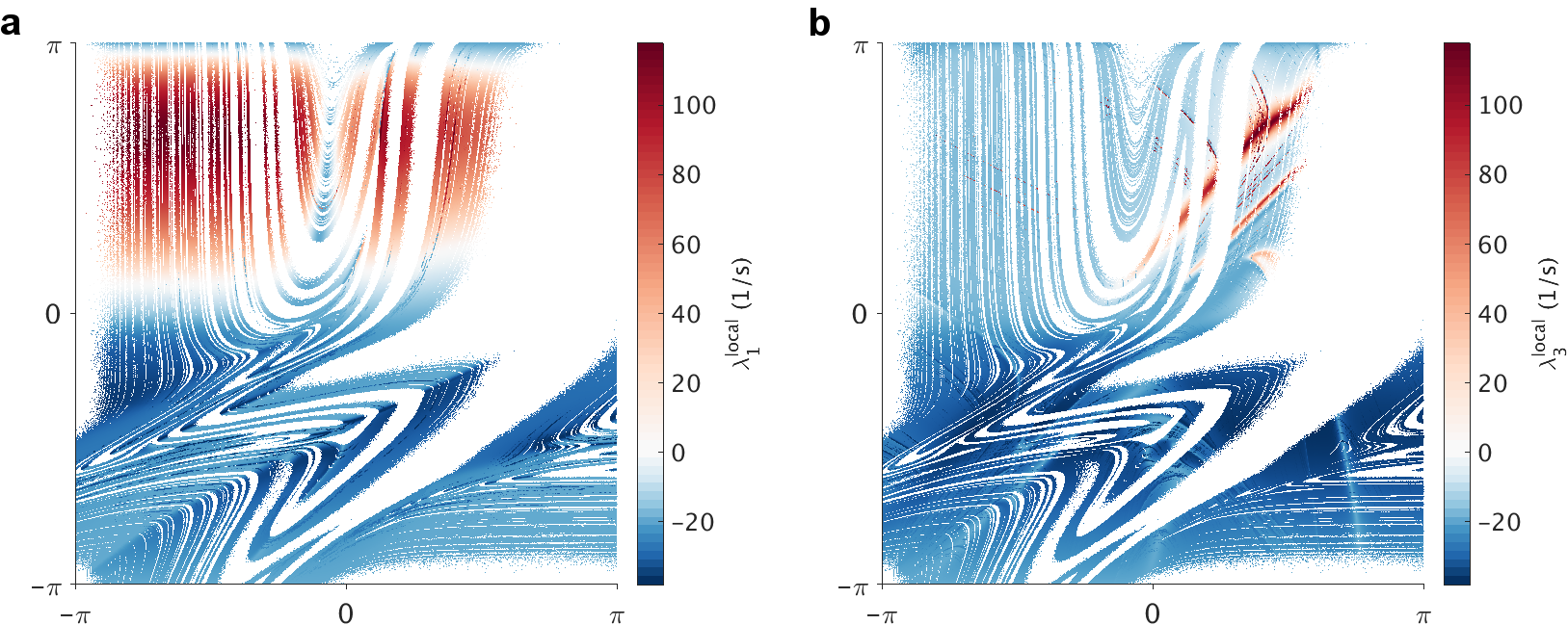} 

\caption{\label{fig:fluxtube-poincare-sections-N3-highRapLLE}\textbf{Local
Lyapunov exponent reveal stable and unstable phase space regions of
chaotic attractor in small networks }$N=3,\,r=4$\textbf{:} Poincaré
section of the phases of neuron 2 and 3 whenever neuron 1 spikes.
The first local Lyapunov exponents (LLE) at each point is color-coded,
red colors indicate local instability, blue indicates local stability.
\textbf{b)} Same as \textbf{a} for third LLE. The second LLE is trivially
zero (neutral direction, not shown). The associated density of states
is depicted in \textbf{Fig.~\ref{fig:fluxtube-poincare-sections-N3}b}
(parameters: $\bar{\nu}=14.5\,\mathrm{Hz}$, $J_{0}=1$, $\tau_{\mathrm{m}}=10\,\mathrm{ms},$
$N=3$, $K=1$, $r=4$).}
\end{figure*}

In the three-dimensional network, the local Lyapunov exponents (LLEs)
can visualize further structural properties of the strange chaotic
attractor. In \textbf{Fig.~\ref{fig:fluxtube-poincare-sections-N3-LLE}a},
colors indicate the first local Lyapunov exponent, when neuron 1 spikes,
plotted at the location of phases of neuron 2 and 3 for rapidness
$r=1$. Red colors indicate a positive LLE, blue colors indicate a
negative LLE. The fine structure of the first LLE is similar to the
density (\textbf{Fig.~\ref{fig:fluxtube-poincare-sections-N3}}),
while the third LLE has a fine structure which dissimilar to the density.
A possible explanation is that the first LLE is smooth along the unstable
manifolds, while the third LLE is smooth along the stable manifolds.
The spatial distribution of the LLEs on the Poincaré surface is determined
by the topology in combination with the single neuron properties.
In the displayed network, neuron 1 is only connected to neuron 2,
therefore the LLEs mainly reflect the derivative of the phase response
curve evaluated at the respective value of neuron 2. This can also
be observed at a higher rapidness (rapidness $r=1$ \textbf{Fig.~\ref{fig:fluxtube-poincare-sections-N3-highRapLLE}}).

\section{Chaos and dynamical entropy rate in structured network topologies}

\paragraph{Figure 4 a-e of main paper: }

\begin{figure*}
\includegraphics[clip,width=1\columnwidth]{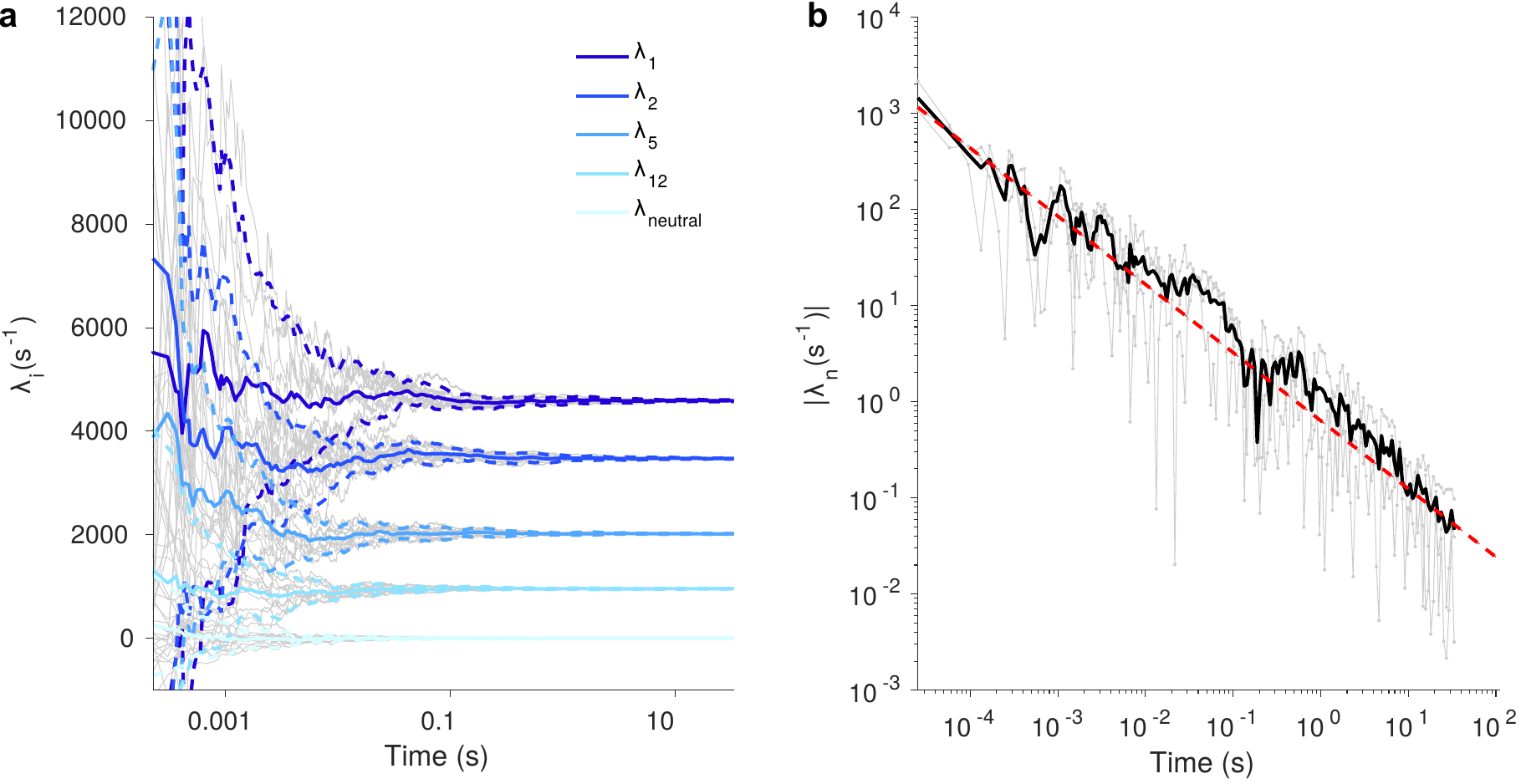} 

\caption{\label{fig:convergence-multilayer-network}\textbf{Lyapunov exponents
for cortical column model convergence over orders of magnitude:} \textbf{a)}
gray lines: some Lyapunov exponents for ten different initial conditions,
solid color lines: averages, dotted color lines: averages $\pm$ double
standard errors. $r=10$, where the largest Lyapunov exponent is maximal.
\textbf{b)} Convergence of the neutral Lyapunov exponent. Grey lines:
absolute value of neutral Lyapunov exponent for three initial conditions,
black line: average of absolute values of neutral Lyapunov exponents
across initial conditions, red line: power law fit using the Levenberg-Marquardt
algorithm with exponent $-0.7$.}
\end{figure*}

To test whether spike onset also exerts a strong influence on the
phase space structure in more realistic network topologies, we used
a previously established multilayered model of a cortical column with
77169 neurons and around 285 million synapses \citep{key-Potjans}.
Numbers of neurons per population and inter-layer wiring probabilities
were taken from a numerical model of a cortical column based on anatomically
measured synapse counts and connection probabilities between different
cortical layers \citep{key-Potjans}. The cortical column model consists
of four layers (layers 2/3, 4, 5 and 6) each with an excitatory and
an inhibitory population. The number of neurons in each layer and
the connection probability between layers are stated in Table 1. According
to these numbers, the connectivity for each population and each projection
between the populations is generated as a directed sparse Erd\H{o}s\textendash Rényi
random graph (Table \ref{tab:currents-1}). We calculated also for
this large network the Lyapunov spectrum making use of a parallelized
implementation of the semi-analytic calculation described in section
\ref{sec:Numerical-convergence}. The sparseness of the connectivity
was utilized for the efficient storage of the coupling matrices, the
updates of the postsynaptic neurons and the matrix multiplications
of the orthonormal system with the sparse single spike Jacobians implemented
in custom code written in Julia and C++ making use of the Automatically
Tuned Linear Algebra Software (ATLAS) for matrix multiplications in
the Gram\textendash Schmidt procedure and the Message Passing Interface
(MPI) for the parallel implementation of the simulations. For the
reorthonormalization, we chose a parallel recursive blocked version
of the Gram\textendash Schmidt procedure \citep{key-Gram-Schmidt}.
Constant external input currents were adapted to obtain a desired
global firing rate, for a fair comparison across different values
of AP onset rapidness $r$. The convergence of the Lyapunov spectrum
across different initial conditions is shown in \textbf{Fig.~\ref{fig:convergence-multilayer-network}}.
The shaded lines correspond to different initial conditions, dashed
lines indicate standard error of the mean. Even for such large networks,
the numerically precise event-based implementation fully converges.
The neutral Lyapunov exponent converges to zero across multiple orders
of magnitude (\textbf{Fig.~\ref{fig:convergence-multilayer-network}b}).
The dynamical entropy rate per spike of the multilayered network is
considerably smaller than in random networks. A reason for that might
be the occurrence of very high firing rate in some neurons with low
firing rates in other neurons, possibly because the balance inequality
(Eq.~\ref{eq:theta-balance_inequality}) is not satisfied. The dynamical
entropy rate per spike is known to decrease for high firing rates
(See \textbf{Fig.~\ref{fig:lambda_max_K-r}}).

\begin{table}
\begin{tabular}{|c|c|c|c|c|c|c|c|c|c|}
\hline 
Probability & \multicolumn{9}{c|}{from}\tabularnewline
\hline 
\hline 
\multirow{9}{*}{to} &  & L2/3e & L2/3i & L4e & L4i & L5e & L5i & L6e & L6i\tabularnewline
\cline{2-10} \cline{3-10} \cline{4-10} \cline{5-10} \cline{6-10} \cline{7-10} \cline{8-10} \cline{9-10} \cline{10-10} 
 & L2/3e & 0.101  & 0.169  & 0.044  & 0.082  & 0.032 & 0  & 0.008  & 0\tabularnewline
\cline{2-10} \cline{3-10} \cline{4-10} \cline{5-10} \cline{6-10} \cline{7-10} \cline{8-10} \cline{9-10} \cline{10-10} 
 & L2/3i & 0.135  & 0.137  & 0.032  & 0.052  & 0.075  & 0 & 0.004 & 0\tabularnewline
\cline{2-10} \cline{3-10} \cline{4-10} \cline{5-10} \cline{6-10} \cline{7-10} \cline{8-10} \cline{9-10} \cline{10-10} 
 & L4e & 0.008  & 0.006  & 0.050  & 0.135  & 0.007 & 0.0003  & 0.045  & 0\tabularnewline
\cline{2-10} \cline{3-10} \cline{4-10} \cline{5-10} \cline{6-10} \cline{7-10} \cline{8-10} \cline{9-10} \cline{10-10} 
 & L4i & 0.069 & 0.003 & 0.079 & 0.160 & 0.003 & 0  & 0.106 & 0\tabularnewline
\cline{2-10} \cline{3-10} \cline{4-10} \cline{5-10} \cline{6-10} \cline{7-10} \cline{8-10} \cline{9-10} \cline{10-10} 
 & L5e & 0.10 & 0.062 & 0.051  & 0.006 & 0.083 & 0.373  & 0.020  & 0\tabularnewline
\cline{2-10} \cline{3-10} \cline{4-10} \cline{5-10} \cline{6-10} \cline{7-10} \cline{8-10} \cline{9-10} \cline{10-10} 
 & L5i & 0.055 & 0.027 & 0.026  & 0.002 & 0.060  & 0.316 & 0.009  & 0\tabularnewline
\cline{2-10} \cline{3-10} \cline{4-10} \cline{5-10} \cline{6-10} \cline{7-10} \cline{8-10} \cline{9-10} \cline{10-10} 
 & L6e & 0.016 & 0.007 & 0.021  & 0.017  & 0.057 & 0.020  & 0.040  & 0.225\tabularnewline
\cline{2-10} \cline{3-10} \cline{4-10} \cline{5-10} \cline{6-10} \cline{7-10} \cline{8-10} \cline{9-10} \cline{10-10} 
 & L6i & 0.036 & 0.001  & 0.003 & 0.001 & 0.028 & 0.008 & 0.066 & 0.144\tabularnewline
\hline 
\end{tabular}~~~~~~~~~~~%
\begin{tabular}{|c|c|c|}
\hline 
 & number of neurons & external input \tabularnewline
\hline 
L2/3e & 20683 & 3.6923\tabularnewline
\hline 
L2/3i & 5834 & 12.272\tabularnewline
\hline 
L4e & 21915 & 4.5737 94.0675\tabularnewline
\hline 
L4i & 5479 & 16.5517 106.0455\tabularnewline
\hline 
L5e & 4850 & 19.6825\tabularnewline
\hline 
L5i & 1065 & 85.1521\tabularnewline
\hline 
L6e & 14395 & 9.6156 99.1084\tabularnewline
\hline 
L6i & 2948 & 34.0002 123.491\tabularnewline
\hline 
\end{tabular}

\caption{\label{tab:currents-1}Wiring probabilities and number of neurons
per population in cortical column model.}
\end{table}

\paragraph{Figure 4 f,~g of main paper: }

We also studied the dynamics of networks with a structured microscopic
architecture to corroborate our main results. We used second order
networks (SONETS), which are random networks where two synapse motifs
and connection probabilities are varied, while keeping higher order
structures random \citep{key-Nykamp}. This is achieved by using dichotomized
Gaussian random variables with a desired covariance structure to generate
the topology \citep{key-Macke}. In an Erd\H{o}s\textendash Rényi
graph only the connection probability $p=K/(N-1)=P(A_{ij}=1)$ is
fixed, hence for any two synapses, the joint probability of being
connected is $P(A_{ij}=1,\,A_{kl}=1)=p^{2}$. In SONETS, also the
joint probability of two connections $P(A_{ij}=1,\,A_{kl}=1)=p^{2}(1+\alpha_{x})$
is fixed, where $\alpha_{x}$ is the respective motif frequency $\alpha_{x}=\{\alpha_{\mathrm{reciprocal}},\alpha_{\mathrm{converging}},\alpha_{\mathrm{diverging}},\alpha_{\mathrm{chain}}\}$.
$\alpha_{\mathrm{converging}}$ is proportional to the variance of
the indegree $K_{\mathrm{in}}$, $\alpha_{\mathrm{diverging}}$ is
proportional to the variance of the outdegree $K_{\mathrm{out}}$
and $\alpha_{\mathrm{chain}}$ is proportional to the covariance of
the indegree $K_{\mathrm{in}}$ and the outdegree $K_{\mathrm{out}}$.
Using this approach, we interpolated excitatory-excitatory connectivity
from a random graph to a graph with the second order motif structure
found experimentally in superficial cortical layers \citep{key-Song}.
Again we fixed the average network firing rate by adapting the external
current $I_{\mathrm{ext}}$. We used excitatory-inhibitory networks
with $N_{I}=2000$, $N_{E}=8000$, $\eta=0.9$, $\varepsilon=0.3$.
The excitatory-excitatory adjacency matrix was interpolated between
a directed Erd\H{o}s\textendash Rényi graph and a SONET with the experimentally
found motif structure. While varying second order motifs can generally
change the largest Lyapunov exponent, dynamical entropy and the attractor
dimensionality by a factor of $\approx3$, we demonstrate that the
importance of action potential onset rapidness prevails.

\section{Experimentally measured action potential (AP) onset rapidness in
cortical circuits\label{sec:Experimentally-measured-action}}

\paragraph{}

Action potential initiation is an important bottleneck for cortical
information transmission. Only the information a neuron encodes in
its spike train can be used by its local network, by subsequent processing
stages and, ultimately, to guide behavior. Experimental studies estimated
that the spiking output of a cortical neuron contains twenty- to one
hundred fold less information about the synaptic input than its membrane
potential \citep{key-Polavieja2005}. This might not come as a surprise
because the membrane potential carries more information about the
dense stream of incoming postsynaptic potentials than the temporally
sparse sequence of outgoing action potentials. However, the finding
highlights the gatekeeping function of the action potential generation
mechanism for the information transmission: it decides which aspects
of the membrane potential are reflected in the outgoing spike train.

In this paragraph, we want to relate the rapidness in the idealized
mathemtically tractable rapid theta model with realistic values of
spike onset rapidness based on effective neurons models fitted to
experimental data.

Experimental findings revealed a surprisingly broad encoding bandwidth
of cortical neurons:\textbf{ }high-frequency input components of a
stimulus immersed in noise are reliably encoded in the outgoing spike
trains up to frequencies of several hundred Hertz. This has been first
reported in acute slice preparations of regular-spiking layer 5 pyramidal
cells of the rat somatosensory cortex for mean-modulated fluctuations
\citep{key-K=0000F6ndgen2008} and was also later found for variance-modulated
fluctuations later \citep{key-Boucsein}. Independent studies confirmed
the broad bandwidth using a protocol with weaker stimulation strength
both in the time and frequency domain \citep{key-Silberberg,key-Higgs,key-Tchumatchenko,key-Ilin}.

\paragraph{Theoretical predictions}

Theoretical studies predict that a rapid spike onset\textbf{ }is necessary
for the ultra-fast response in a feedforward architecture. Ensembles
of neurons with instantaneous spike onset, such as leaky integrate-and-fire
neurons \citep{key-Lapicque1907}, can transmit signals in the variance
channel unattenuated for arbitrarily high frequencies \citep{key-Brunel1999,key-Lindner2001}.
For the mean-modulation channel, however, the output amplitude declines
$\propto\frac{1}{\sqrt{f}}$ \citep{key-Brunel1999}. More generally,
high spike onset rapidness increases the population encoding bandwidth.
This relationship between broad encoding bandwidth and high action
potential (AP) onset rapidness was first directly demonstrated in
the exponential integrate-and-fire model, whose AP onset rapidness
is changeable \citep{key-Fourcaud-Trocm=0000E92003}. A strong influence
of spike onset on high-frequency encoding was also predicted by Naundorf
and colleagues \citep{key-Naundorf2005,key-Naundorf2006}. Wei and
Wolf confirmed this analytically using a mathematically tractable
piecewise linear neuron model, which allows an analytical calculation
of the frequency response for different AP onset rapidness \citep{key-wei-2011}.
In numerical simulations of multi-compartment conductance-based models,
which reproduce the initiation of spikes in the axon initial segment,
fast AP onset at the initiation site was necessary for encoding high
frequencies \citep{key-Ilin}.

\paragraph*{Experimental confirmations}

These theoretical predictions on the importance of rapid AP onset
for high-frequency encoding were confirmed in several experiments.\textbf{
}In a recent study, different ways of decreasing the AP onset rapidness
all impaired the ability to encode high-frequency stimulus components
into the spike train \citep{key-Ilin}. In this experiment, the AP
onset rapidness was decreased in slices of rat visual cortex first
by decreasing extracellular sodium concentration by partially substituting
NaCl by choline chloride in the extracellular solution. Secondly,
the effective density of voltage-gated sodium channels (Na$_{V}$)
was reduced by blocking voltage-gated sodium channels using small
concentrations of tetrodotoxin (TTX) locally at the site of the axon
initial segment (AIS). Both manipulations had the effect of impairing
the high-frequency encoding. A similar impaired high-frequency encoding
was observed in neurons from juvenile animals (P9 - P13), which naturally
have a slower AP onset \citep{key-Ilin}. However, the underlying
mechanism of the rapid spike onset is still a largely open question
of interest for both biophysical modeling and electrophysiological
experiments.

\paragraph*{\textbf{Experimentally measured AP onset rapidness}}

While it is experimentally challenging to measure the transmembrane
potential at the axon initial segment, somatic measurements of the
membrane potential are routinely performed. Mathematical neuron models
of different complexity have been fitted to the membrane potential
for a time-varying input stimulus. A univariate integrate-and-fire
type model, that has been fitted to experimental data with surprisingly
accurate results is the exponential integrate-and-fire neuron \citep{key-Fourcaud-Trocm=0000E92003}. 

\begin{equation}
\frac{dV}{dt}=\frac{1}{\tau_{\mathrm{m}}}\left(E-V+\triangle_{T}\exp\left(\frac{V-V_{T}}{\triangle_{T}}\right)\right),\label{eq:rapid-ExpIF}
\end{equation}

$\triangle_{T}$ is a paramter to vary the instability at the spike
onset which is called spike slope factor. Small $\triangle_{T}$ corresponds
to large values of spike onset rapidness.

Various experiments fitted recorded voltage traces based on time-varying
injected currents \citep{key-Badel-2008,key-Badel-2008b,key-Harrison2015,key-Kaufmann2016,key-Doose2016,key-Pozzorini2015,key-Hert=0000E4g}
(see table \ref{tab:measured-rapidness}). We used two methods to
convert the experimentally measured values of spike onset rapidness
of theses fitted exponential integrate-and-fire models into spike
onset rapidness of the rapid theta model. In the first method, EIF
and rapid theta model have the same slope $\frac{d\dot{V}}{dV}$ at
the unstable fixed point (in case of no external input). In order
to do so, we calculated the slope of the $\dot{V}-V-$curve of the
fitted EIF neuron for $\dot{V}=0$ and converted to dimensionless
variables by multiplying by $\tau_{m}$: 
\[
r_{\dot{V}=0}=\frac{d\dot{V}}{dV}(\dot{V}=0)\cdot\tau_{\mathrm{m}}
\]
In the second method, the EIF model and rapid theta model have the
same curvature at the vertex, where $\dot{V}$ takes its minimum value,
which is in case of the rapid theta model also the glue point $V_{\mathrm{G}}$.
The value $r_{V_{G}}$ is thus given by:

\[
r_{V_{G}}=\frac{V_{T}\text{\textminus}E}{\text{\ensuremath{\Delta}}T}
\]

Table \ref{tab:measured-rapidness} gives both $r_{\dot{V}=0}$ and
$r_{V_{G}}$ converted from the EIF model fitted to experimental data
recorded in various cell types.

\begin{table}
\begin{tabular}{|l|l|l|l|l|l|l|l|}
\hline 
$\text{\ensuremath{\Delta}}T$ (mV) & $V_{T}\text{\textminus}E$ (mV) & $r_{V_{G}}$ & $r_{\dot{V}=0}$ & neurons & species & cell type/location & Reference\tabularnewline
\hline 
\hline 
$1.2\pm0.4$ & $12.6$ & $10.5$ & 12.1 & $n=14$ & mouse & layer 5 pyramidal neurons (pn) & \citep{key-Badel-2008}\tabularnewline
\hline 
$1.34\pm0.550$ & $29.8$ & $22.2$ & $24.5$ & $n=31$ & \multirow{4}{*}{male Wistar rats} & layer 2/3 pn & \citep{key-Harrison2015}\tabularnewline
\cline{1-5} \cline{2-5} \cline{3-5} \cline{4-5} \cline{5-5} \cline{7-8} \cline{8-8} 
$1.28\pm0.394$ & $23.2$ & $18.2$ & 20.2 & $n=29$ &  & layer 4 pn & \citep{key-Harrison2015}\tabularnewline
\cline{1-5} \cline{2-5} \cline{3-5} \cline{4-5} \cline{5-5} \cline{7-8} \cline{8-8} 
$1.35\pm0.523$ & $20.1$ & $14.9$ & 16.8 & $n=29$ &  & slender-tufted layer 5 pn & \citep{key-Harrison2015}\tabularnewline
\cline{1-5} \cline{2-5} \cline{3-5} \cline{4-5} \cline{5-5} \cline{7-8} \cline{8-8} 
$1.16\pm0.479$ & $15.7$ & $13.5$ & 14.8 & $n=47$ &  & thick-tufted layer 5 pn & \citep{key-Harrison2015}\tabularnewline
\hline 
$1.00\text{\ensuremath{\pm}}0.35$ & $20.4$ & $20.4$ & 22.6 & $n=5$ & male B6CBF1 mice & thick-tufted layer 5 pn & \citep{key-Kaufmann2016}\tabularnewline
\hline 
\multicolumn{4}{|l|}{only histogram} & $n=10$ & male C57Bl/6J mouse & layer 5 pn & \citep{key-Pozzorini2015}\tabularnewline
\hline 
\multicolumn{4}{|l|}{only histogram} & $n=6$ &  & GABAergic cortical interneurons & \citep{key-Badel-2008b}\tabularnewline
\hline 
\end{tabular}

\caption{\label{tab:measured-rapidness}Experimentally measured values of action
potential onset rapidness, where $r_{\dot{V}=0}=\frac{d\dot{V}}{dV}(\dot{V}=0)\cdot\tau_{\mathrm{m}}$
and $r_{V_{G}}=\frac{V_{T}\text{\textminus}E}{\text{\ensuremath{\Delta}}T}$}
\end{table}

These experimental findings unanimously indicate high values of spike
onset rapidness obtained by both methods ($r\in[10.5,22.6]$) - more
than one order of magnitude higher than the classical theta model
with $r=1$ \citep{Skey-thetaneuron}. This is consistent with earlier
reports of high spike onset rapidness \citep{key-Naundorf2005,key-Naundorf2006,key-Badel-2008,key-Badel-2008b}
and consistent with a strong voltage dependence of the escape rate
in spike response models \citep{key-Pozzorini2015}. The experimentally
fitted values of spike onset rapidness are inside the sparse chaos
regime we found in balanced networks of rapid theta neurons (Figure
5 c and d of main paper). Our scaling analysis shows that the peak
rapidness $r_{\mathrm{peak}}$ is independent of network size $N$
for large networks but depends on membrane time constant, coupling
strength, firing rate and the mean number of synapses per neuron.
As these parameters do not vary by orders of magnitude across the
cortex, we propose that many cortical circuits operate in the sparse
chaos regime.

How well streams of spikes from one circuit can control spiking dynamics
in the subsequent circuit limits its ability to encode information.
It is presumably harder to control very chaotic networks by input
spike trains. Therefore, we conjecture that high spike onset rapidness
facilitates network state control and information transmission of
subsequent circuits.

\section{Minimal example for Julia \label{sec:Minimal-example-for}}

The following code for Julia (\url{www.julialang.org}) Version 1.0
demonstrates the event-based simulations by calculating the Poincaré
sections displayed in \textbf{Fig.~\ref{fig:fluxtube-poincare-sections-N3}a}.
Performant code for calculating Lyapunov spectra is available upon
request.

\begin{minted}[mathescape]{julia}
using PyPlot
function poincare()
Ncalc = 10^7                               # number of spikes in calculation
A = 0 .<[0 0 0;1 0 1;0 1 0]                # define connectivity matrix
phi = rand(3)                              # initialize neurons
pAll = Float64[]
    for s = 1:Ncalc
        pMax,j = findmax(phi)              # find next spiking neuron j
        dt = pi/2-pMax                     # calculate next spike time
        phi.+= dt                          # evolve phases till next spike time
        p = A[:,j]                         # postsynaptic neurons
        phi[p] = atan.(tan.(phi[p]).-1)    # update postsynaptic neurons
        phi[j] = -pi/2                     # reset spiking neuron to -$\pi/2$
       j==1 && append!(pAll,phi[2:3])      # save neuron 2 & 3 whenever neuron 1 spikes
    end
     plot(2pAll[1:2:end], 2pAll[2:2:end],",k",alpha=0.01); axis("off")
end
\end{minted}

\section{Scale invariances of the theta model}

We note that the rapid theta in the parametrization introduced in
the main text has several scale-invariances. This means, certain reparametrizations
of pairs of the network parameters $\tau_{\textnormal{\ensuremath{\mathrm{m}}}}$,
$J_{0}$ and $I$ give rise to exactly the same network dynamics with
identical spike times. We now demonstrate these scale-invariances
and describe their implications for the interpretation of the results.
We first consider the QIF model:

\[
\tau_{\mathrm{m}}\frac{\textrm{d}V}{\textrm{d}t}=V^{2}+I_{\mathrm{ext}}
\]
where $I_{\mathrm{ext}}=I_{0}\sqrt{K}=\tau_{\mathrm{m}}\sqrt{K}J_{0}\nu_{\textrm{w}}$.
So

\[
\tau_{\mathrm{m}}\frac{\textrm{d}V}{\textrm{d}t}=V^{2}+\tau_{\mathrm{m}}\sqrt{K}J_{0}\nu_{\textrm{w}}
\]

\subsection{Scaling $J_{0}$ and $\nu_{\textrm{w}}$ by factor $\alpha>0$}

\[
\tau_{\mathrm{m}}\frac{\textrm{d}V}{\textrm{d}t}=V^{2}+\alpha^{2}\tau_{\mathrm{m}}\sqrt{K}J_{0}\nu_{\textrm{w}}
\]

Rescaling voltage and time using $V=\alpha\hat{V}$and $\hat{t}=\alpha t$,
with $d\hat{t}=\alpha dt$ :

\[
\tau_{\mathrm{m}}\alpha\frac{\textrm{d}\hat{V}}{\textrm{d}\hat{t}/\alpha}=\alpha^{2}\hat{V}^{2}+\alpha^{2}\tau_{\mathrm{m}}\sqrt{K}J_{0}\nu_{\textrm{w}}
\]

After dividing both sides by $\alpha^{2}:$

\[
\tau_{\mathrm{m}}\frac{\textrm{d}\hat{V}}{\textrm{d}\hat{t}}=\hat{V}^{2}+\tau_{\mathrm{m}}\sqrt{K}J_{0}\nu_{\textrm{w}}
\]

Thus, we arrived at identical dynamics for $\hat{V}$after rescaling
time. Note that when receiving input spikes, the voltage experiences
a jump of $-\frac{\alpha J_{0}}{\sqrt{K}}$(due to the scaling of
$J_{0}).$ Given the voltage rescaling $V=\alpha\hat{V}$ this corresponds
to a jump of $-\frac{J_{0}}{\sqrt{K}}$ in the rescaled voltage $\hat{V}$.
This scale invariance arises from the homogeneity of the QIF voltage
dynamics, where all terms scale consistently under the transformations
$V\rightarrow\alpha\hat{V}$ and $t\rightarrow\alpha\hat{t}$. Specifically,
if $F(V)=V^{2}$, then rescaling $V$ by $\alpha$ results in $\alpha^{2}\hat{V}^{2}$,
and the $\frac{dV}{dt}$ term also scales by $\alpha^{2}$ after the
time rescaling. Moreover, as for QIF, reset and threshold are $-\infty$
and $\infty$ respectively, voltage can be rescaled. This invariance
generally does not hold for neuron models whose dynamics lack this
homogeneity or who have a finite reset $V_{re}$ or threhold $V_{th}$.
In the case, of finite $V_{re}$ or $V_{th}$, these also have to
be rescaled accordingly to find the identical network dynamics.

\subsection{Scaling $\tau_{\mathrm{m}}$ and $J_{0}$ by a Factor $\beta>0$}

Consider scaling $\tau_{\mathrm{m}}$ to $\beta\tau_{\mathrm{m}}$
and $J_{0}$to $\beta J_{0}$:

\[
\beta\tau_{\mathrm{m}}\frac{\textrm{d}V}{\textrm{d}t}=V^{2}+\beta^{2}\tau_{\mathrm{m}}\sqrt{K}J_{0}\nu_{\textrm{w}}
\]

Rescaling voltage:$V=\beta\hat{V}:$

\[
\beta^{2}\tau_{\mathrm{m}}\frac{\textrm{d}\hat{V}}{\textrm{d}t}=\beta^{2}\hat{V}^{2}+\beta^{2}\tau_{\mathrm{m}}\sqrt{K}J_{0}\nu_{\textrm{w}}
\]

Dividing by $\beta^{2}:$

\[
\tau_{\mathrm{m}}\frac{\textrm{d}\hat{V}}{\textrm{d}t}=\hat{V}^{2}+\tau_{\mathrm{m}}\sqrt{K}J_{0}\nu_{\textrm{w}}
\]

Note that time is not rescaled.

\subsection{Scaling $\tau_{\mathrm{m}}$ and $\nu_{\textrm{w}}$ by a Factor
$\gamma>0$ and $\frac{1}{\gamma}$}

Consider scaling $\tau_{\mathrm{m}}$ to $\gamma\tau_{\mathrm{m}}$
and $\nu_{\textrm{w}}$to $\frac{\nu_{\textrm{w}}}{\gamma}$ :

\[
\gamma\tau_{\mathrm{m}}\frac{\textrm{d}V}{\textrm{d}t}=V^{2}+\gamma\tau_{\mathrm{m}}\sqrt{K}J_{0}\frac{\nu_{\textrm{w}}}{\gamma}
\]

Rescaling time: $t=\gamma\hat{t}$, so $dt=\gamma d\hat{t}$:

\[
\tau_{\mathrm{m}}\frac{\textrm{d}V}{\textrm{d}\hat{t}}=V^{2}+\tau_{\mathrm{m}}\sqrt{K}J_{0}\nu_{\textrm{w}}
\]

Thus, the dynamics of $V$ as a function of the rescaled time $\hat{t}$are
identical to the original dynamics of $V$ as a function of $t$.

In conclusion, we find three interrelated scale-invariances of the
QIF model. The last scale-invariance is generic and also exists for
other integrate-and-fire type models, as it relies on rescaling the
time, related to the arbitrariness in which units time is measured
($\tau_{\mathrm{m}}$ is the only timescale of the system and could
be in milliseconds, seconds or hours). The other two rely on the homogeneity
of the QIF dynamics.

\subsection{Scale invariances of the rapid theta model}

The same three scale invariances also hold for the rapid theta model

\begin{equation}
\tau_{\mathrm{m}}\frac{\mathrm{d}V}{\mathrm{d}t}=\begin{cases}
a_{\mathrm{S}}(V-V_{\mathrm{G}})^{2}+\tau_{\mathrm{m}}\sqrt{K}J_{0}\nu_{\textrm{w}} & \text{for }V\le V_{\mathrm{G}}\\
a_{\mathrm{U}}(V-V_{\mathrm{G}})^{2}+\tau_{\mathrm{m}}\sqrt{K}J_{0}\nu_{\textrm{w}} & \text{for }V>V_{\mathrm{G}}.
\end{cases}\label{eq:rapid-theta-eq-V-3}
\end{equation}

Let's define center the voltage $\tilde{V}=V-V_{G}$. The original
equation becomes:

\begin{equation}
\tau_{\mathrm{m}}\frac{\mathrm{d}\tilde{V}}{\mathrm{d}t}=\begin{cases}
a_{\mathrm{S}}\tilde{V}^{2}+\tau_{\mathrm{m}}\sqrt{K}J_{0}\nu_{\textrm{w}} & \text{for }\tilde{V}\le0\\
a_{\mathrm{U}}\tilde{V}^{2}+\tau_{\mathrm{m}}\sqrt{K}J_{0}\nu_{\textrm{w}} & \text{for }\tilde{V}>0.
\end{cases}\label{eq:rapid-theta-eq-V-3-1}
\end{equation}

We observe that the piecewise equation for $\tilde{V}$obeys the same
homogeneity.

Now considering multiplying $J_{0}$ and $\nu_{\textrm{w}}$ by a
factor $\alpha$:

\begin{equation}
\tau_{\mathrm{m}}\frac{\mathrm{d}\tilde{V}}{\mathrm{d}t}=\begin{cases}
a_{\mathrm{S}}\tilde{V}^{2}+\alpha^{2}\tau_{\mathrm{m}}\sqrt{K}J_{0}\nu_{\textrm{w}} & \text{for }\tilde{V}\le0\\
a_{\mathrm{U}}\tilde{V}^{2}+\alpha^{2}\tau_{\mathrm{m}}\sqrt{K}J_{0}\nu_{\textrm{w}} & \text{for }\tilde{V}>0.
\end{cases}\label{eq:rapid-theta-eq-V-3-1-1}
\end{equation}

Then rescaling voltage $\tilde{V}=\alpha\hat{V}$and rescaling time
$\hat{t}=\alpha t$, so $d\hat{t}=\alpha dt$ gives

\begin{equation}
\alpha^{2}\tau_{\mathrm{m}}\frac{\mathrm{d}\tilde{V}}{\mathrm{d}t}=\begin{cases}
a_{\mathrm{S}}\alpha^{2}\tilde{V}^{2}+\alpha^{2}\tau_{\mathrm{m}}\sqrt{K}J_{0}\nu_{\textrm{w}} & \text{for }\tilde{V}\le0\\
a_{\mathrm{U}}\alpha^{2}\tilde{V}^{2}+\alpha^{2}\tau_{\mathrm{m}}\sqrt{K}J_{0}\nu_{\textrm{w}} & \text{for }\tilde{V}>0.
\end{cases}\label{eq:rapid-theta-eq-V-3-1-1-1}
\end{equation}

which yields the initial dynamics with rescaled time. The other two
scale invariances also hold for the rapid theta model. For the second
invariance (scaling $\tau_{m}$ and $J_{0}$ by $\beta$), the derivation
follows analogously to the QIF case, applied separately to both pieces
of the piecewise definition, leveraging the homogeneity on both branches.
Similarly, the third invariance (scaling $\tau_{m}$ and $\nu_{w}$)
holds due to the time rescaling argument, which is independent of
the specific form of voltage dynamics.

\end{document}